\documentclass{aa}  

\usepackage[]{natbib}
\usepackage{txfonts}
\usepackage{gensymb}
\usepackage{amsmath}
\usepackage{booktabs}
\usepackage{color}
\usepackage{caption}
\usepackage[export]{adjustbox}
\usepackage{graphicx,subfigure}
\usepackage{xcolor}
\usepackage{dblfloatfix}

\graphicspath{{figures/}} 

\begin{document} 

\title{SDSS-IV/SPIDERS: A Catalogue of X-Ray Selected AGN Properties}
\subtitle{Spectral Properties and Black Hole Mass Estimates for SPIDERS SDSS DR14 \textbf{Type 1} AGN}

\author{D. Coffey\inst{\ref{inst1}} \and M. Salvato\inst{\ref{inst1}} \and A. Merloni\inst{\ref{inst1}} \and Th. Boller\inst{\ref{inst1}} \and K. Nandra\inst{\ref{inst1}} \and T. Dwelly\inst{\ref{inst1}} \and J. Comparat\inst{\ref{inst1}} 
\and A. Schulze\inst{\ref{inst2}}
\and A. Del Moro \inst{\ref{inst1}} 
\and D. P. Schneider\inst{\ref{inst3}}}

\authorrunning{D. Coffey et al.}
\titlerunning{Properties of SPIDERS DR14 Quasars}

\institute{
Max-Planck-Institut f\"{u}r extraterrestrische Physik, Giessenbachstra{\ss}e 1, 85748 Garching, Germany\\
\email{coffeydg@mpe.mpg.de}\label{inst1}
\and
National Astronomical Observatory of Japan, Mitaka, Tokyo 181-8588, Japan\label{inst2}
\and
Department of Astronomy and Astrophysics, and the Institute for
Gravitation and the Cosmos, The Pennsylvania State University, 525 Davey Laboratory, University Park, PA 16802, USA\label{inst3}}

\date{Accepted 8 April 2019}

 \abstract
{This work presents the catalogue of optical spectral properties 
for all X-ray selected SPIDERS active galactic nuclei (AGN) 
up to SDSS DR14.
SPIDERS (SPectroscopic IDentification of \textit{eROSITA} Sources)
is an SDSS-IV programme that is currently conducting optical spectroscopy 
of the counterparts to the X-ray selected sources 
detected in the \textit{ROSAT} all-sky survey and the 
\textit{XMM-Newton} slew survey in the footprint of the 
Extended Baryon Oscillation Spectroscopic Survey (eBOSS).
The SPIDERS DR14 sample is the largest sample of 
X-ray selected AGN with optical spectroscopic follow-up to date.
The catalogue presented here is based on a clean sample 
of 7344 2RXS ($\rm \bar{z}=0.5$)
and 1157 \textit{XMM-Newton} slew survey ($\rm \bar{z}=0.4$)
type 1 AGN with spectroscopic coverage of the H$\beta$ and/or MgII
emission lines.
Visual inspection results for each object in this sample are available
from a combination of literature sources and the SPIDERS group,
which provide both reliable redshifts and source classifications.
The spectral regions around the H$\rm\beta$ and MgII emission lines
have been fit in order to measure both line and continuum properties, 
estimate bolometric luminosities,
and provide black hole mass estimates 
using the single-epoch (or photoionisation) method.
The use of both H$\rm\beta$ and MgII allows 
black hole masses to be estimated up to z\,$\simeq$\,2.5. 
A comparison is made between the spectral properties and black hole mass estimates 
derived from H$\rm\beta$ and MgII using the subsample of objects 
which have coverage of both lines in their spectrum.
These results have been made publicly available 
as an SDSS-IV DR14 value added catalogue.}


\maketitle

\section{Introduction}\label{intro}

\noindent
A crucial requirement for understanding AGN
evolution and demographics
is the ability to select a sample for study 
in a complete and unbiased way.
X-ray emission has been frequently used for AGN selection,  
and can distinguish the high energy emission 
associated with mass accretion by a black hole (BH)
from inactive galaxies and stars.
Combining wide-area X-ray surveys 
with the ability to classify large numbers of objects spectroscopically via 
the Sloan Digital Sky Survey 
\citep[SDSS,][]{2000AJ....120.1579Y,2006AJ....131.2332G}
provides a powerful tool for the study of AGN.

SPIDERS (SPectroscopic IDentification of \textit{eROSITA} 
Sources; PIs Merloni and Nandra) is an 
SDSS-IV \citep{2017AJ....154...28B} 
eBOSS \citep{2016AJ....151...44D} subprogramme
that is currently conducting optical spectroscopy of extragalactic X-ray detections 
in wide-area \textit{ROSAT} and \textit{XMM-Newton} surveys
\citep{2017MNRAS.469.1065D}.
Lying at the bright end of the X-ray source population, 
these sources will also be detected by \textit{eROSITA}
\citep{2012arXiv1209.3114M, 2016SPIE.9905E..1KP}.
The current SDSS DR14 \citep{2018ApJS..235...42A} SPIDERS sample
is a powerful resource for the multiwavelength analysis of AGN.
This work aims to capitalise on the wealth of information already available
by providing detailed optical spectral measurements,
as well as estimates of BH masses and Eddington ratios.

An accurate measurement of the central supermassive black hole (SMBH) mass
is necessary for the study of AGN 
and their coevolution with their host galaxies.
BH mass has been found to scale 
with a number of host galaxy spheroid properties;
stellar velocity dispersion
\citep[the $\rm M_{BH}-\sigma$ relation, e.g.][]{2000ApJ...539L..13G, 2001ApJ...547..140M, 2002ApJ...574..740T},
stellar mass \citep[e.g.][]{1998AJ....115.2285M},
and luminosity \citep[e.g.][]{1995ARA&A..33..581K}.
These correlations suggest a symbiotic evolution 
of SMBHs and their host galaxies.

Reverberation mapping (RM) has been used to measure the 
approximate radius of the broad-line region (BLR) in AGN
\citep[e.g.][]{1972ApJ...171..467B, 1982ApJ...261...35C, 1982ApJ...255..419B, 1993PASP..105..247P, 2015PASP..127...67B, 2015ApJS..216....4S}.
This technique involves measuring the time delay between 
variations in the continuum emission,
which is expected to arise from the accretion disk,
and the induced variations in the broad emission lines.
It was found that different emission lines have different time delays,
which is expected if the BLR is stratified, 
with lines of lower ionisation being emitted 
further from the central ionising source \citep[e.g.][]{1986ApJ...305..175G}.
For example, the high ionisation line CIV$\rm \lambda$1549 
has a shorter time delay than H$\rm \beta$ \citep{2000ApJ...540L..13P}.

The RM effort has also revealed a tight relationship between the 
continuum luminosity and the radius of the BLR \citep{2000ApJ...533..631K, 2006ApJ...644..133B, 2009ApJ...697..160B}.
Therefore, by using the measured luminosity as a proxy for the BLR radius,
and measuring the BLR line-of-sight velocity
from the width of the broad emission lines,
BH masses can be estimated from a single spectrum
\citep{2002ApJ...571..733V, 2002MNRAS.337..109M, 2006ApJ...641..689V, 2011ApJ...742...93A, 2012ApJ...753..125S, 2013BASI...41...61S}.
This approach is known as the single-epoch, or photoionisation, method.

Since H$\beta$ is the most widely studied RM emission line
it is therefore considered to be the most reliable line to use for single-epoch mass estimation. 
In addition, AGN H$\beta$ emission lines typically exhibit 
a clear inflection point between the broad and narrow line components,
making the virial full width at half maximum (FWHM) measurement 
relatively straightforward (see section~\ref{bl_decomp}).
The MgII line width correlates well with that of H$\beta$ 
(see section~\ref{compare_mass}),
and therefore MgII has also been used for single-epoch mass estimation
\citep[e.g.][]{2002MNRAS.337..109M}.
For SDSS spectra, either H$\beta$ or MgII 
is visible in the redshift range 0\,$\leq$\,z\,$\lesssim$\,2.5. 

At higher redshifts, 
the broad, high-ionisation line CIV$\lambda1549$ is available.
The CIV line width does not correlate strongly with that of low ionisation lines
\citep[e.g.][]{2005MNRAS.356.1029B, 2012MNRAS.427.3081T}
and this, along with the presence of a large blueshifted component
\citep[e.g.][]{2002AJ....124....1R}
makes it difficult to employ CIV for mass estimation.
A number of calibrations have been developed 
which aim to improve the mass estimates derived from CIV
\citep[][]{2012ApJ...759...44D, 2013MNRAS.434..848R, 2013ApJ...770...87P, 2017MNRAS.465.2120C},
however, whether CIV can provide reliable mass estimates when compared with 
low ionisation lines is still a subject of debate
\citep[see][]{2018MNRAS.478.1929M}.

This paper is organised as follows:
the selection of a reliable subsample of sources 
to be used for optical spectral fitting 
is discussed in section~\ref{data}.
Section~\ref{spec_fit} describes the method used 
to fit the H$\beta$ and MgII emission line regions.
The methods for estimating BH mass and bolometric luminosity 
are discussed in section~\ref{estimate_mass}.
The X-ray flux measurements used in this work are discussed in section~\ref{xray_flux}.
Section~\ref{access} describes where the catalogue containing the results of this work can be accessed.
A comparison between the UV and optical spectral fitting results 
is given in section~\ref{compare_uv_opt}.
Section~\ref{prop} provides a discussion of the sample properties, 
and finally, section~\ref{limits} includes a discussion of the reliability and limitations of
the fitting procedure.
 In order to facilitate a direct comparison with previous studies based on X-ray surveys,
a concordance flat $\rm \Lambda CDM$ cosmology was adopted
where $\Omega_{\rm M}$=0.3, $\Omega_{\Lambda}$=1-$\Omega_{\rm M}$,
and H$_{0}$=70\,km\,s$^{-1}$\,Mpc$^{-1}$.

\section{Preparing the Input Catalogue}\label{data}

\subsection{X-ray Data}

\noindent
The \textit{ROSAT} sample used in this work 
is part of the second \textit{ROSAT} all-sky survey (2RXS) 
catalogue \citep{2016A&A...588A.103B},
which has a limiting flux of $\rm \sim10^{-13}\,erg\,cm^{-2}\,s^{-1}$,
which corresponds to a luminosity of $\rm \sim10^{43}\,erg\,s^{-1}$ at z = 0.5.
Compared to the first \textit{ROSAT} data release \citep{1999A&A...349..389V},
the 2RXS catalogue is the result of an improved detection algorithm,
which uses a more detailed background determination 
relative to the original \textit{ROSAT} pipeline.
A full visual inspection of the 2RXS catalogue has been performed,
which provides a reliable estimate of its spurious source content
\citep[see][]{2016A&A...588A.103B}.
The first \textit{XMM-Newton} slew survey catalogue
release 1.6
\citep[XMMSL1;][]{2008A&A...480..611S} was also used in this work.
This catalogue includes observations made by 
the European Photon Imaging Camera (EPIC) pn detectors 
while slewing between targets,
and has a limiting flux of $\rm 6\times10^{-13}\,erg\,cm^{-2}\,s^{-1}$
in the soft band,
which corresponds to a luminosity of $\rm 5.8\times10^{44}\,erg\,s^{-1}$ at z=0.5.

\begin{figure*}[h]
	\centering
                \includegraphics[width=0.65\textwidth]{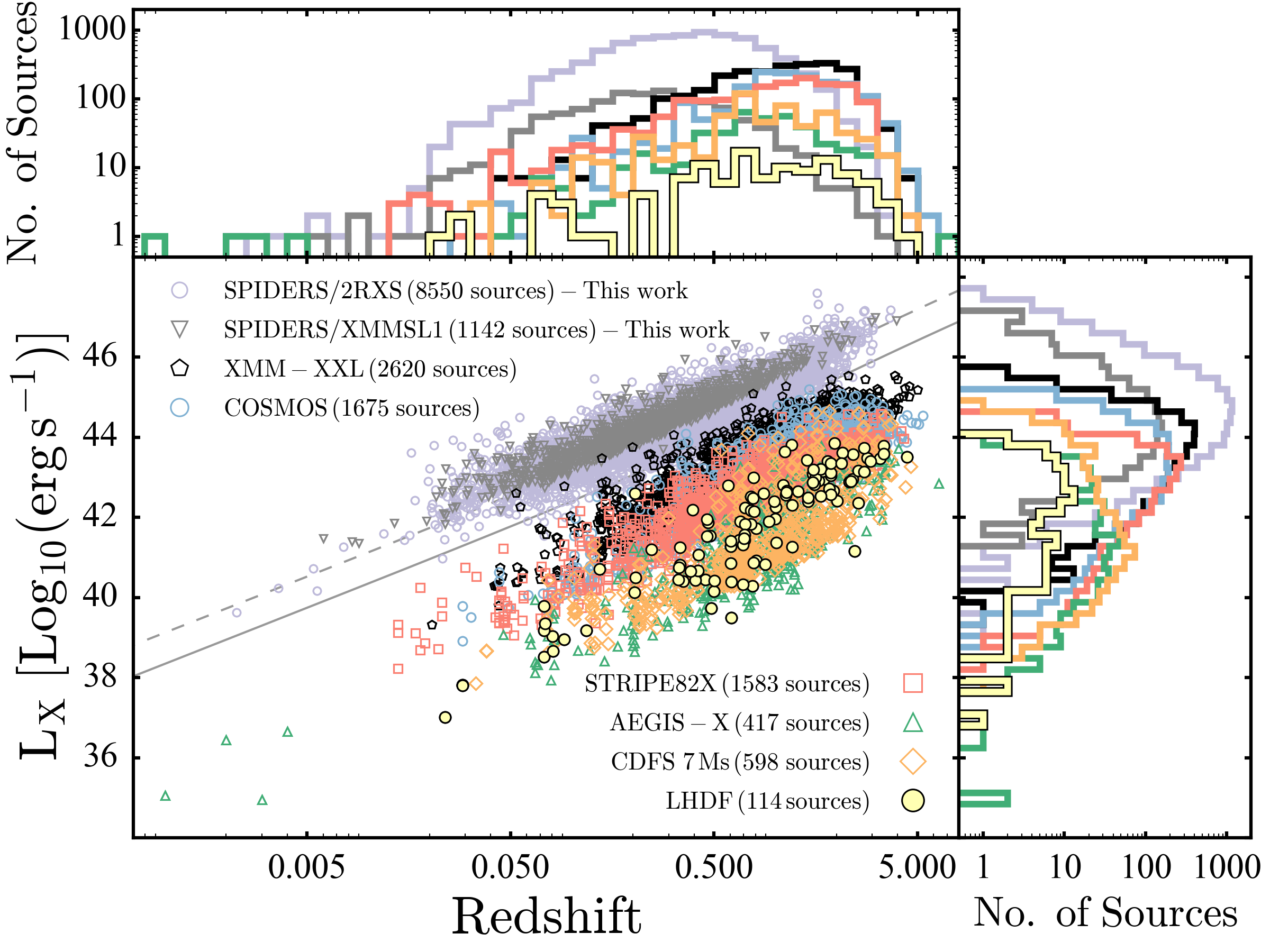}
\caption{\footnotesize
Soft X-ray luminosity versus spectroscopic redshift
for the samples presented in this work
and the following previously published X-ray selected samples;
XMM-XXL \citep{2016MNRAS.457..110M, 2016MNRAS.459.1602L},
CDFS \citep{2017ApJS..228....2L},
STRIPE82X \citep{2016ApJ...817..172L},
COSMOS \citep{2016ApJ...817...34M, 2016ApJ...830..100M, 2016ApJ...819...62C},
AEGIS-X \citep{2015ApJS..220...10N}, and the 
Lockman Hole deep field (LHDF) \citep{2008A&A...479..283B, 2012ApJS..198....1F}.
For each sample, the 0.5-2 keV luminosities are shown,
except for the 2RXS sample, where the 0.1-2.4 keV luminosities are shown,
and the XMMSL1 sample, 
where the 0.2-2 keV luminosities from \citet{2008A&A...480..611S} are shown.
The detection limit for the 2RXS and XMMSL1 samples 
are shown by the solid and dashed grey lines respectively.
The X-ray luminosities for the 2RXS sample are derived from the classical 
flux estimates described in section~\ref{2RXS_flux},
however it is noted here that some low count rate 
2RXS sources do not have flux estimates.
For sources that were detected in both 2RXS and XMMSL1,
only the XMMSL1 luminosities are shown.
Sources classified as stars have not been included in this figure.}
\label{compare_samples}
\end{figure*}

\subsection{The SPIDERS Programme}\label{spiders_prog}

The SPIDERS programme has been providing SDSS spectroscopic observations 
of 2RXS and XMMSL1 
sources\footnote{The SPIDERS programme targets both point-like and extended X-ray sources. This work focuses on the counterparts to point-like X-ray detections, 
which are predominantly AGN,
and therefore, the samples discussed in the subsequent paragraphs 
are derived from the SPIDERS-AGN programme.}
in the eBOSS footprint. 
Before the start of the eBOSS survey in 2014, 
the SPIDERS team compiled a sample of X-ray selected spectroscopic targets
and submitted this sample for spectroscopic follow-up using the BOSS spectrograph 
as part of the eBOSS/SPIDERS subprogramme
\citep[see][for further details on the SPIDERS programme]{2017MNRAS.469.1065D}.
As of the end of eBOSS in February 2019,
the eBOSS/SPIDERS survey has covered a sky area of $\rm 5321 \, deg^{2}$.
The SDSS DR14 SPIDERS sample presented in this work
covers an area of $\rm \sim2200 \, deg^{2}$ 
($\rm \sim 40\%$ of the final eBOSS/SPIDERS area).

The spectroscopic completeness achieved by the SPIDERS survey 
as of SDSS DR14 in the eBOSS area is 
$\sim 53\%$ for the sample as a whole, 
$\sim 63 \%$ considering only high-confidence X-ray detections 
(see section~\ref{clean_sample}),
and $\sim 87 \%$ considering sources with high-confidence X-ray detections and 
optical counterparts with magnitudes in the nominal survey limits 
($\rm 17\leq m_{\rm Fiber2,i}\leq22.5$).
Outside the eBOSS area, the spectroscopic completeness of this sample is lower:
$\sim 28\%$ for the sample as a whole, 
$\sim 39 \%$ considering only high-confidence X-ray detections,
and $\sim 57 \%$ considering sources with high-confidence X-ray detections and 
optical counterparts with magnitudes in the nominal survey limits.
The spectroscopic completeness of the SDSS DR16 SPIDERS sample 
inside and outside the eBOSS area is expected to be similar 
to that of the sample presented here.

In addition to those targeted during eBOSS/SPIDERS, 
a large number of 2RXS and XMMSL1 sources 
received spectra during the SDSS-I/II
\citep[2000-2008;][]{2000AJ....120.1579Y}
and the SDSS-III \citep{2011AJ....142...72E}
BOSS \citep[2009-2014;][]{2013AJ....145...10D} surveys.
This paper includes spectra obtained 
by eBOSS/SPIDERS up to DR14 (2014-2016) 
as well as spectra from SDSS-I/II/III.

\subsection{Identifying IR Counterparts}\label{nway_match}

To identify SPIDERS spectroscopic targets,
the Bayesian cross-matching algorithm ``NWAY'' \citep{2018MNRAS.473.4937S}
was used to select AllWISE \citep{2013yCat.2328....0C} 
infrared (IR) counterparts for the 2RXS and XMMSL1 
X-ray selected sources in the BOSS footprint.
The AllWISE catalogue consists of data obtained 
during the two main survey phases of the 
Wide-field Infrared Survey Explorer mission
\citep[WISE;][]{2010AJ....140.1868W}
which conducted an all-sky survey 
in the 3.4, 4.6, 12, and 22 $\rm \mu m$ bands 
(magnitudes in these bands are denoted [W1], [W2], [W3], and [W4] respectively).
The matching process used the colour-magnitude priors 
[W2] and [W2-W1] \citep[see][]{2017MNRAS.469.1065D}
which, at the depth of the 2RXS and XMMSL1 surveys, 
can distinguish between the correct counterparts and chance associations.
These colours would not be efficient if the 2RXS survey was much deeper
\citep[see][for a complete discussion]{2018MNRAS.473.4937S}.
The resulting 2RXS and XMMSL1 catalogues
with AllWISE counterparts contained 53455 and 4431 sources respectively.
AllWISE positions were then matched to 
photometric counterparts, where available, in SDSS.

\subsection{Comparison with Previous X-ray Surveys}

Figure~\ref{compare_samples} displays the 
sources in the 2RXS and XMMSL1 samples
which have spectroscopic redshifts and measurements of the soft X-ray flux.
For comparison, a series of previously published X-ray selected samples 
that have optical spectroscopic redshifts are also shown.
The large number of sources present in the 2RXS and XMMSL1 samples 
motivated the optical spectroscopic analysis discussed in the following sections.

\begin{figure*}[h]
\centering
\includegraphics[width=0.49\textwidth]{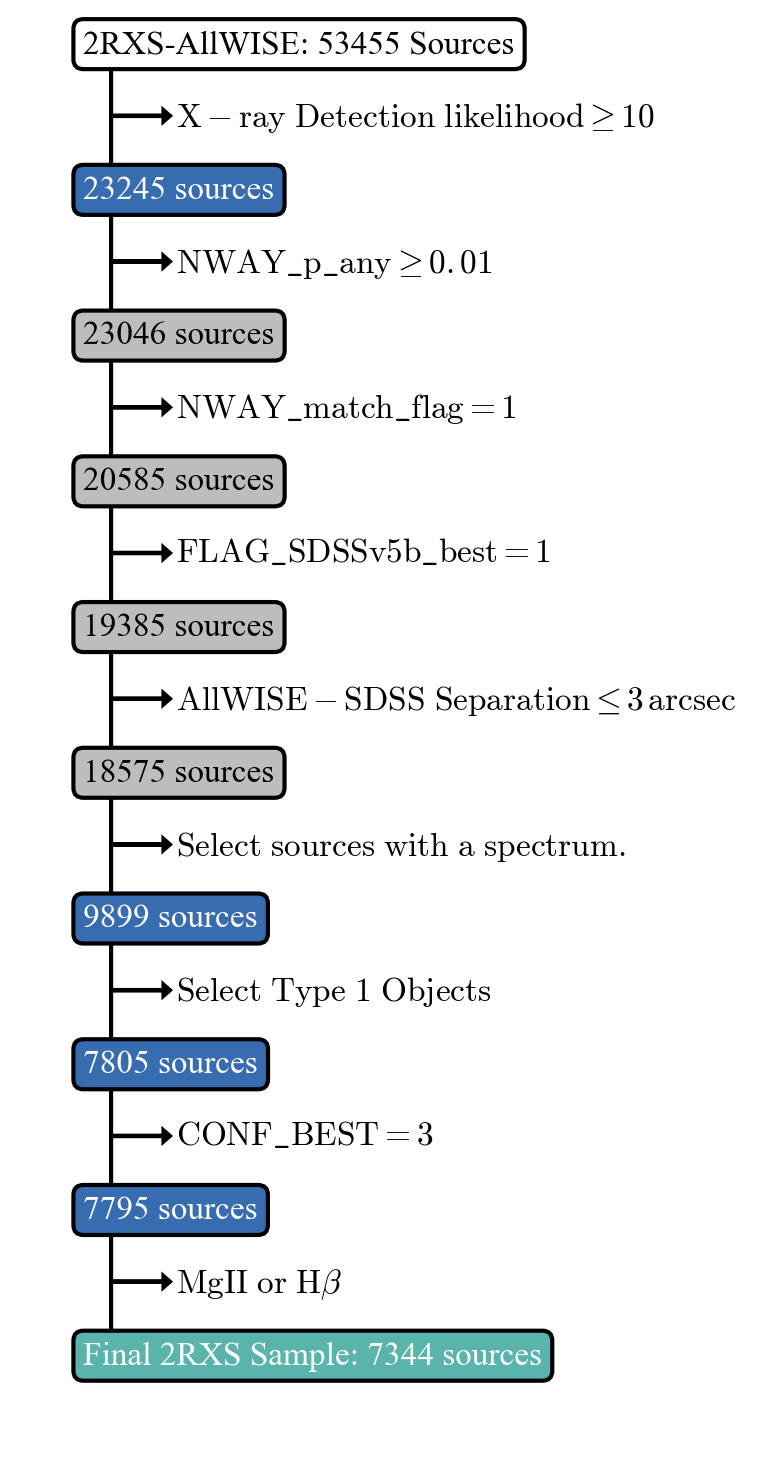}
\includegraphics[width=0.49\textwidth]{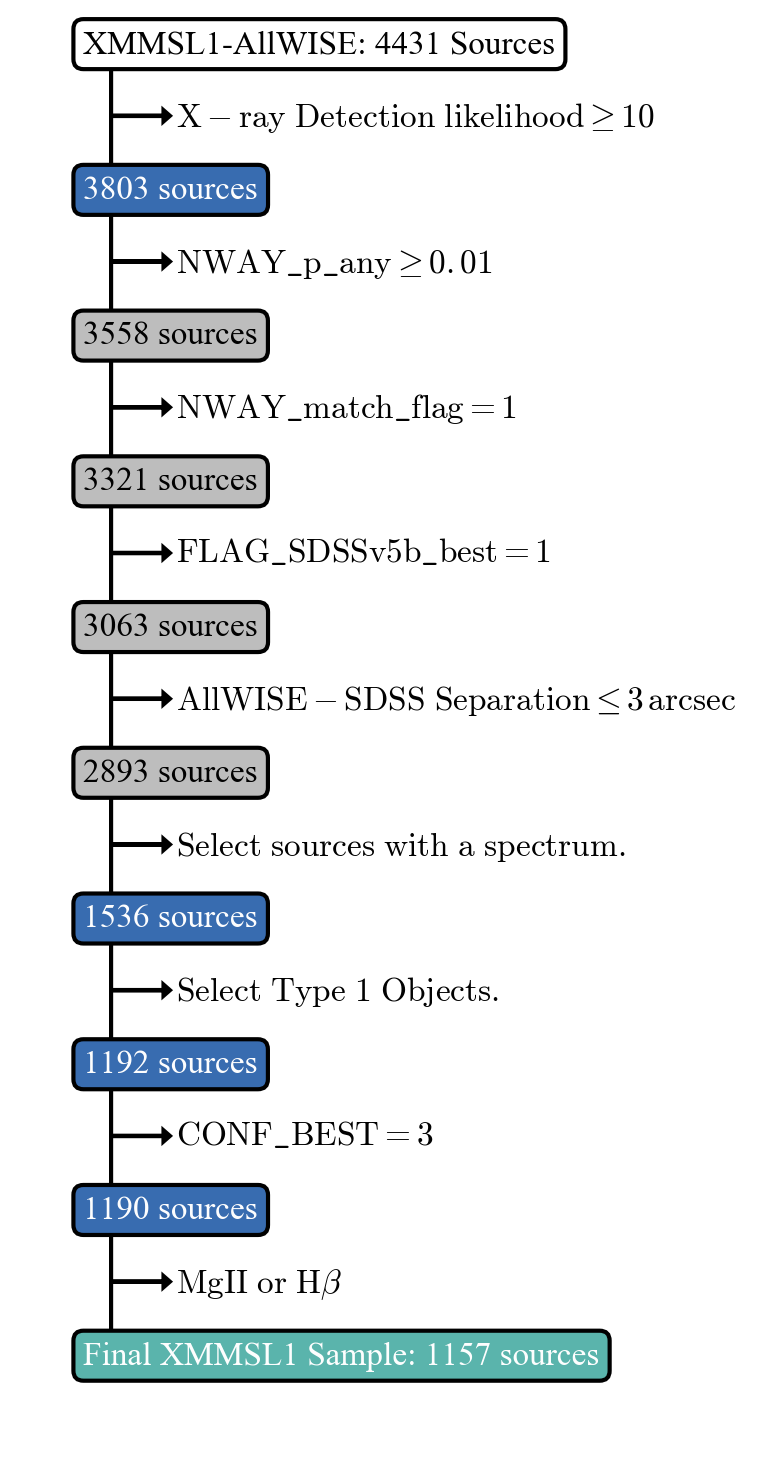}
\caption{\footnotesize 
Sequence of quality cuts applied to the 2RXS and XMMSL1 samples 
to produce the subsample used for spectral analysis.
The starting points (2RXS-AllWISE and XMMSL1-AllWISE)
are the full samples of 2RXS and XMMSL1 selected sources 
with AllWISE IR counterparts in the BOSS footprint
(see section~\ref{nway_match}).
The steps in grey are those that have been discussed 
in \citet{2017MNRAS.469.1065D}.}
\label{cuts}
\end{figure*}

\subsection{Selecting a Reliable Subsample}\label{clean_sample}

\noindent
The selection of SPIDERS spectroscopic targets 
was discussed in detail by \citet{2017MNRAS.469.1065D}.
This section summarises the selection steps 
discussed in detail by \citet{2017MNRAS.469.1065D}
and describes the additional cuts made in this work to 
select a sample for spectral analysis.
The sequence of selection criteria used and the resulting sample size 
are shown in figure~\ref{cuts}.

2RXS sources with an X-ray detection likelihood (EXI\_ML)
$\leq \,$10 were excluded since these detections are considered 
highly uncertain with a spurious fraction 
$\geq20\%$ \citep[see][]{2016A&A...588A.103B}. 
XMMSL1 sources with an X-ray detection likelihood 
(XMMSL\_DET\_ML\_B0) $\leq \,$10 were also excluded.
\citet[][figure 1]{2018MNRAS.473.4937S} show the 
distribution of flux with detection likelihood for both samples.
These cuts returned 23245/53455 2RXS and 3803/4431 XMMSL1 sources.

The following cuts, which were described in detail in \citet{2017MNRAS.469.1065D},
have also been applied to the sample:

\begin{itemize}
\item For each X-ray source, \citet{2018MNRAS.473.4937S} give 
the probability, p\_any, that a reliable counterpart exists among the 
possible AllWISE associations. Sources with p\_any < 0.01 were removed.
This returned 23046/23245 2RXS and 3558/3803 XMMSL1 sources. \\

\item For each X-ray source, the most probable AllWISE counterpart was chosen
by selecting sources with match\_flag=1.
This returned 20585/23046 2RXS and 3321/3558 XMMSL1 sources. \\

\item For each AllWISE counterpart, the brightest SDSS-DR13 photometric source within the AllWISE matching radius was selected using FLAG\_SDSSv5b\_best=1.
This returned 19385/20585 2RXS and 3063/3321 XMMSL1 sources. \\

\item Cases where the AllWISE-SDSS separation 
exceeded 3\,arcsec were removed.
This returned 18575/19385 2RXS and 2893/3063 XMMSL1 sources. \\
\end{itemize}

\noindent
The results of these constraints are displayed in grey in figure~\ref{cuts}.
As shown above, sources with match\_flag=1 were targeted;
however, for 14\% of the 2RXS sample and 10\% of the XMMSL1 sample,
more than one counterpart was highly likely.
This implies that either the counterpart association was not reliable, 
or that the X-ray detection was the result of emission from multiple sources.
These sources were not included in the discussion of 
optical spectral properties as a function of X-ray properties in section~\ref{aox_section}.
After selecting the brightest SDSS-DR13 \citep{2017ApJS..233...25A} 
photometric source, there were 15 cases where two unique 2RXS sources 
were matched to the same AllWISE/SDSS counterpart 
and 3 cases where two unique XMMSL1 sources 
were matched to the same AllWISE/SDSS counterpart. 
These sources were also removed.

\begin{figure*}
\centering
\includegraphics[width=0.49\textwidth]{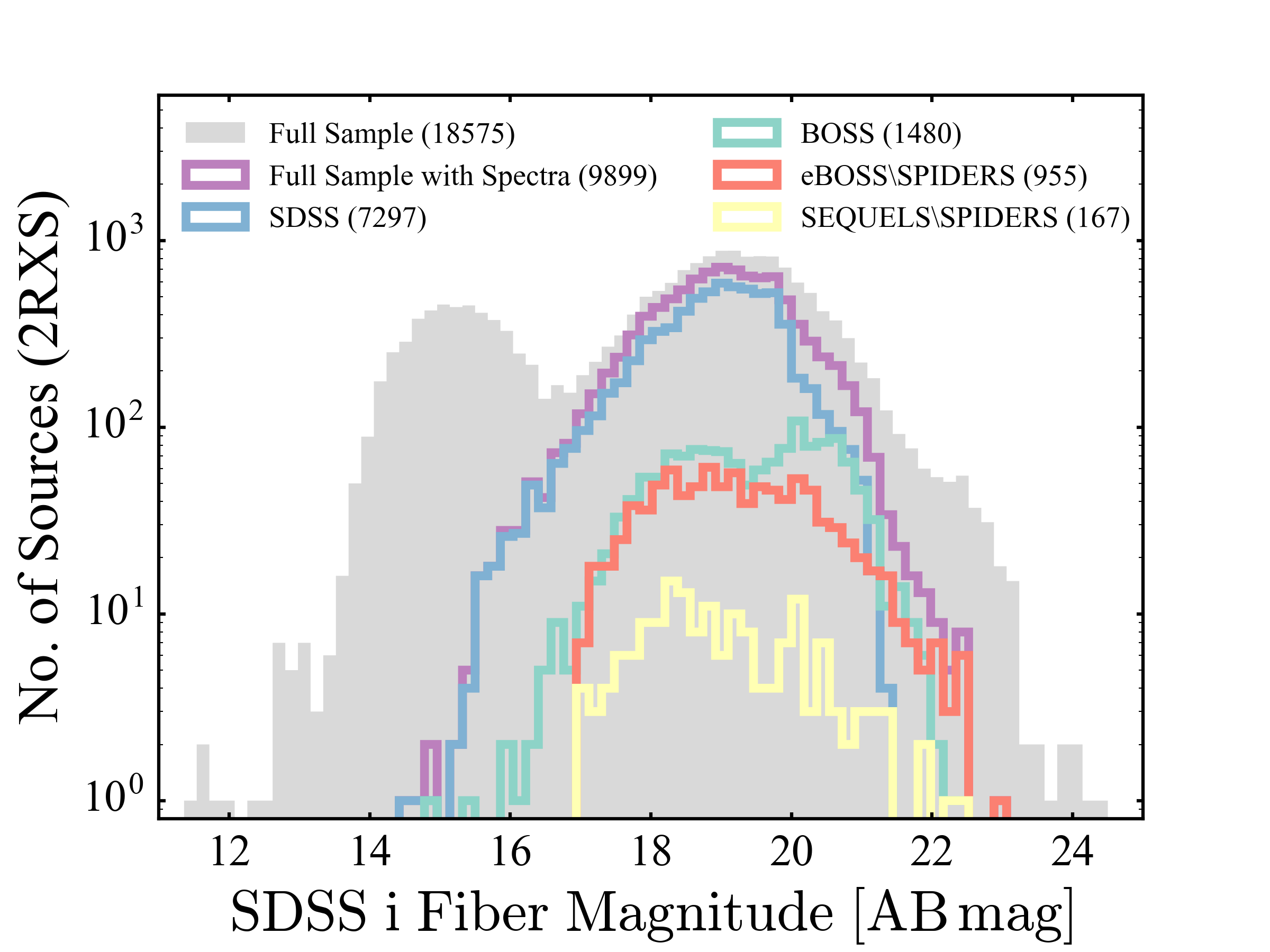}
\includegraphics[width=0.49\textwidth]{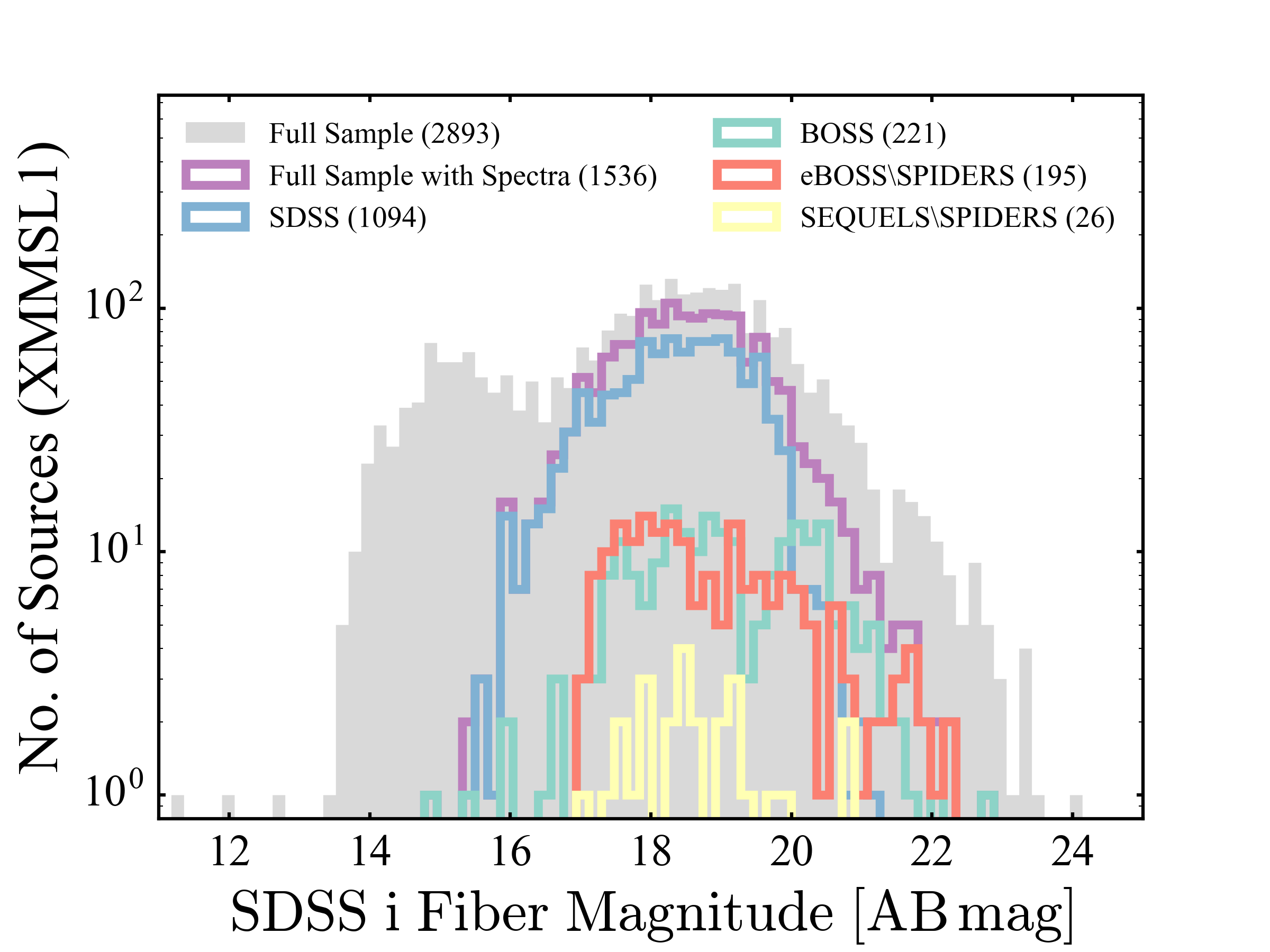}
\caption{\footnotesize 
Distribution of i-band fiber magnitudes (fiber2Mag).
The coloured curves represent all of the sources with spectra,
and the survey from which the spectra were taken.
The grey histogram displays the X-ray sources with a reliable SDSS photometric counterpart, including stars 
which cannot be targets for spectroscopy due to their brightness.}
\label{origin_spec}
\end{figure*}

Of these samples with reliable SDSS photometric counterparts, 
8777 2RXS and 1315 XMMSL1 sources have received spectra during SDSS-I/II/III 
while 1122 2RXS and 221 XMMSL1 sources have received spectra 
during the SPIDERS programme (including SEQUELS), resulting in a sample of 
9899 2RXS and 1536 XMMSL1 sources with spectra as of DR14.
The distribution of SDSS i band fiber2 magnitudes for this sample 
(showing the different spectroscopic programmes)
is presented in figure~\ref{origin_spec}.
Due to targeting constraints (as discussed in section~\ref{spiders_prog}), 
the sample completeness is much lower 
outside of the nominal magnitude limits for the survey 
($\rm 17\leq m_{\rm Fiber2,i}\leq22.5$ for eBOSS).

\subsection{Source Classification}\label{source_classification}

\noindent
Visual inspection results for each object in this sample 
are available from a combination of literature sources 
\citep{2007AJ....133..313A, 2010AJ....139..390P, 2010AJ....139.2360S, 2017A&A...597A..79P} and the SPIDERS group.
The SPIDERS visual inspection 
\citep[see][for further details]{2017MNRAS.469.1065D}
provides a visual confirmation of the SDSS pipeline redshift and object classification.
The results of this inspection include 
a flag indicating the confidence of the redshift, ``CONF\_BEST'',
which can take the values 3 (highly secure), 
2 (uncertain), 1 (poor/unusable), 0 (insufficient data).
A confirmation of the source classification 
was also added during the visual inspection, 
which uses the categories QSO, broad absorption line QSO (BALQSO), 
blazar, galaxy, star, and none.
\citet{2007AJ....133..313A} provide the broad line AGN (BLAGN) and narrow line AGN (NLAGN) classifications, which are defined based on the presence or absence of 
broad (FWHM\,>\,1000$\rm \,km\,s^{-1}$) permitted emission lines.

The main goal of this work is to analyse the type 1 AGN in the SPIDERS sample,
and therefore only sources that have been classified via their optical spectra 
as either ``BLAGN'' or ``QSO'' were selected for spectroscopic analysis.  
This returned 7805/9899 2RXS and 1192/1536 XMMSL1 sources.
Since the categories ``BLAGN'' and ``QSO'' are based on different classification criteria,
there will be some overlap between the two sets of sources.
Therefore, no distinction will be made between the two categories;
instead, both sets of objects will be considered type 1 AGN in this work.

\subsection{Contamination from Starburst Galaxies}

\noindent
Although our sample probes luminosity ranges
typically associated with AGN emission,
starburst galaxies are also powerful X-ray sources
and may be present as contaminants in our AGN sample.
The X-ray emission from starburst galaxies is expected to 
originate from a number of energetic phenomena 
including supernova explosions and X-ray binaries
\citep[e.g.][]{2002A&A...382..843P}.
Therefore, the X-ray emission from starburst galaxies 
can be expected to be correlated with the star formation rate (SFR).
Using their sample of luminous infrared galaxies, 
and a sample of nearby galaxies from \citet{2003A&A...399...39R},
\citet{2011A&A...535A..93P} found that the total SFR
is related to the soft X-ray luminosity as follows:

\begin{equation}\label{SFR}
\rm SFR_{IR+UV}\,(M_{\odot}\,yr^{-1}) = 3.4 \times 10^{-40}\,L_{0.5-2keV} \, (erg \, s^{-1})
\end{equation}

\noindent
\citet{2015A&A...579A...2I}, figure 3, show the specific 
SFR for the COSMOS \citep{2007ApJS..172....1S} 
and GOODS \citep{2004ApJ...600L..93G} surveys 
for a series of redshift bins in the range $\rm 0.2 <  z < 1.4$.
The peak of the redshift distribution of the 2RXS/XMMSL1 samples 
presented in this work is $\sim0.25$.
Therefore, assuming that the COSMOS/GOODS sample in the redshift bin 0.2-0.4
is a good representative of the 2RXS/XMMSL1 samples, 
the upper limit on the SFR that can be expected 
for galaxies in our sample is $\rm \sim50\,M_{\odot}\,yr^{-1}$.
According to equation~\ref{SFR}, this corresponds to a soft X-ray luminosity of 
$\rm \sim10^{41}\,erg\,s^{-1 }$,
which is below the lower range probed by our samples 
$\rm (\sim10^{42} \, erg \, s^{-1}$, see figure~\ref{compare_samples}).

\subsection{Redshift Constraints}\label{selection}

\noindent
Using the ``CONF\_BEST'' flag, sources with uncertain redshift or spectral classification 
(identified during the visual inspection of the sample) were also removed.
This process resulted in a sample of 7795/7805 2RXS sources 
and 1190/1192 XMMSL1 sources.
In the spectral fitting procedure (described in section~\ref{spec_fit}), 
the H$\beta$ and MgII lines were fit independently.
Sources with H$\beta$ and MgII present in their optical spectrum 
were selected using the following logic:

\begin{equation}\nonumber
\begin{split}
\rm H_{\beta}: \rm (SN\_MEDIAN\_ALL > 5) \,\,\&\&\,\, \\
\rm (( (INSTRUMENT == SDSS) \,\,\&\&\,\, (0 < Z\_BEST < 0.81) )\,\,||\,\, \\
\rm ( (INSTRUMENT == BOSS) \,\,\&\&\,\, (0 < Z\_BEST < 1.05)))
\end{split}
\end{equation}

\begin{equation}\nonumber
\begin{split}
\rm MgII: \rm (SN\_MEDIAN\_ALL > 5) \,\,\&\&\,\, \\
\rm (( (INSTRUMENT == SDSS) \,\,\&\&\,\, (0.45 < Z\_BEST < 2.1) )\,\,||\,\, \\
\rm ( (INSTRUMENT == BOSS) \,\,\&\&\,\, (0.38 < Z\_BEST < 2.5)) )
\end{split}
\end{equation}

\noindent
Different redshift ranges have been used because the 
BOSS spectrograph has a larger wavelength coverage than the SDSS spectrograph.
In some cases, parts of the fitting region will have been redshifted 
out of the SDSS/BOSS spectrograph range \citep{2013AJ....146...32S}, 
and therefore will not be fit.
However, the redshift limits where chosen
so that both samples contain the broad lines used for estimating BH mass.
Sources with a median signal-to-noise ratio (S/N) less than or equal to five 
per resolution element were excluded from the spectral analysis 
since for these sources the broad line decomposition 
and resulting BH mass estimates may be unreliable
\citep[see][]{2009ApJ...692..246D,2011ApJS..194...45S}.

\begin{figure}
\centering
\includegraphics[width=0.49\textwidth]{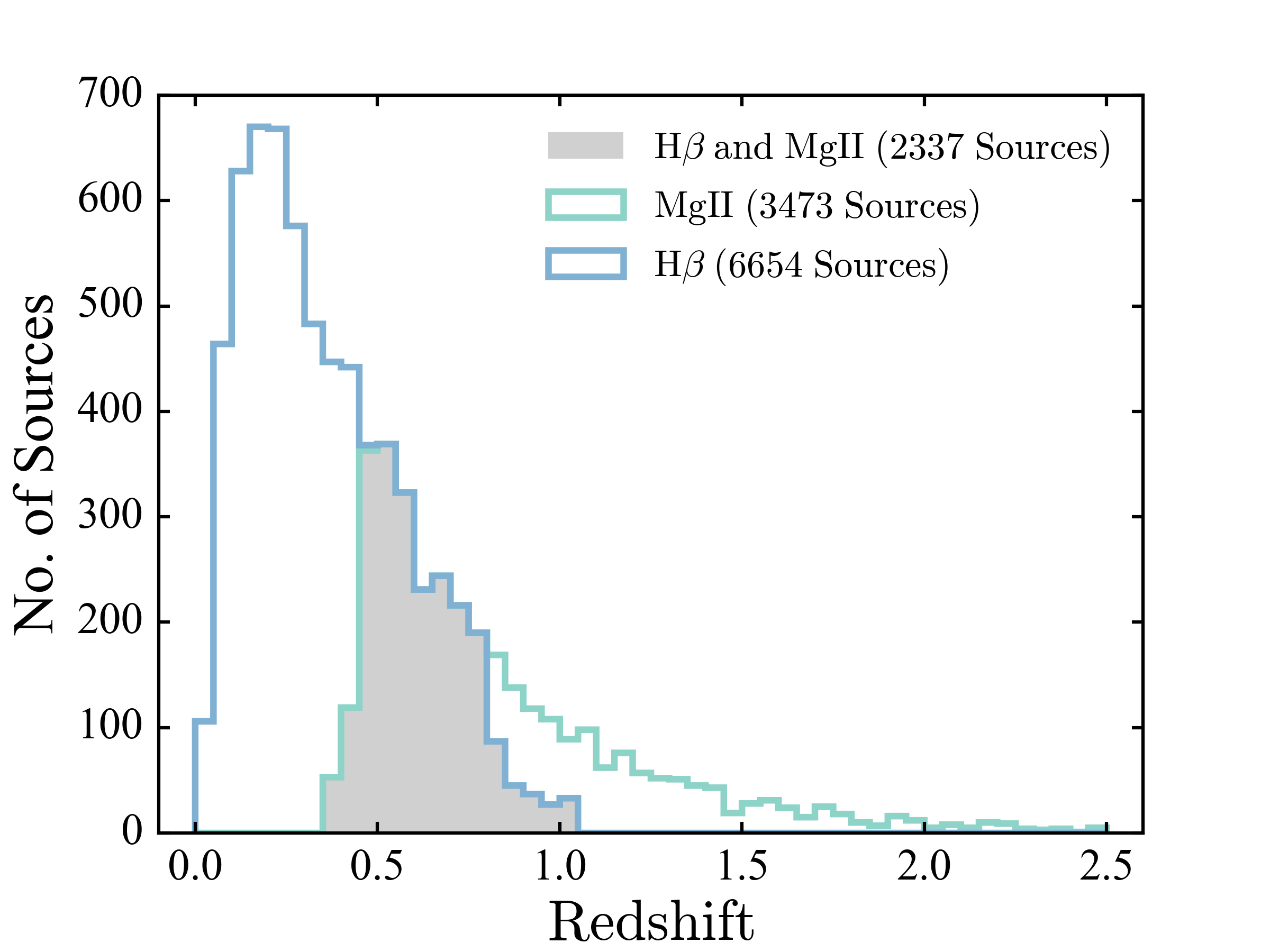}
\caption{\footnotesize
Redshift distribution of the sample of type 1 AGN with coverage of $\rm H\beta$ 
and/or MgII.}
\label{clean_subsample}
\end{figure}

Table~\ref{spec_sample} lists the numbers of sources 
with spectral coverage of either H$\beta$ or MgII, 
while figure~\ref{clean_subsample} shows the redshift distribution of these sources.
There are 711 cases where the same optical counterpart 
was detected by both 2RXS and XMMSL1.
The final combined (2RXS and XMMSL1) sample for spectral analysis  
contains 7790 unique type 1 sources.

\begin{table}
\caption{\footnotesize 
The coverage of the H$\beta$ and MgII emission lines 
in the two samples used in this work.
There are 711 sources which were detected in both 
the 2RXS and XMMSL1 surveys. 
The ``total'' row lists the total number of unique sources 
obtained from combining the 2RXS and XMMSL1 samples.}
\centering  
\begin{tabular}{l c c c c c}      
\toprule 
\midrule

\multicolumn{1}{c}{}
& \multicolumn{1}{c}{MgII}
& \multicolumn{1}{c}{H$\beta$}
& \multicolumn{1}{c}{H$\beta$ and MgII}
& \multicolumn{1}{c}{H$\beta$ or MgII}
\\
\midrule
2RXS & 3310 & 6268 & 2234 & 7344 \\ 
XMMSL1 & 314 & 1070 & 227 & 1157 \\
Total & 3473 & 6654 & 2337 & 7790 \\
\midrule
\bottomrule \\
\end{tabular}
\label{spec_sample}
\end{table}

\section{Spectral Analysis}\label{spec_fit}

\noindent
A series of scripts have been written to perform spectral fits
using the MPFIT least-squares curve fitting routine 
\citep{2009ASPC..411..251M}.
Each spectrum was corrected for Milky Way extinction 
using the extinction curve from \citet{1989ApJ...345..245C},
and the dust map from \citet{1998ApJ...500..525S},
with an R$\rm_{V}$=3.1.
No attempt has been made to estimate and correct for 
the intrinsic (host) extinction of each source\footnote{
Also note that extinction laws (e.g. Calzetti and Prevot) 
are based on samples of nearby SB and irregular type galaxies. 
Due to the lack of nearby passive galaxies,
an extinction law for these galaxy types is not yet available.}.
Measured line widths were corrected for the resolution of the \
SDSS/BOSS spectrographs.
The H$\beta$ and MgII emission line regions 
were fit independently using similar
methods described in the following 
sections\footnote{For each model parameter,
the 1-sigma uncertainties from MPFIT were adopted.}.

\subsection{Iron Emission Template}\label{fe}

\noindent
AGN typically exhibit FeII emission consisting of 
a large number of individual lines 
across the optical and UV regions of the spectrum.
These lines appear to be blended, 
probably due to the motion of the gas from which they are emitted,
and the magnitude of this broadening varies significantly from source to source.
The presence of FeII emission in the optical and UV portions of the spectrum can be
a significant complication when attempting to accurately measure line profiles.
Therefore it is crucial that the model used to derive line widths for BH mass measurements also accounts for the nearby FeII emission.

\begin{figure}[h!]
	\centering
                \includegraphics[width=0.49\textwidth]{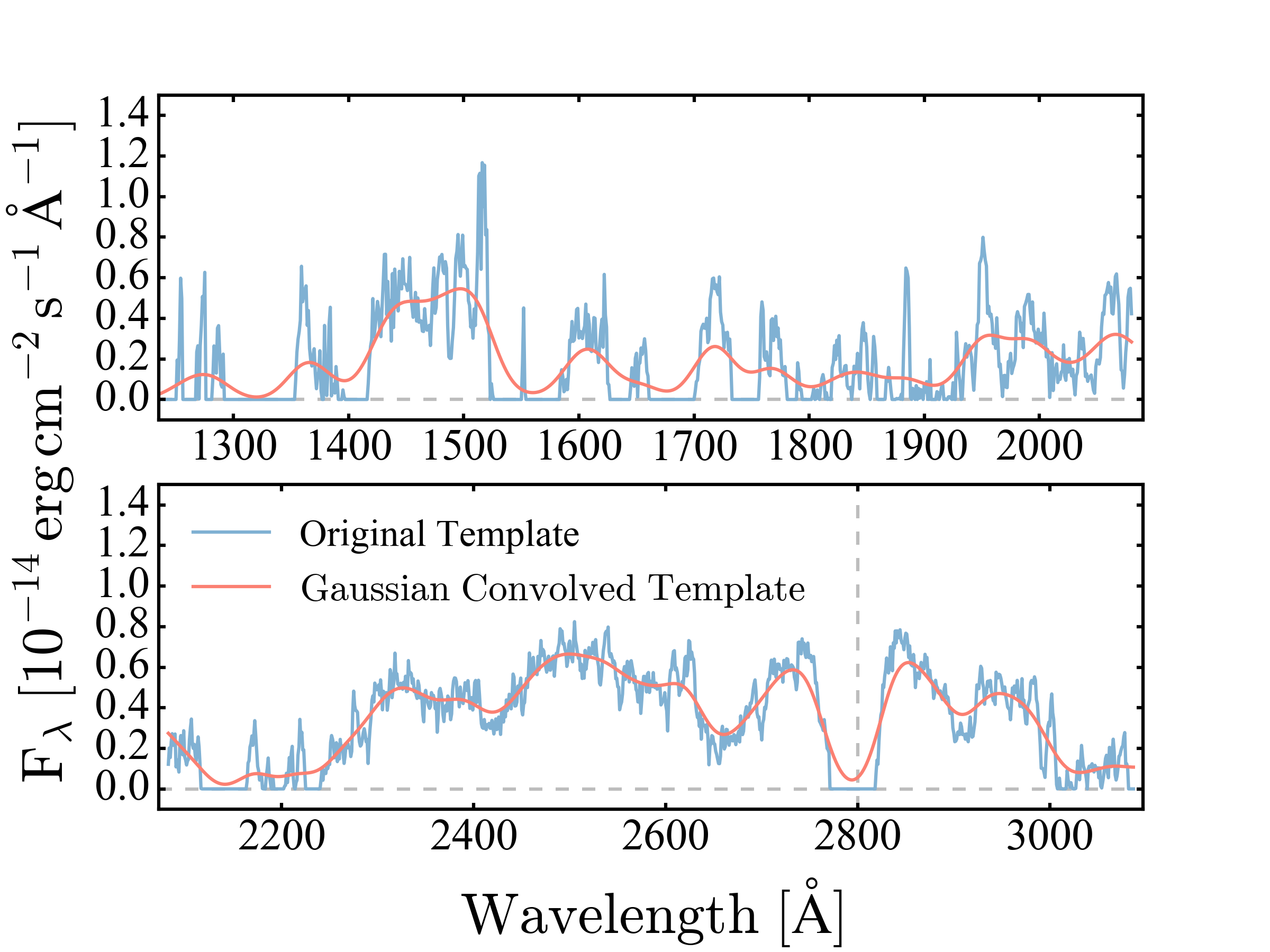}\\
                \includegraphics[width=0.49\textwidth]{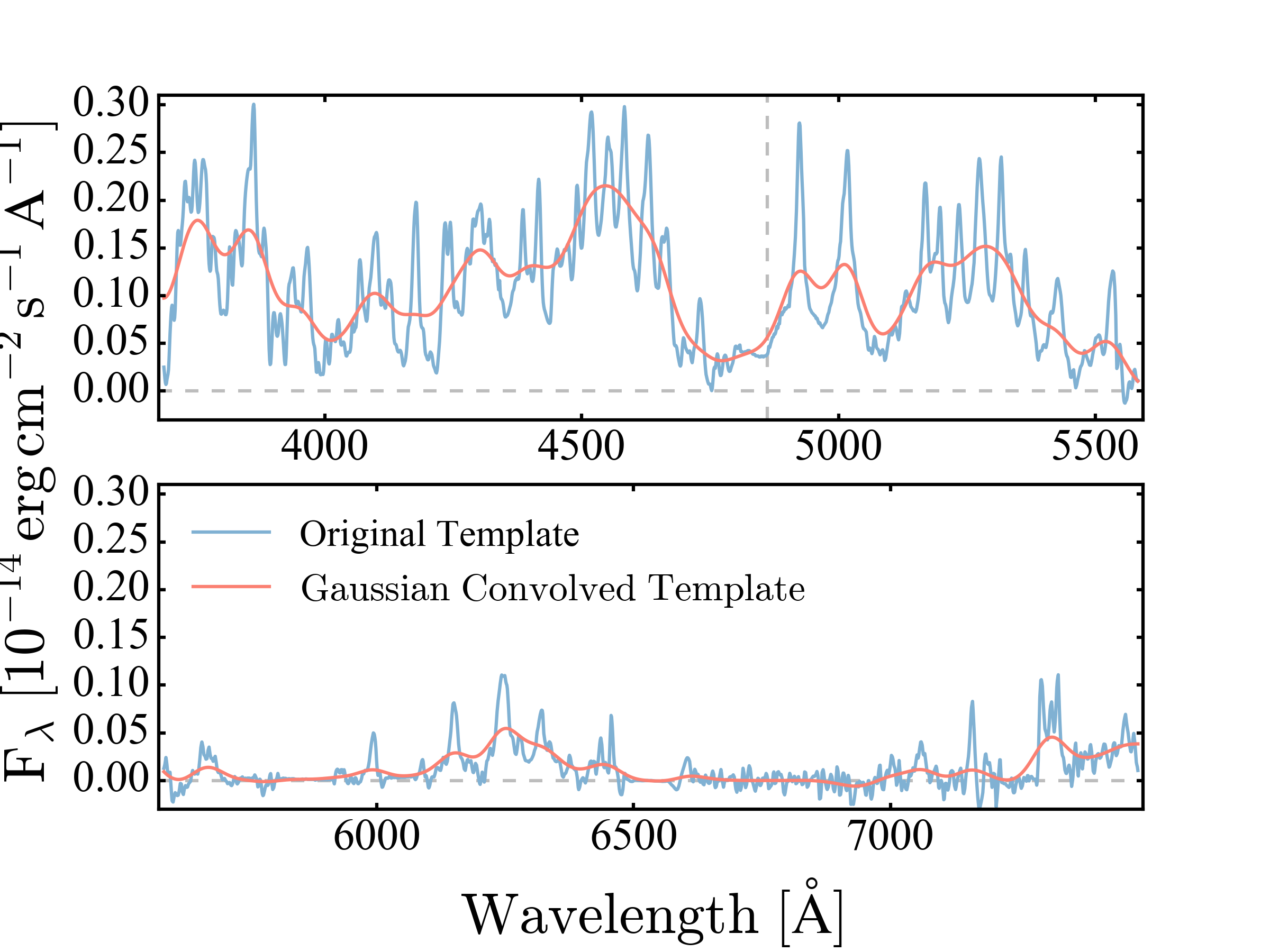}
\caption{\footnotesize
Upper panel: The \citet{2001ApJS..134....1V} FeII template 
which was used when fitting the MgII emission line region.
Lower panel: The \citet{1992ApJS...80..109B} FeII template
which was used when fitting the H$\beta$ emission line region. 
In both cases, the original template is shown (blue) 
along with the same template convolved with a Gaussian with FWHM\,=\,4000\,km\,s$^{-1}$ (red).
The vertical dashed lines correspond to the position of MgII and H$\beta$ 
at 2800$\AA$ and 4861$\AA$.}
\label{fe_template}
\end{figure}

Figure~\ref{fe_template} shows the two FeII templates used in this work;
the \citet{2001ApJS..134....1V} and \citet{1992ApJS...80..109B}
templates used for the UV and optical regions of the spectrum, respectively. 
Both of these templates have been derived from 
the narrow line Seyfert 1 galaxy I Zwicky 1
which, due to its bright FeII emission and narrow emission lines, 
is an ideal candidate for generating the FeII template.
In order to model the observed blending of the FeII emission,
the templates were convolved with a Gaussian 
whose width was included as a free parameter in the fitting procedure.

\begin{figure*}[h!]
	\centering
                \includegraphics[width=0.49\textwidth]{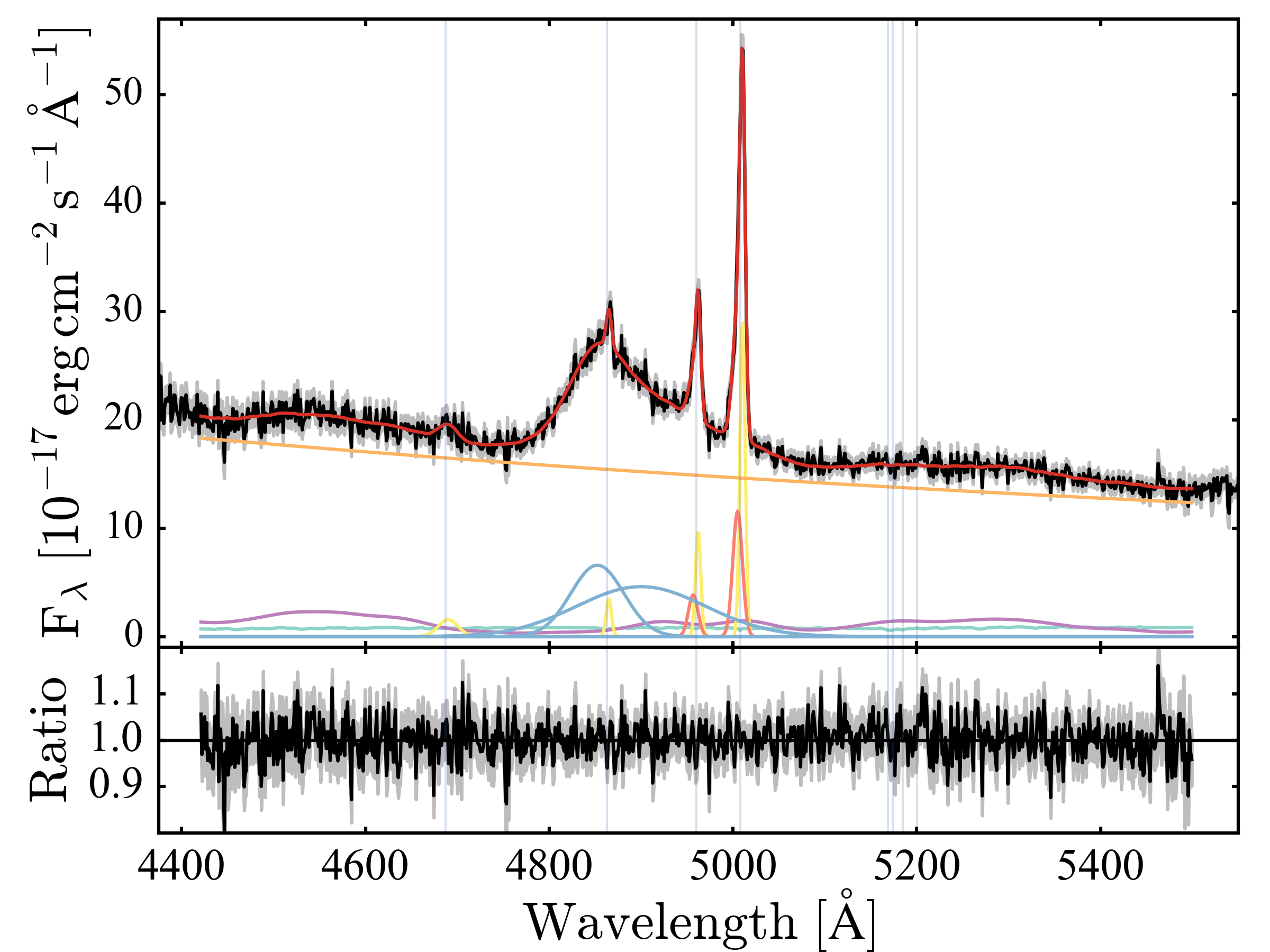}
                \includegraphics[width=0.49\textwidth]{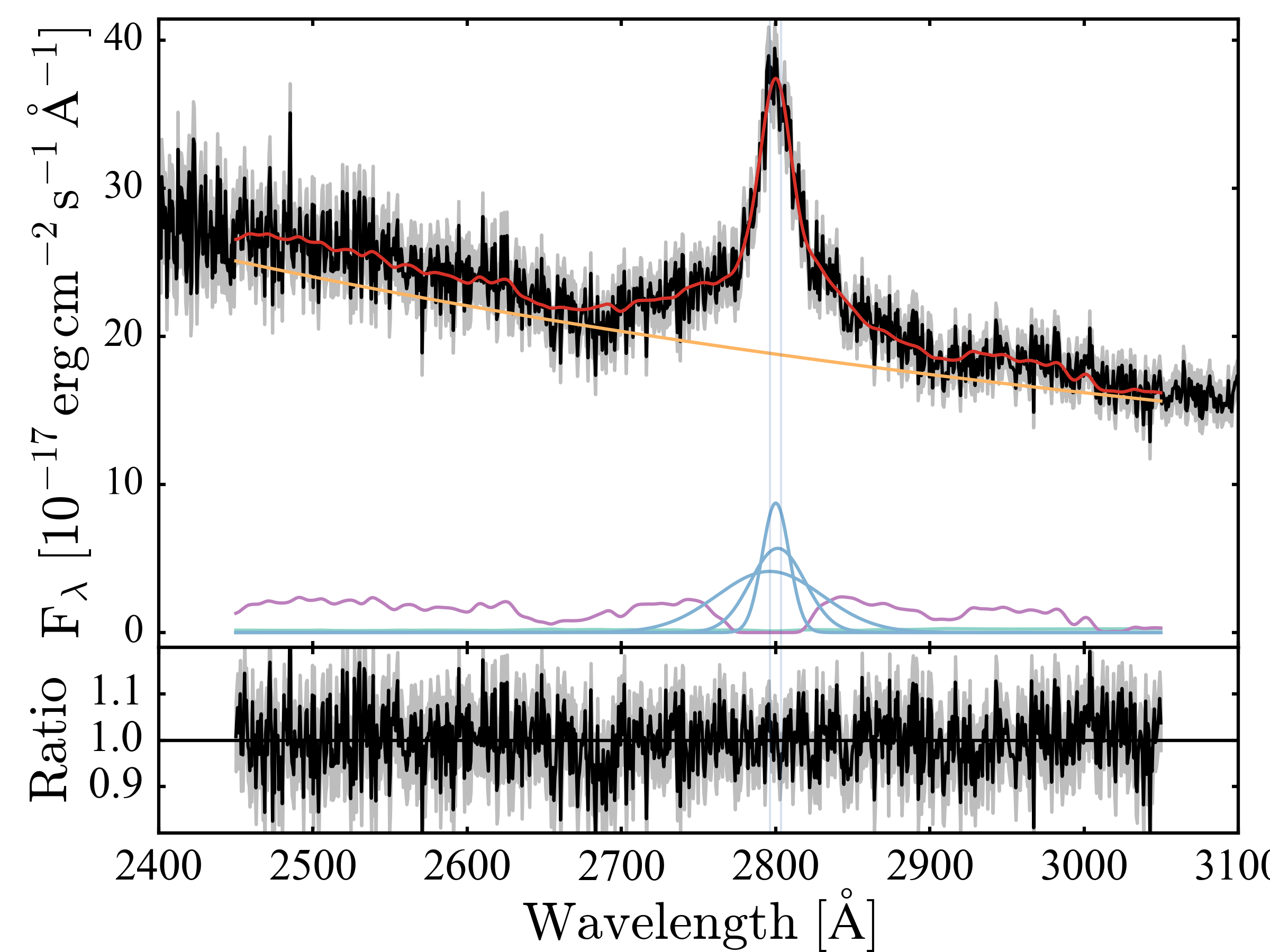}
\caption{\footnotesize
Examples of model fits to the H$\beta$ 
(left panel; plate=1159, MJD=52669, fiber=470) 
and MgII (right panel; plate=423, MJD=51821, fiber=250) spectral regions.
The model components are colour-coded as follows;
power law (orange), iron emission (violet), broad lines (blue), 
narrow lines (yellow), [OIII] shifted wings (red), and the total model (red).
The panels beneath the spectra show the data/model ratio.}
\label{examples}
\end{figure*}

\subsection{H$\beta$}\label{HB_fit_method}

\noindent
The region from 4420-5500\,$\AA$ was fit for each spectrum. 
The continuum model consisted of a power law,
a galaxy template, and the \citet{1992ApJS...80..109B} FeII emission template.
The FeII template was convolved with a Gaussian while fitting, 
and the width of this Gaussian, along with the normalisation of the template 
were included as free parameters in the fit (see section~\ref{fe}).
Previous spectral analyses of AGN spectra have assumed 
an early-type galaxy component in the model \citep{2017MNRAS.472.4051C}.
Following this method, we use an early-type SDSS galaxy
template\footnote{template number 24 on \\ http://classic.sdss.org/dr5/algorithms/spectemplates/}  in the fit, and the normalisation of this template as well as the 
normalisation and slope of the power law were also free parameters.
The use of a single, early-type galaxy template is an approximation, 
however, it is considered to be justified since 
AGN are typically found to reside in bulge dominated galaxies
\citep[e.g.][]{2005ApJ...627L..97G, 2007ApJ...660L..19P},
and the spectroscopic fiber collects emission mostly from 
the bulge (which is characterised by an old stellar population)
and the active nucleus.

The [OIII]$\lambda$4959 and [OIII]$\lambda$5007 narrow lines
were each fit with two Gaussians, one used to fit the narrow core, 
and an additional Gaussian to account for the presence of 
blue-shifted wings which are often detected \citep{2005AJ....130..381B}. 
A single Gaussian was used to fit the HeII$\lambda$4686 emission line.
To avoid overfitting the H$\beta$ line, the fitting process was run four times, 
with one, two, three, and four\footnote{Three broad Gaussians are used in addition to a single narrow component to account for the three distinct broad components 
that are expected to be present
(see section~\ref{VBC_section})
in at least some sources \citep{2010MNRAS.409.1033M}.} 
Gaussian components used to fit the H$\beta$ line.
For each fit, the velocity width and peak wavelength of one of the Gaussian components was fixed to that of [OIII]$\lambda$4959 and [OIII]$\lambda$5007 
in order to aid the identification of the narrow H$\beta$ component.
The normalisation ratio of the [OIII]$\lambda$4959 and [OIII]$\lambda$5007
lines was fixed to the expected value of 1:3 \citep[e.g.][]{2000MNRAS.312..813S}.
The best-fit model was then selected using the Bayesian information criterion 
\citep[BIC,][]{1978AnSta...6..461S},
which can be written as 

\begin{equation}\nonumber
\rm BIC = ln(n)k+\chi^{2} 
\end{equation}

\noindent
where n is the number of data points, k is the number of model parameters,
and $\rm \chi^{2}$ is the chi-square of the fit.
The preferred model is that with the lowest BIC.
An example of a fit to the H$\beta$ spectral region 
is shown in the left panel of figure~\ref{examples}.

\subsubsection{Broad Line Decomposition}\label{bl_decomp}

\begin{figure}
\centering
\includegraphics[width=0.49\textwidth]{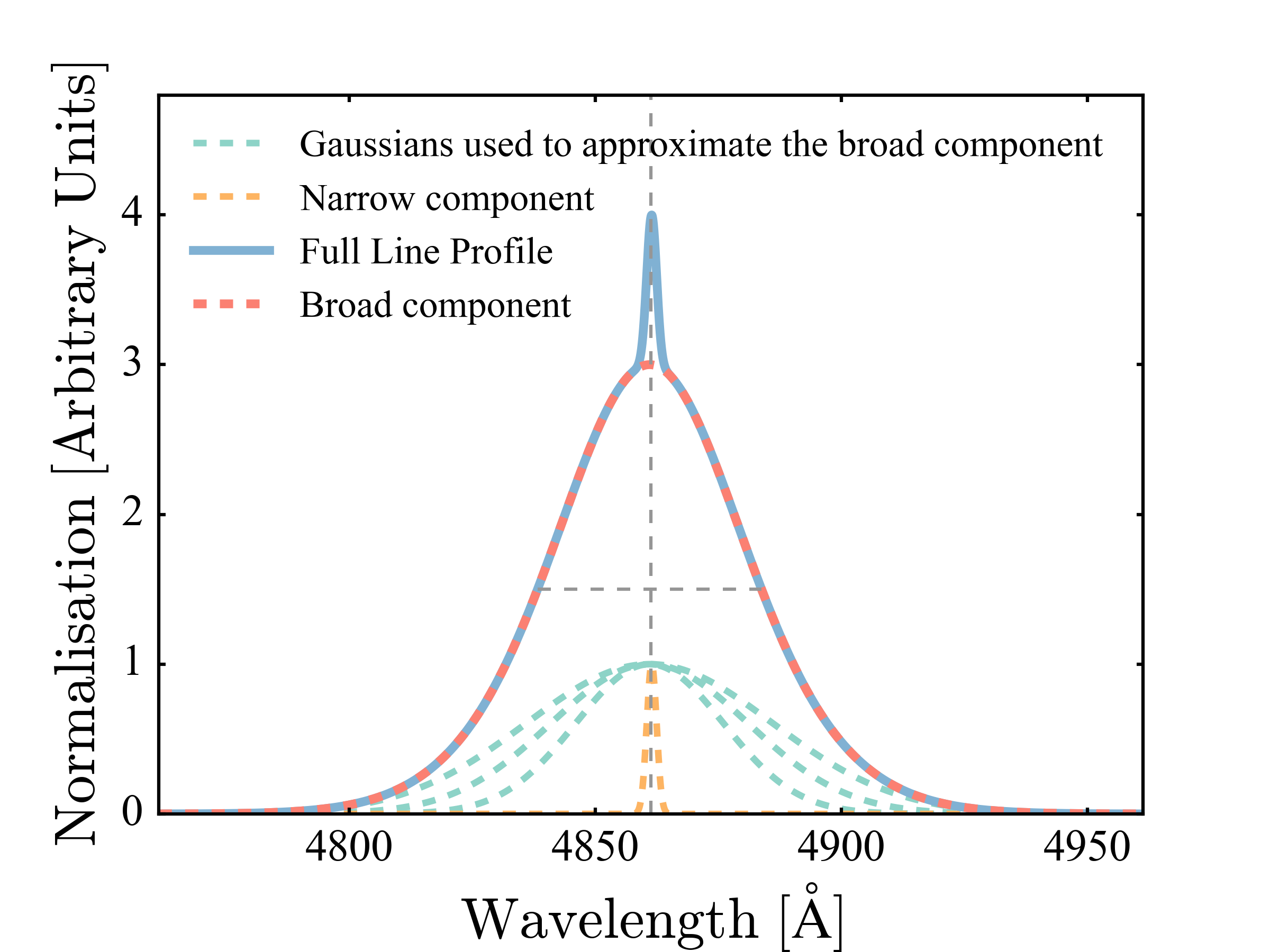}
\caption{\footnotesize 
An example of the decomposition of a typical AGN H$\beta$ line 
(plate=7276, MJD=57061, fiber=470).
The horizontal dashed line represents the FWHM used for BH mass estimation.
The vertical dashed line indicates the rest-frame wavelength of H$\beta$.
See section~\ref{bl_decomp} for further details.}
\label{line_decomp}
\end{figure}

\noindent
The narrow H$\beta$ and [OIII] components 
are required to have widths $\rm \leq800\,km\,s^{-1}$.
Any of the additional Gaussians used to fit MgII and H$\beta$
with FWHM\,$\rm > 800\,km\,s^{-1}$ are considered ``broad''.
This threshold of 800$\rm \, km\, s^{-1}$ is taken from the approximate division between broad and narrow FWHM distributions in the lower panels of figure~\ref{line_width}.
The virial FWHM used for BH mass estimation 
is the FWHM of the line profile defined by the 
sum of these broad Gaussian components
(see figure~\ref{line_decomp}).
A major challenge with using the single-epoch method for estimating BH mass 
is decomposing the broad and narrow components of the line
in order to measure the virial FWHM.
Figure~\ref{line_decomp} presents an example 
of the decomposition of a broad H$\beta$ line.
In this case, the narrow H$\beta$ core can be easily distinguished 
and removed before measuring the virial FWHM.
However, there are many cases 
where the broad and narrow components are blended,
making it difficult to successfully identify the appropriate virial FWHM.
There are also cases where there is a clear distinction between
two broad line components that are shifted in wavelength 
relative to each other (known as ``double-peaked emitters''). 
How one should interpret 
the single-epoch BH mass estimates 
for these unusual objects is uncertain 
(also see section~\ref{reliability}).

\subsection{MgII}\label{mgII_fit}

\noindent
The region from 2450-3050$\AA$ was fit for each spectrum. 
As in the case of the H$\beta$ fits, a power law,
an early-type galaxy template 
\citep[5\,Gyr old elliptical galaxy;][]{1998ApJ...509..103S, 2007ApJ...663...81P},
and the \citet{2001ApJS..134....1V} FeII emission template
were used to fit the continuum.
Again, the FeII template normalisation, and width of the 
Gaussian smoothing applied to the template, 
were included as free parameters in the fit.
The MgII line is a doublet; however, due to the close spacing 
and virial broadening of the lines,
it usually appears as a single broad component in AGN spectra.
The narrow MgII line cores are usually not observed in AGN spectra,
therefore the MgII profile was fit using three broad Gaussians.
An example of a fit to the MgII spectral region 
is presented in the right panel of figure~\ref{examples}.

\begin{table*}[h]
\caption{\footnotesize 
BH mass calibrations used in this work.
A, B, and C are the calibration constants for single-epoch mass estimation
(see equation~\ref{mass_est}).}
\centering  
\begin{tabular}{l c c c c c}      
\toprule 
\midrule

\multicolumn{1}{c}{}
& \multicolumn{1}{c}{A}
& \multicolumn{1}{c}{B}
& \multicolumn{1}{c}{C}
& \multicolumn{1}{c}{Reference}
\\
\midrule
MgII, L$_{3000\AA}$ & 1.816 & 0.584 & 1.712 & \citet{2012ApJ...753..125S} \\
H$_{\beta}$, L$_{5100\AA}$ & 0.91 & 0.5 & 2 & \citet{2006ApJ...641..689V} \\ 
H$_{\beta}$, L$_{5100\AA}$ & 0.895 & 0.52 & 2 & \citet{2011ApJ...742...93A} \\
\midrule
\bottomrule \\
\end{tabular}
\label{calib}
\end{table*}

\begin{figure*}[h!]
\begin{center}
\includegraphics[width=0.33\textwidth]{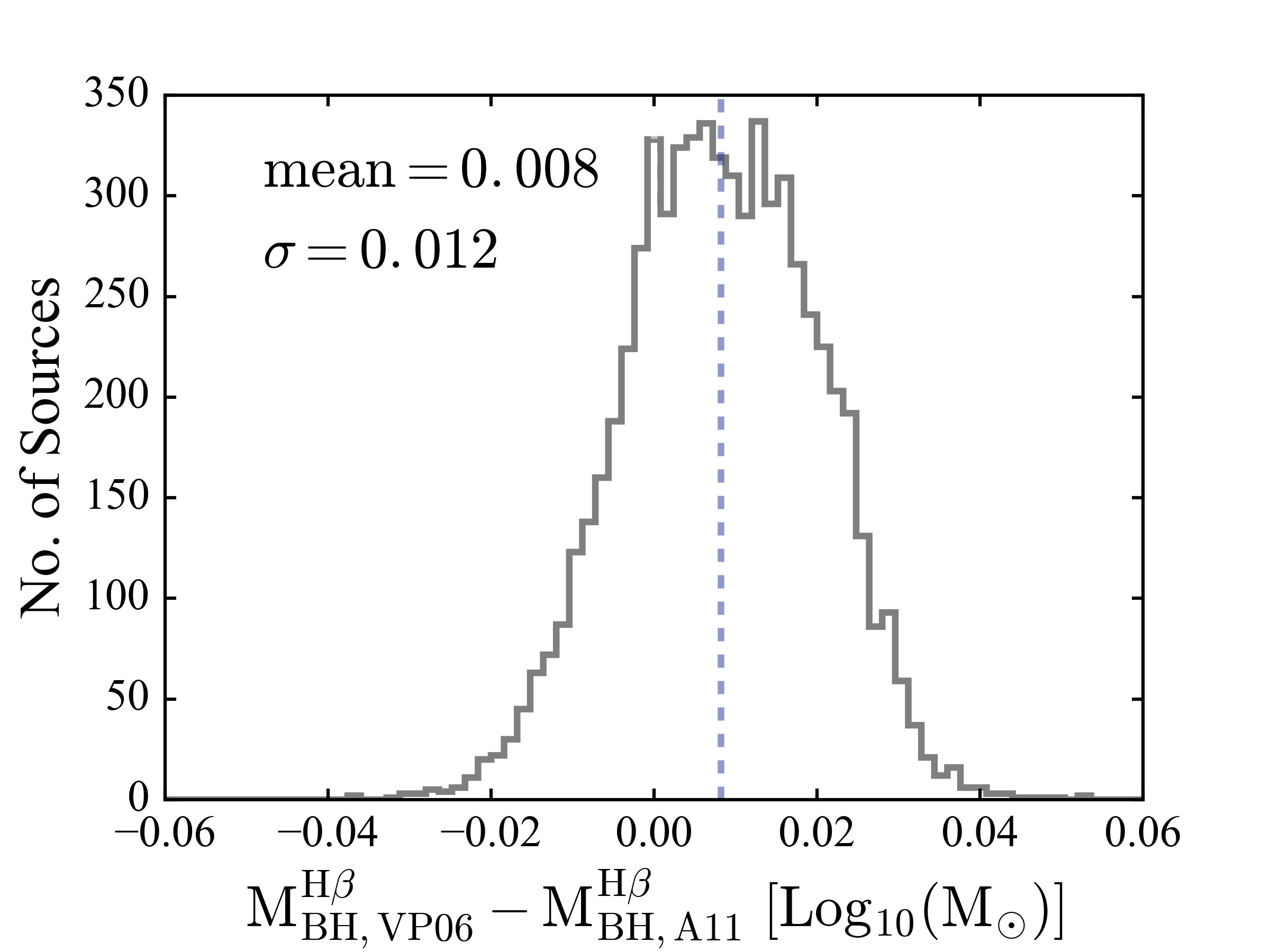}
\includegraphics[width=0.33\textwidth]{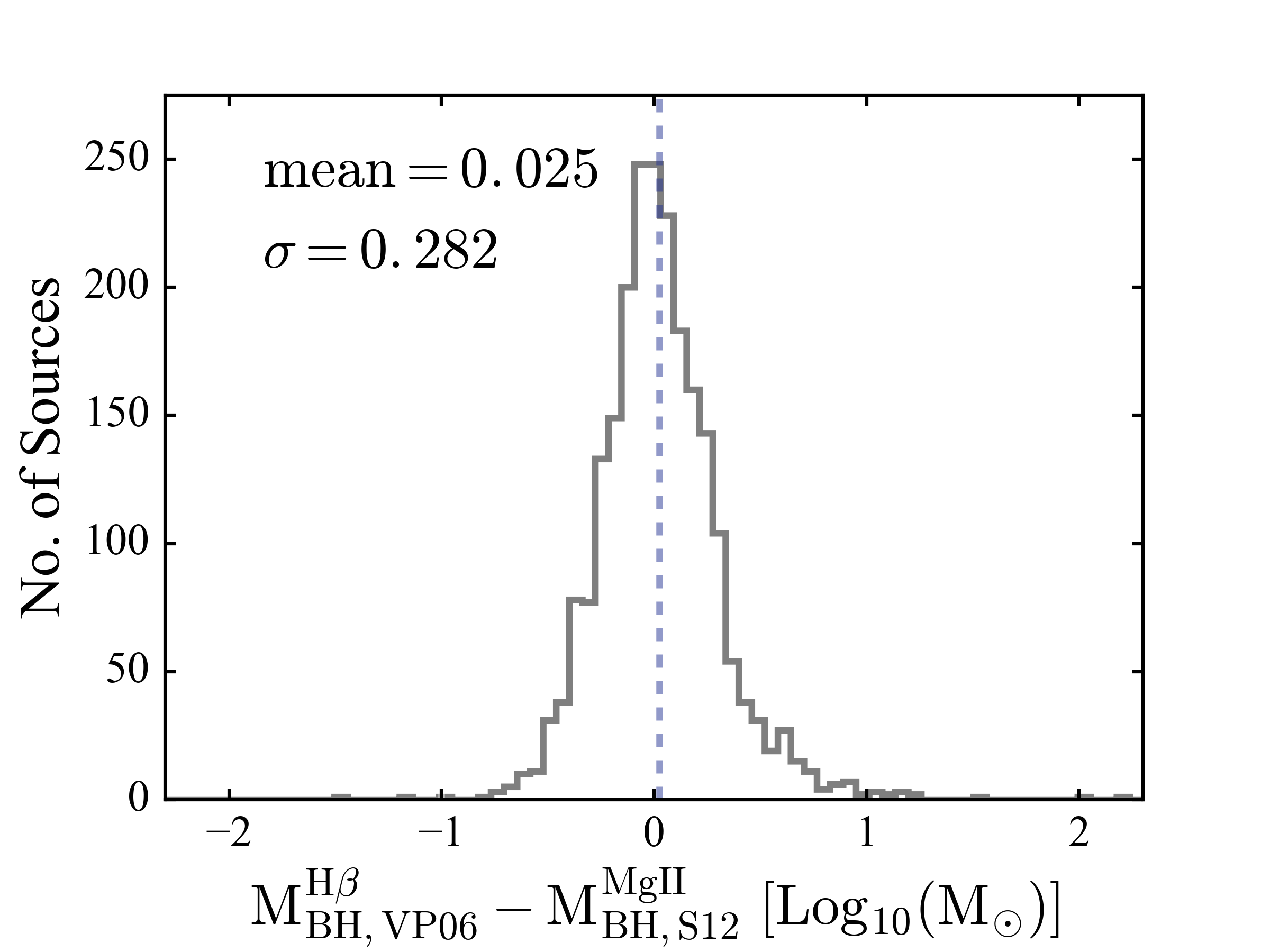}
\includegraphics[width=0.33\textwidth]{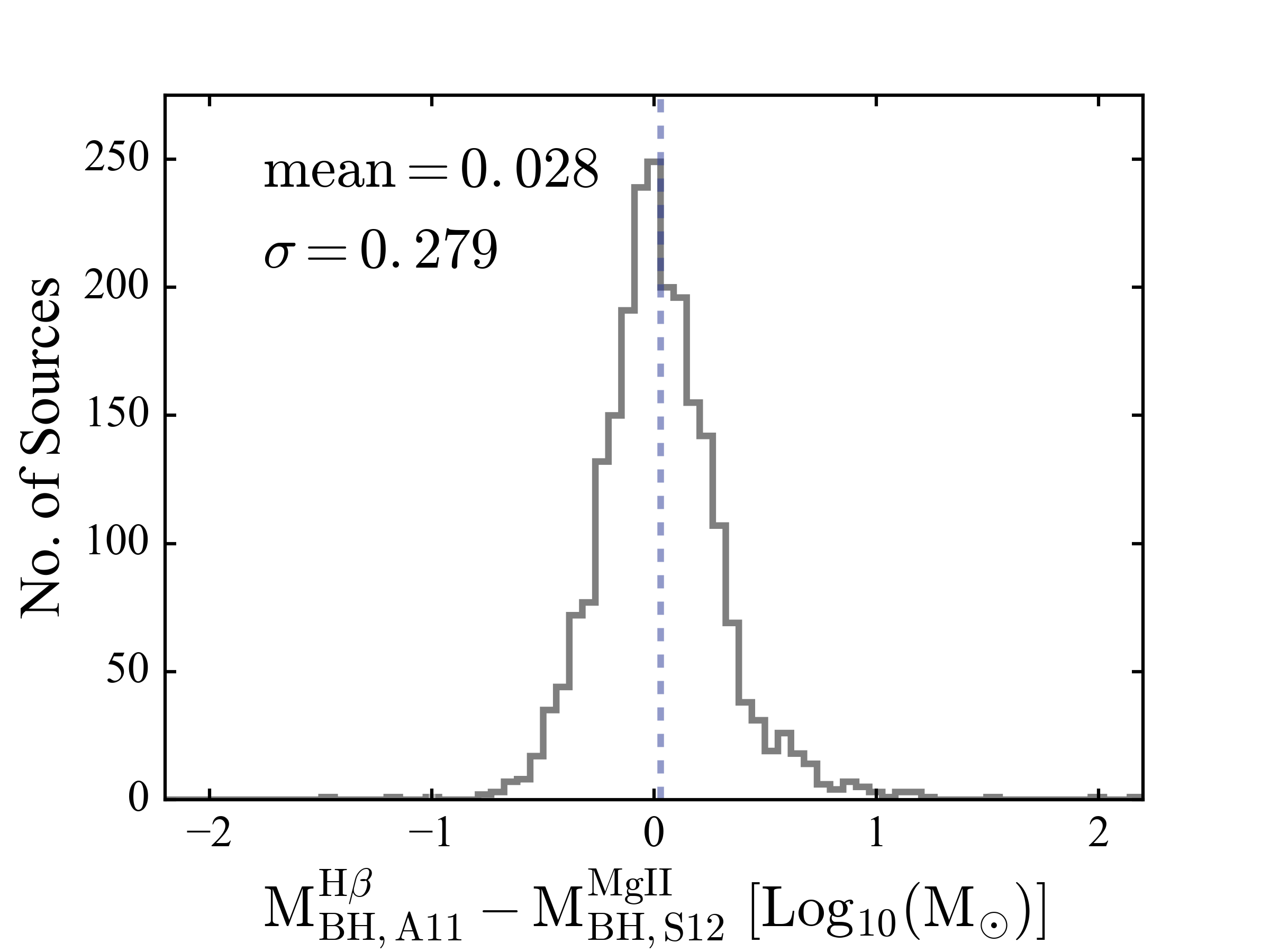}
\caption{\footnotesize \label{mass_hist}
Differences between the BH mass calibrations used in this work
(see section~\ref{estimate_mass} for further details).
The vertical blue lines indicate the mean value of each distribution.
The standard deviation, $\rm \sigma$, of each distribution is also shown.} 
\end{center}
\end{figure*}

\section{Bolometric Luminosity and BH Mass Estimation}\label{estimate_mass}

\noindent
Bolometric luminosities were estimated from 
the monochromatic luminosities 
using the bolometric corrections from 
\citet{2006ApJS..166..470R, 2011ApJS..194...45S}:

\begin{equation}\nonumber
\begin{split}
\rm L_{Bol} = 5.15 \, L_{3000\AA} \\
\rm L_{Bol} = 9.26 \, L_{5100\AA}
\end{split}
\end{equation}

\noindent
These bolometric corrections have been derived using mean AGN SEDs;
however, \citet{2006ApJS..166..470R} note that 
using a bolometric correction resulting from a single mean SED 
can result in bolometric luminosities with inaccuracies up to 50$\%$.

Under the assumption that the BLR gas is virialised, 
the single-epoch method can be used to estimate BH mass as follows:

\begin{equation}\label{mass_est}
\rm log\left(\frac{M_{BH}}{M_{\odot}}\right) = A + B\,log\left(\frac{\lambda L_{\lambda}}{10^{44}\,erg\,s^{-1}}\right)+C\,log\left(\frac{FWHM}{km\,s^{-1}}\right)
\end{equation}

\noindent
where $\rm L_{\lambda}$ is the monochromatic luminosity
at wavelength $\lambda$,
and FWHM is the full width at half maximum of the broad component of the emission line.
A, B, and C are constants that are calibrated using RM results
and vary depending on which line is used. 

Over the years, many groups have provided 
calibrations of equation~\ref{mass_est} for MgII and H$\beta$.
In this work, the calibrations from 
\citet{2006ApJ...641..689V} and \citet{2011ApJ...742...93A} are used for H$\beta$.
\citet{2006ApJ...641..689V} based their work on 
an updated study of the $\rm R_{BLR} - L$ relationship 
\citep{2005ApJ...629...61K, 2006ApJ...644..133B} and 
a reanalysis of the RM mass estimates \citep{2004ApJ...613..682P}
and therefore presented an improved mass calibration relative to previous studies.
\citet{2011ApJ...742...93A} provide a mass calibration that is based on the 
$\rm R_{BLR} - L$ relationship from \citet{2009ApJ...705..199B}.
\citet{2006ApJ...641..689V} and \citet{2011ApJ...742...93A} 
both provide similar calibrations for single-epoch H$\beta$ mass estimation,
as can be seen from the left panel of figure~\ref{mass_hist}.

The \citet{2012ApJ...753..125S} calibration is used in this work for MgII.
This calibration is based on a sample of 
60 high-luminosity ($\rm L_{5100\AA} > 10^{45.4} erg \, s^{-1}$)
quasars in the redshift range 1.5 - 2.2.
\citet{2012ApJ...753..125S} use the \citet{2006ApJ...641..689V} 
mass estimates as a reference when determining their MgII calibration.
The centre and right panels of figure~\ref{mass_hist}
show the comparison between the \citet{2012ApJ...753..125S} MgII calibration
and the $\rm H \beta$ calibrations from 
\citet{2006ApJ...641..689V} and \citet{2011ApJ...742...93A}.
These calibrations agree reasonably well,
with the standard deviation $\rm \sigma \simeq 0.3$ in both cases,
which is likely due to the fact that these BH mass estimates were derived 
using two different emission lines.
A list of the three BH mass calibrations used in this work 
is given in Table~\ref{calib}.

BH masses were computed for each of these calibrations 
and are included in the catalogue (see section~\ref{columns}). 
BH masses were only estimated for sources with a detected 
broad line component (see section~\ref{bl_decomp}).
These BH mass estimates were then used to estimate 
the Eddington luminosity and the Eddington ratio

\begin{equation}\nonumber
\begin{split}
\rm  L_{Edd}=4\pi c G M_{BH} m_{p}/\sigma_{T} \\
\rm \lambda_{E} = L_{Bol}/L_{Edd}
\end{split}
\end{equation}

\noindent
where c is the speed of light, 
G is the gravitational constant,
M$\rm_{BH}$ is the BH mass,
m$\rm_{p}$ is the proton mass,
and $\rm \sigma_{T}$ is the Thomson scattering cross-section.

\begin{figure}[]
\begin{center}
\includegraphics[width=0.49\textwidth]{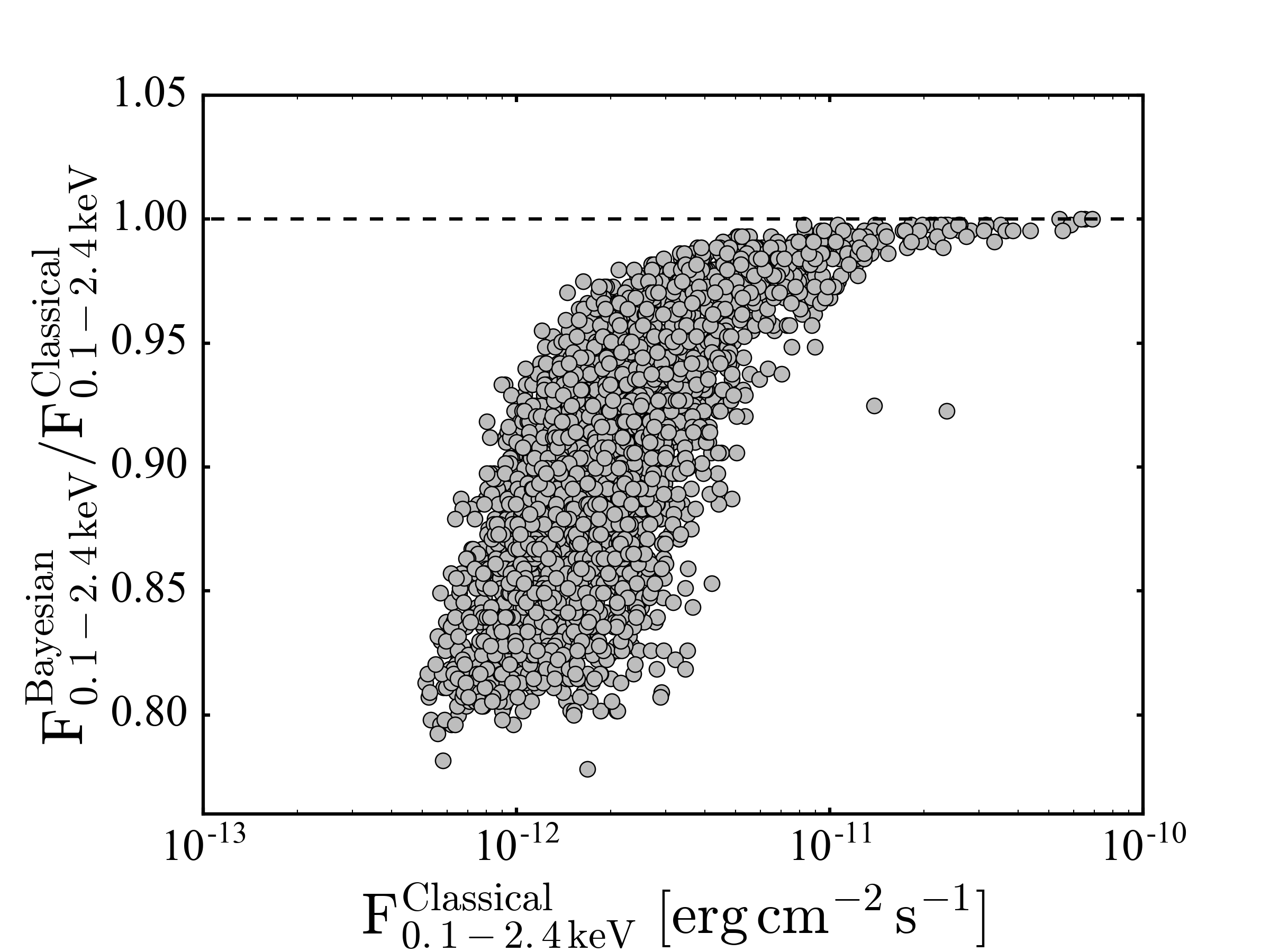}
\caption{\label{bay_class_flux} \footnotesize 
Comparison between the classical and Bayesian methods 
for estimating 2RXS fluxes. The deviation from a ratio of one at fainter fluxes 
results from the attempt to correct for the effect of the Eddington bias
(see section~\ref{2RXS_flux} for further details).} 
\end{center}
\end{figure}

\begin{figure*}[h!]
	\centering
                \includegraphics[width=0.49\textwidth]{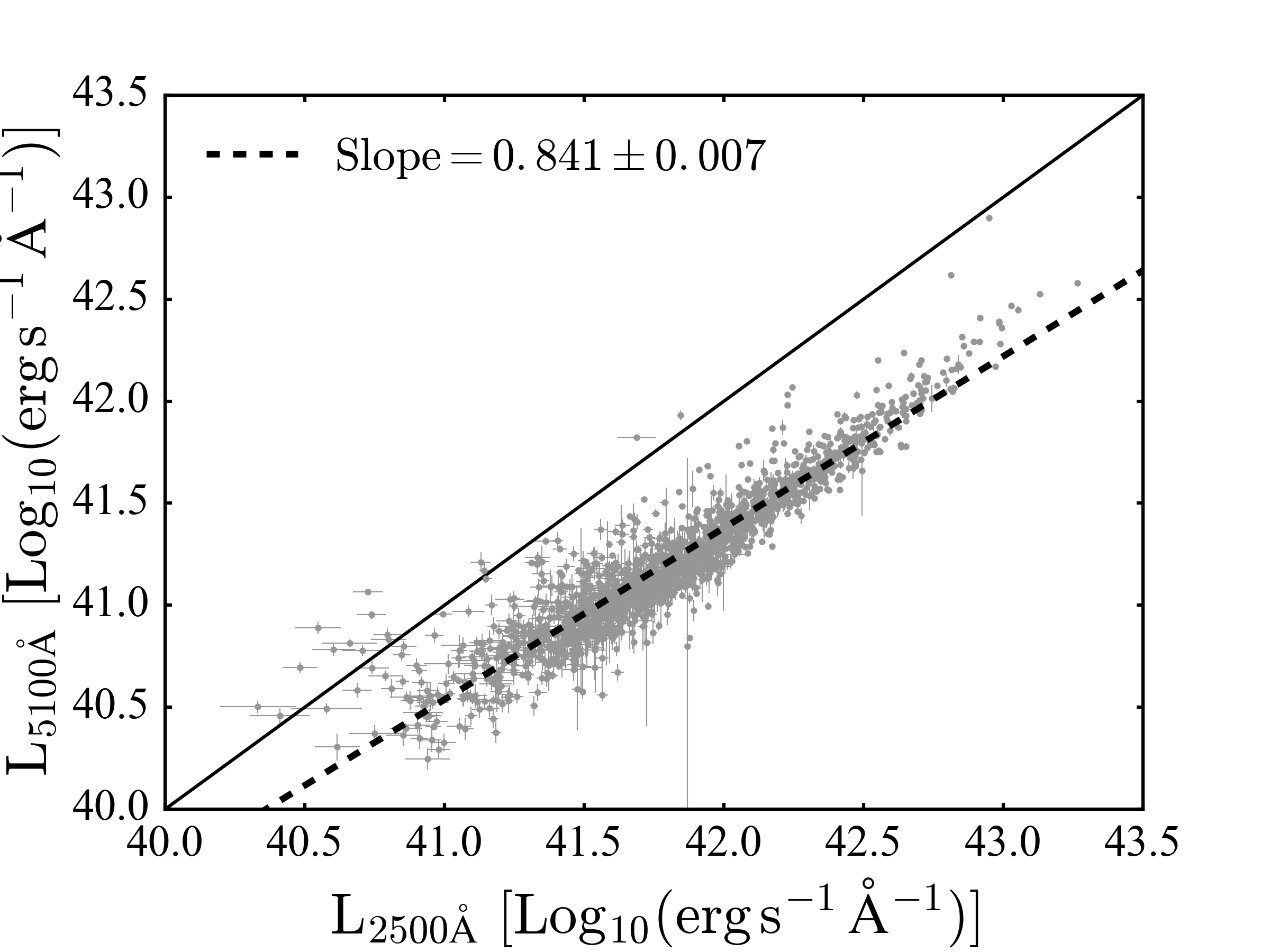}
                \includegraphics[width=0.49\textwidth]{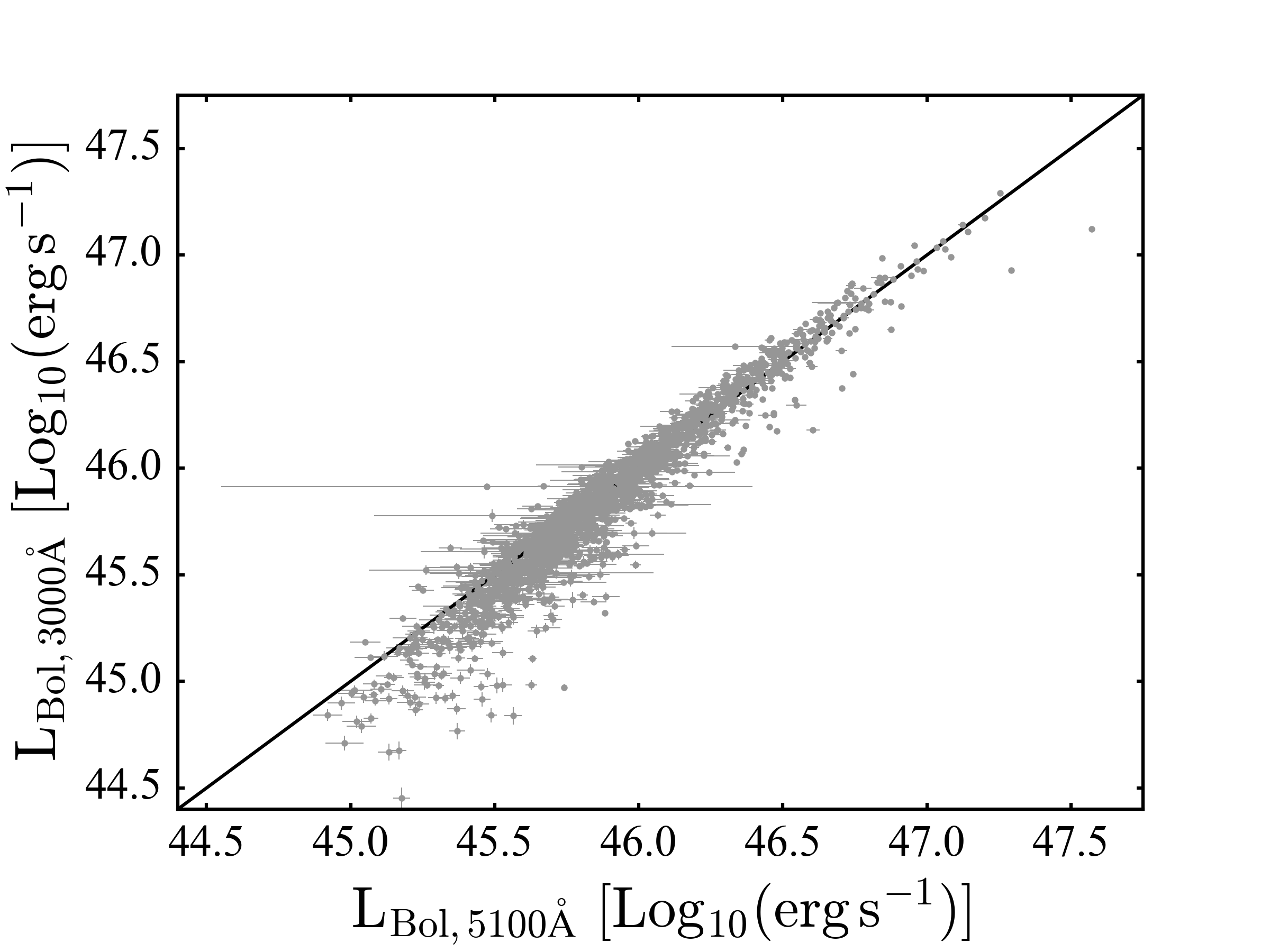}
                \includegraphics[width=0.49\textwidth]{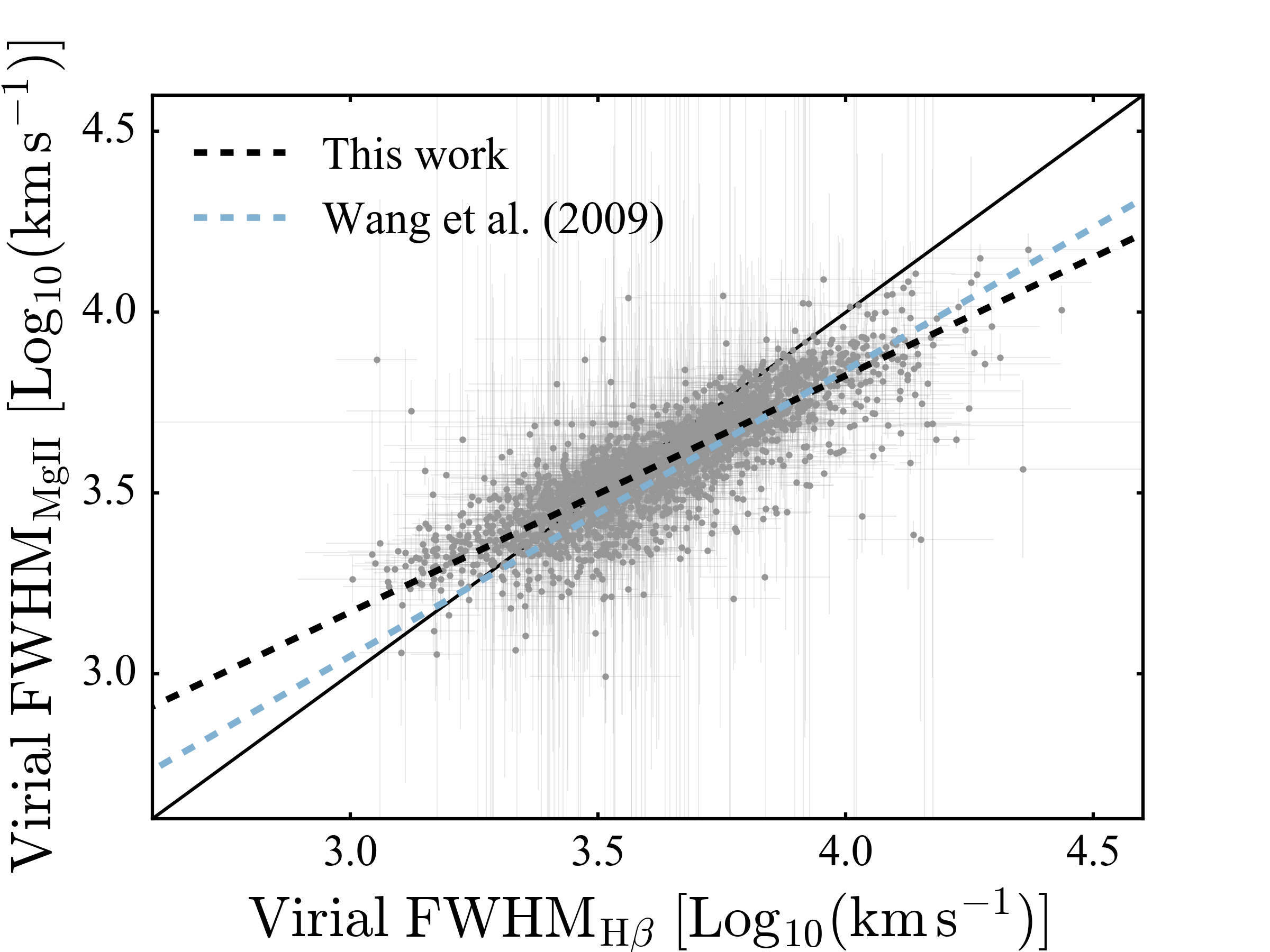}
                \includegraphics[width=0.49\textwidth]{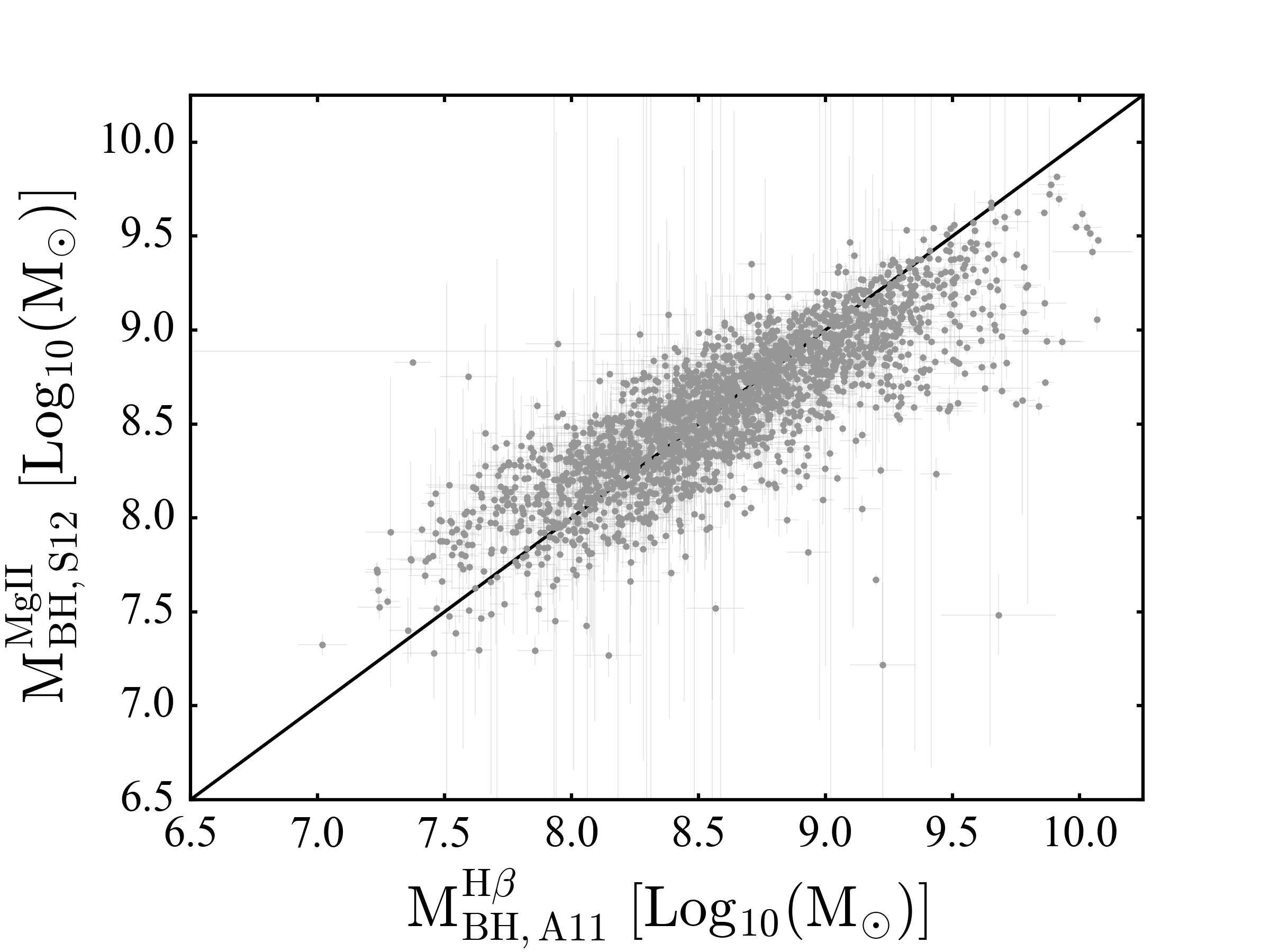}
\caption{\footnotesize 
Comparison between the results obtained from fitting 
the H$\beta$ and MgII regions of the spectrum.
Upper left: 5100$\AA$ and 2500$\AA$ monochromatic luminosities.
Upper right: Bolometric luminosities estimated from the
5100$\AA$ and 3000$\AA$ monochromatic luminosities.
Lower left: Virial FWHM measured using H$\beta$ and MgII.
Lower right: BH mass estimates derived from H$\beta$ 
(using the \citet{2011ApJ...742...93A} calibration) 
and MgII (using the \citet{2012ApJ...753..125S} calibration).
In each panel, the solid black line is the unity line.}
\label{compare}
\end{figure*}

\section{X-ray Flux Estimates}\label{xray_flux}

\noindent
Since X-ray detections are available for all objects in this sample,
X-ray flux estimates have also been included in the catalogue.
XMMSL1 fluxes in the 0.2-12\,keV range 
from \citet{2008A&A...480..611S} are included.
\citet{2008A&A...480..611S} convert the XMMSL1 count rates 
to fluxes using a spectral model consisting of 
an absorbed power law with a photon index of 1.7 
and N$\rm_{H}=3\times10^{20}cm^{-2}$.
The 2RXS fluxes were estimated using the method outlined below.

\subsection{Estimating 2RXS X-Ray Fluxes}\label{2RXS_flux}

Many of the sources in the 2RXS sample 
have flux measurements close to
the \textit{ROSAT} flux limit ($\sim$10$\rm ^{-13}\,erg\,cm^{-2}\,s^{-1}$).
Therefore, when estimating fluxes for this sample,
it was necessary to correct for the Eddington bias.
This was done by adopting a Bayesian method 
to derive a probability distribution of fluxes based on the known 
distribution of AGN as a function of flux.
Following \citet{1991ApJ...374..344K}, \citet{2009ApJS..180..102L},
and \citet{2011MNRAS.414..992G},
the probability of a source having flux f$\rm_{X}$, 
given an observed number of counts C, is 

\begin{equation}\label{prob_flux}
\rm P(f_{X}, C) = \frac{T^{C}e^{-T}}{C!}\pi(f_{X})
\end{equation}

\noindent where C is the total number of observed source and background counts,
T is the mean expected total counts in the detection cell for a given flux,
and $\rm \pi(f_{X})$ is the prior, which is the 
distribution of AGN per unit X-ray flux interval.
The exact expression for the prior was taken from
\citet{2008MNRAS.388.1205G}, equation 1.

Source and background counts were taken from 
the 2RXS catalogue \citep{2016A&A...588A.103B}.
A flux-count rate conversion factor,
which was required to estimate T in equation~\ref{prob_flux},
was derived using \texttt{XSPEC} \citep{1996ASPC..101...17A}
assuming a model consisting of 
a power law (with $\rm \Gamma=2.4$ following \citet{2017MNRAS.469.1065D})
absorbed by the Milky Way column density.
This method was used to estimate 
the flux in the full ROSAT band (0.1-2.4\,keV)
as well the monochromatic flux at 2\,keV.

The fluxes resulting from the method described above 
with and without applying the prior (termed ``Bayesian'' and ``classical'', respectively)
are compared in Figure~\ref{bay_class_flux}.
The disagreement between the two flux estimates increases with decreasing flux,
which is expected since, without the prior, 
the classical method fails to account for the Eddington bias. 
Low count rate sources in this sample would be assigned 
unrealistically low Bayesian fluxes. 
To avoid this, the flux was left as undetermined
when the Bayesian flux estimate was more than a factor of ten smaller 
than the classical flux estimate.

\section{Accessing the Data}\label{access}

\noindent
The results from the spectral analysis discussed above,  
along with X-ray flux measurements and visual inspection results,
have been made available in an SDSS DR14 value added catalogue
which is available at 
http://www.sdss.org/dr14/data\_access/value-added-catalogs/.
Additionally, an extended version of the catalogue will be maintained at 
http://www.mpe.mpg.de/XraySurveys/SPIDERS/SPIDERS\_AGN/
The column description for the catalogue is given in appendix~\ref{columns}.

\begin{figure*}
	\centering
                \includegraphics[width=0.49\textwidth]{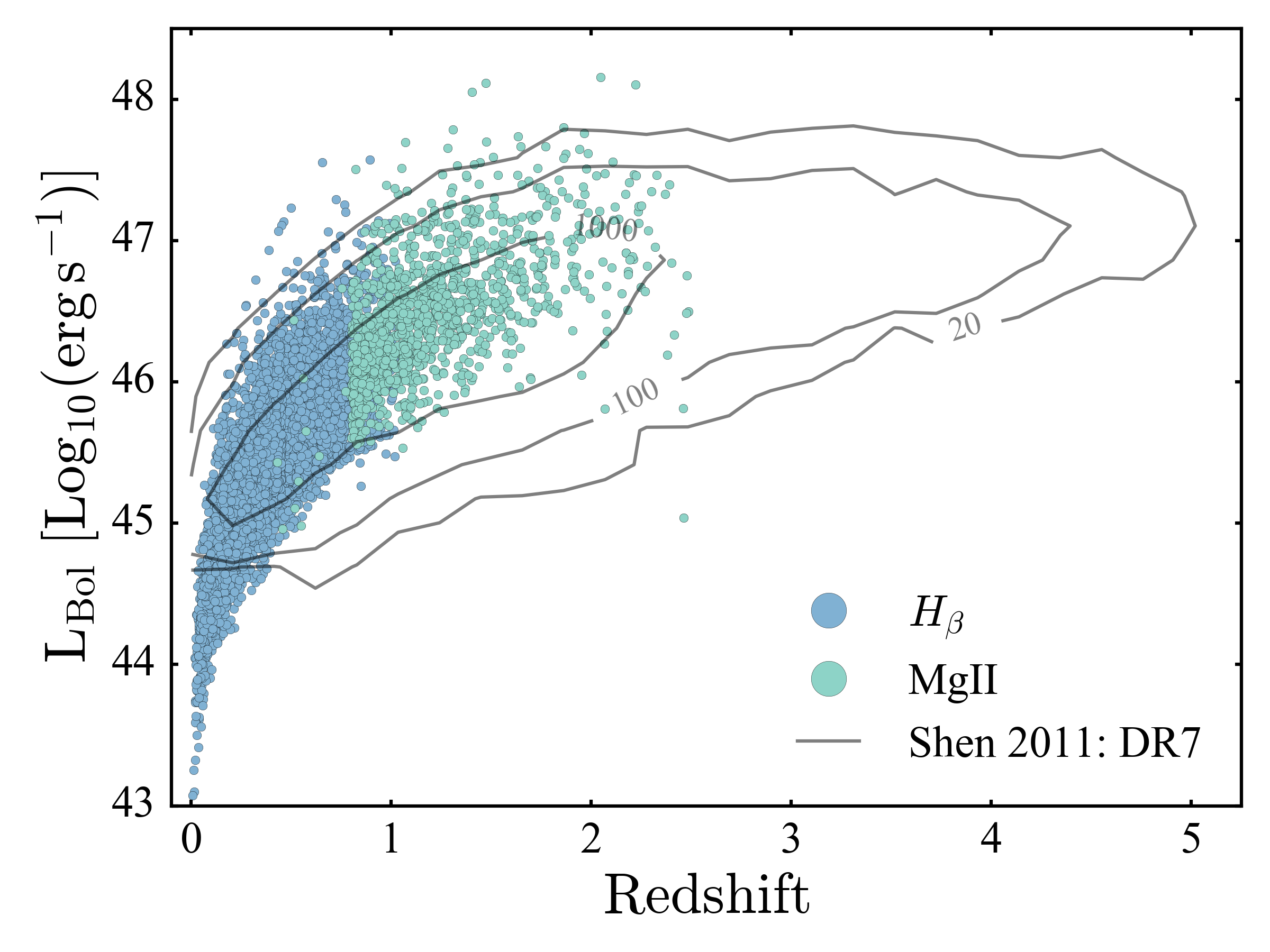}
                \includegraphics[width=0.49\textwidth]{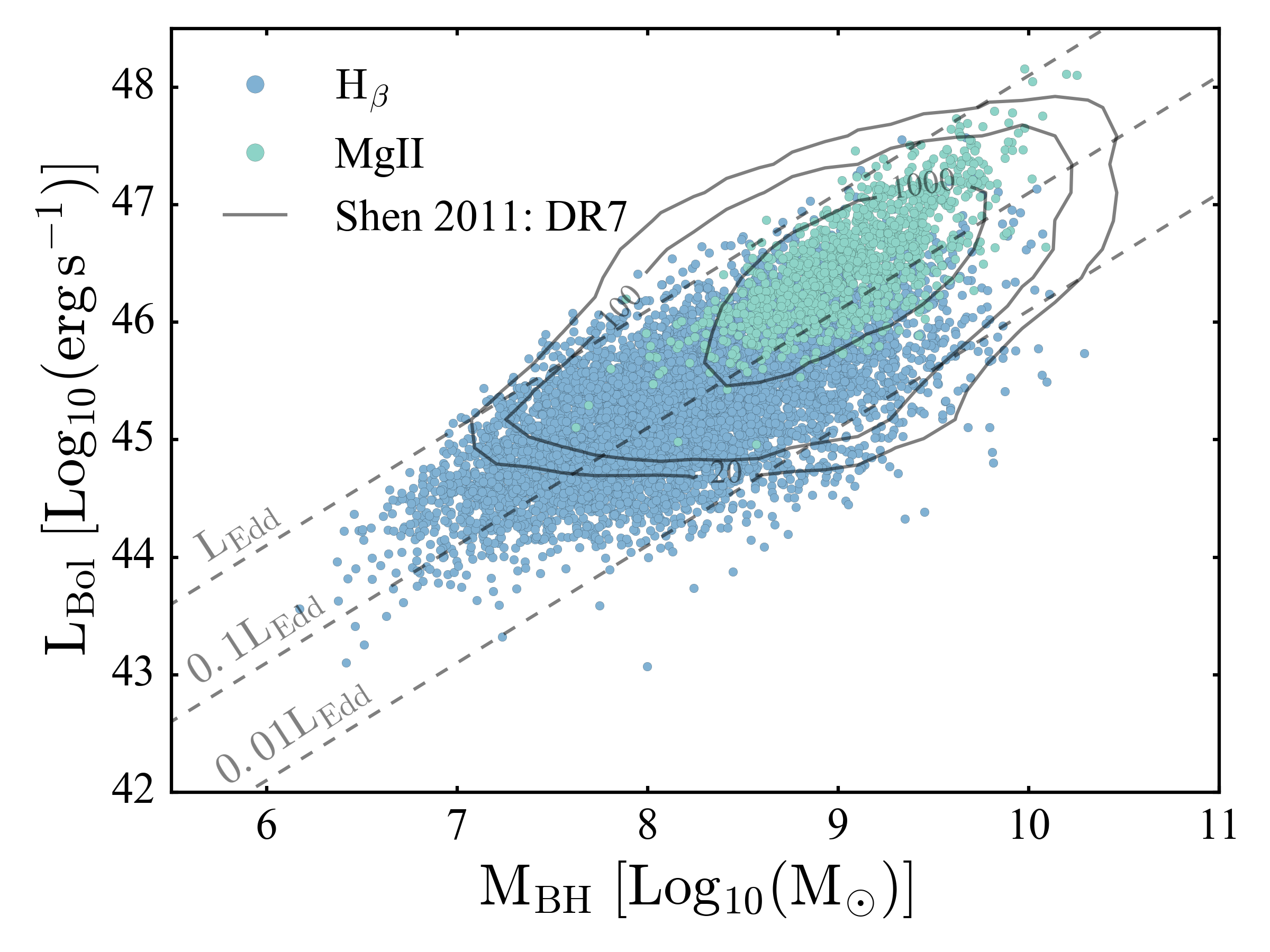}
\caption{\footnotesize 
Bolometric luminosity versus redshift (left panel) and 
bolometric luminosity versus BH mass (right panel)
for the sample presented in this work
and the \citet{2011ApJS..194...45S} sample.
Sources with $\rm H_{\beta}$-derived BH masses are shown in blue,
and sources with MgII-derived BH masses are displayed in green.}
\label{properties}
\end{figure*}

\section{Comparing the UV and Optical Fitting Results}\label{compare_uv_opt}

\noindent
A subsample of sources have spectral measurements available from 
both the MgII and H$\beta$ spectral regions. 
In order to test the consistency of the independent fits to these two regions,
properties measured from both were compared.

\subsection{$L_{2500\AA}-L_{5100\AA}$ Relation}

\noindent
A  subsample of AGN whose spectra cover the rest-frame wavelengths 
2500$\AA$ and 5100$\AA$
was selected using the following criteria:

\begin{equation}\nonumber
\begin{split}
\rm  ( (INSTRUMENT = SDSS) \,\,\&\&\,\, (0.52 < redshift < 0.8) ) \,\, || \\  
\rm ( (INSTRUMENT = BOSS) \,\,\&\&\,\, (0.46 < redshift <1.04) ) 
\end{split}
\end{equation}

\noindent
Of this sample, 1718 sources had reliable measurements of both 
$\rm L_{2500\AA}$ and $\rm L_{5100\AA}$.
The $\rm L_{2500\AA}-L_{5100\AA}$ relation was fit 
using the LINMIX \citep{2007ApJ...665.1489K} package.
LINMIX is a Bayesian linear regression algorithm 
that accounts for uncertainties in both dependent and independent variables,
as well as non-detections.
The upper left panel of figure~\ref{compare}
shows the $\rm L_{2500\AA}-L_{5100\AA}$ distribution and
the best-fit relation

\begin{equation}\label{flux_conversion}
\begin{split}
\rm Log_{10}(L_{5100\AA}) = (0.841\pm 0.007) Log_{10}(L_{2500\AA}) \\
\rm -(6.0 \pm 0.3)
\end{split}
\end{equation}

\noindent
with a regression intrinsic scatter of 0.0151.
The comparison between the estimated bolometric luminosities 
derived from the 3000$\AA$ and 5100$\AA$ monochromatic fluxes
is shown in the upper right panel of figure~\ref{compare}.
Equation~\ref{flux_conversion} can be used to estimate L$\rm_{2500\AA}$
from L$\rm_{5100\AA}$, which allows low redshift sources to be included 
in the $\rm \alpha_{OX}$ analysis discussed in section~\ref{aox_section}.

\subsection{Comparing MgII and H$\beta$ FWHM Measurements}\label{compare_mass}

A  subsample of AGN whose spectra cover the broad
$\rm H\beta$ and MgII emission lines 
was selected using the following criteria

\begin{equation}\nonumber
\begin{split}
\rm  ( (INSTRUMENT = SDSS) \,\,\&\&\,\, (0.45 < redshift < 0.81) ) \,\, || \\  
\rm ( (INSTRUMENT = BOSS) \,\,\&\&\,\, (0.38 < redshift <1.05) ) 
\end{split}
\end{equation}

\noindent
Of this sample, 2323 sources had FWHM measurements
for both H$\beta$ and MgII.
The lower left panel of figure~\ref{compare}
displays the virial FWHM measurements from H$\beta$ and MgII.
The resulting best-fit relation, fit using LINMIX, is

\begin{equation}\nonumber
\begin{split}
\rm Log_{10}(FWHM_{MgII}) = (0.65\pm0.01)\, Log_{10}(FWHM_{H\beta}) \\
\rm +(1.21\pm0.04)
\end{split}
\end{equation}

\noindent
with a regression intrinsic scatter of 0.005.
This deviation from the one-to-one relation has also been observed by 
\citet{2009ApJ...707.1334W}, who reported a slope of $0.81\pm0.02$,
and \citet{2012ApJ...753..125S}, who found a slope of $0.57\pm0.09$.
The single-epoch BH mass relations (equation~\ref{mass_est})
account for the $\rm FWHM_{MgII}-FWHM_{H\beta}$ slope;
when the correct BH mass calibration is used,
the H$\beta$ and MgII virial FWHM measurements 
yield BH masses that are in close agreement
(see the lower right panel of figure~\ref{compare}).

\section{Sample Properties}\label{prop}

\noindent
Figure~\ref{properties} presents the comparison between
this sample and the full sample of SDSS DR7 AGN 
with optical spectral properties measured by \citet{2011ApJS..194...45S}
in the bolometric luminosity-redshift and bolometric luminosity-BH mass planes.
As discussed in section~\ref{intro}, $\rm H_{\beta}$-derived BH masses 
are used where available (shown in blue), 
while MgII-derived masses are used 
for the remaining higher-redshift sources (shown in green).
The left panel of figure~\ref{properties} 
shows that this sample populates 
the low-redshift, high-luminosity region of the parameter space,
which is partially due to the high flux threshold of the X-ray selection.
From the right panel of figure~\ref{properties} 
it can be seen that the sample presented in this work 
appears to be well bounded by the Eddington limit 
at least up to $\rm M_{BH}\simeq 10^{9.5}\,M_{\odot}$.

\subsection{H$\beta$ Line Components}\label{VBC_section}

\begin{figure}
	\centering
                \includegraphics[width=0.49\textwidth]{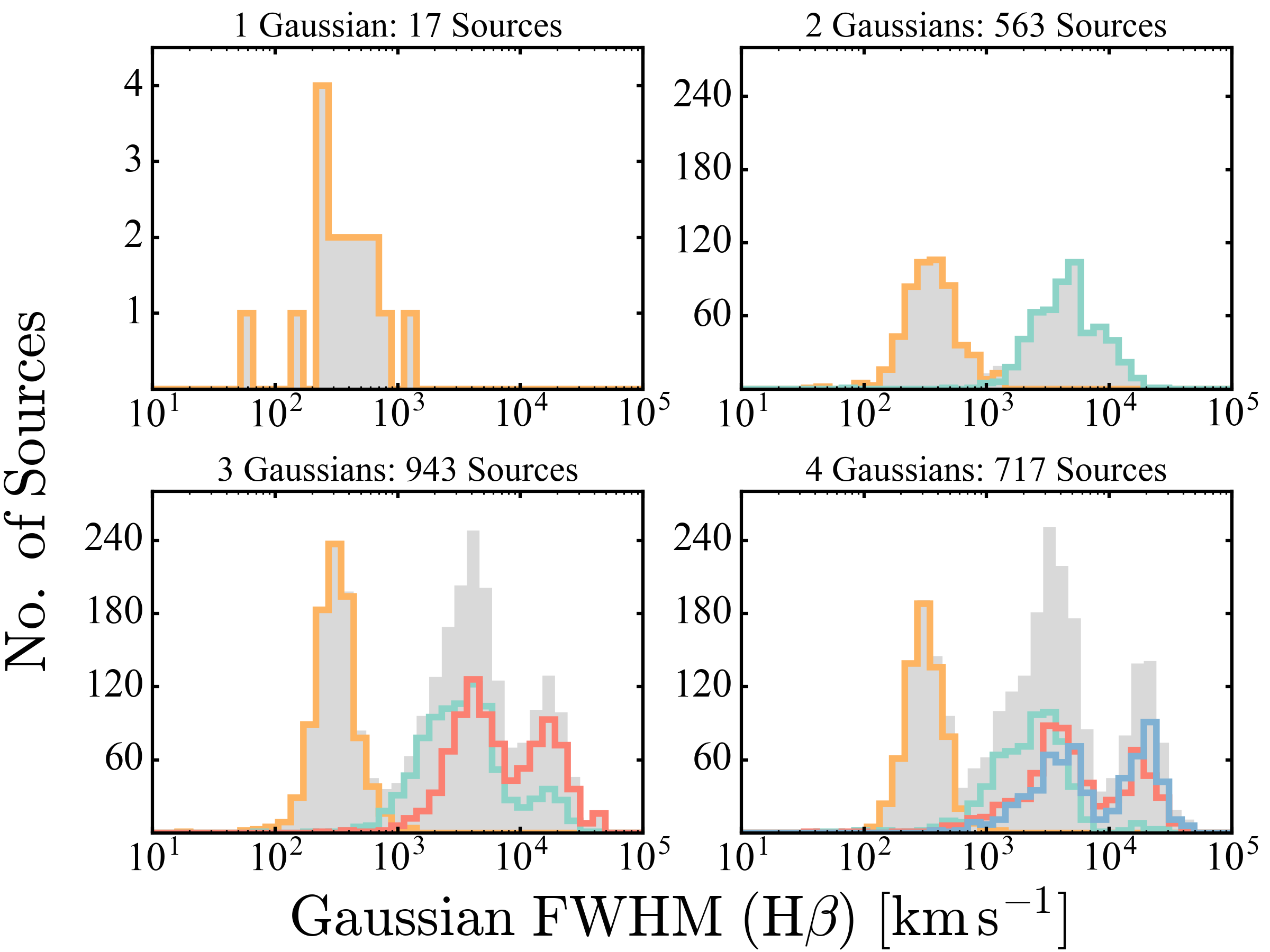}
\caption{\footnotesize 
Distribution of H$\beta$ Gaussian FWHM values.
The panels are split based on the number of Gaussians required to fit the line.
The coloured histograms each represent one of up to four 
possible Gaussians used to fit the H$\rm_{\beta}$ line. 
The grey histograms represent the sum of the individual coloured histograms.}
\label{line_width}
\end{figure}

\noindent
Section~\ref{HB_fit_method} described how the H$\beta$ line profile
was fit with either one, two, three, or four Gaussian components.
Figure~\ref{line_width} displays the resulting distribution of 
$\rm H\beta$ FWHM measurements
(the panels are split based on the 
number of Gaussian components required to fit the line).
There is a clear peak in the distribution at low FWHM
associated with the narrow H$\beta$ core
typically measured at a few hundred $\rm km\,s^{-1}$.
Above $\rm \sim 1000\,km\,s^{-1}$ the distribution is bimodal
(in the two lower panels) with a large number of sources showing evidence for the 
``very broad component'' (VBC) of $\rm H\beta$
at FWHM$\rm \,\geq 10000\,km\,s^{-1}$
also discussed in \citet{2010MNRAS.409.1033M}.

It has been suggested that the VBC
is emitted from a distinct physical region, 
and is possibly the result of line emission from the accretion disk 
\citep[e.g.][]{2009MNRAS.400..924B}.
If the VBC represents emission from the accretion disk,
then a strong VBC may result in a bias towards  
a higher BH mass estimate, since the 
single-epoch method assumes a calibration that is based on 
the luminosity-BLR radius relation.
However, since the kinematics and physical origin of the VBC remains uncertain,
detected VBCs have not been excluded from the broad line profiles 
used to measure the virial FWHM in this analysis 
(as discussed in section~\ref{bl_decomp}).

\begin{figure*}
                \includegraphics[width=0.49\textwidth]{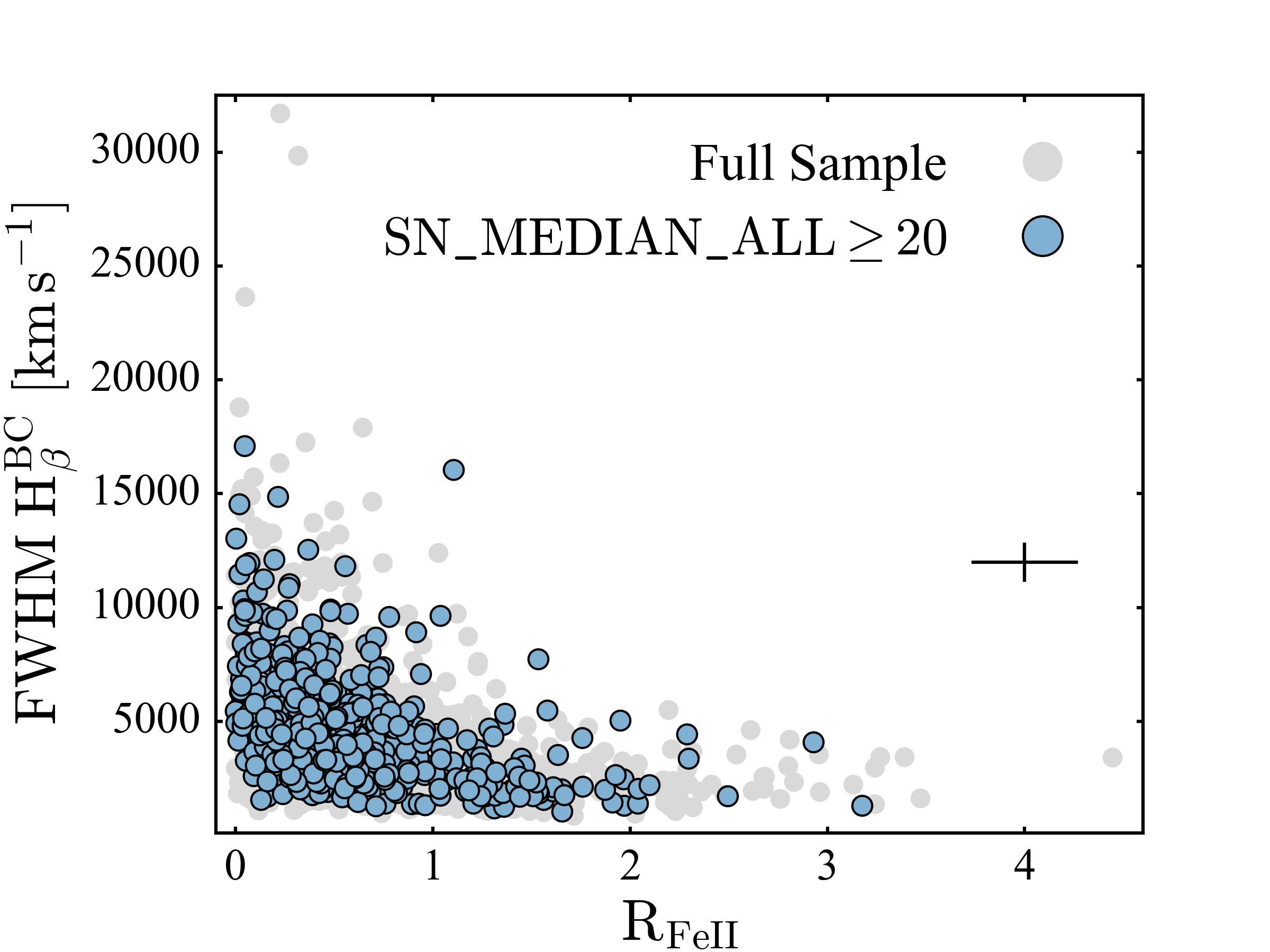}
                \includegraphics[width=0.49\textwidth]{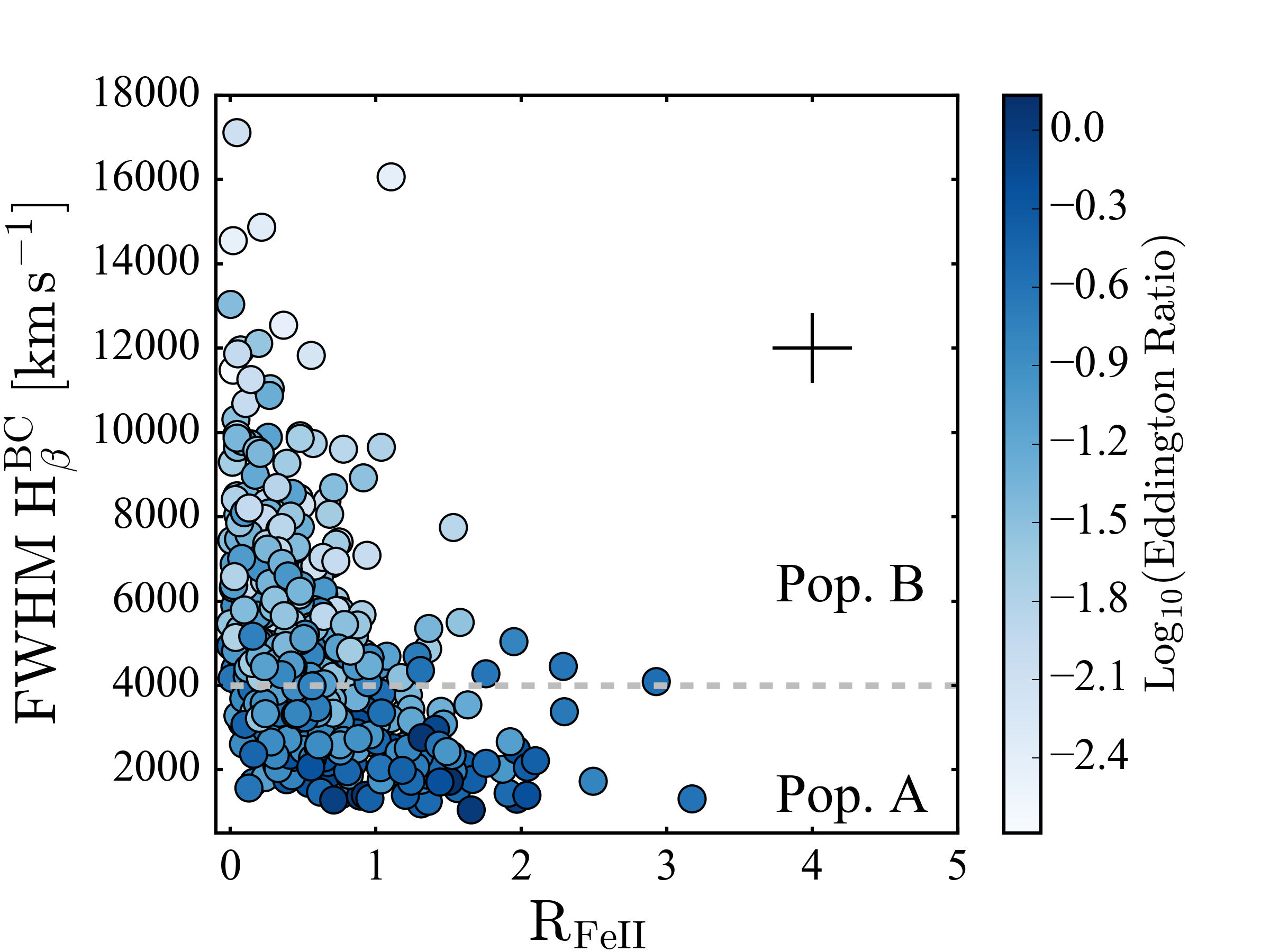}
\caption{\footnotesize 
FWHM of the broad component of H$\beta$ versus R$\rm _{FeII}$.
The left panel displays the sample described in section~\ref{4DE1_section} (grey), 
and sources with a median spectral S/N$\,\geq\,$20 (blue).
The right panel presents the same subsample of high S/N sources, 
colour coded to indicate the trend in Eddington ratio across the distribution.
The grey dashed line marks the division between 
population A ($\rm FWHM\,H_{\beta}^{BC}\leq4000\,km\,s^{-1}$) and 
population B ($\rm FWHM\,H_{\beta}^{BC}\geq4000\,km\,s^{-1}$) sources.
The size of the typical uncertainties, multiplied by a factor of five,
in both variables for the high-S/N subsample is also shown.}
\label{4DE1}
\end{figure*}

\subsection{This Sample in the 4D Eigenvector 1 Context}\label{4DE1_section}

\noindent
The 4D Eigenvector 1 (4DE1) system 
\citep{1992ApJS...80..109B, 2000ApJ...536L...5S, 2011BaltA..20..427S}
aims to define a set of parameters that uniquely account for AGN diversity.
Two main 4DE1 parameters are
the FWHM of the broad component of H$\beta$ ($\rm FWHM \, H_{\beta}^{BC}$)
and the strength of the FeII emission relative to that of H$\beta$, defined as 

\begin{equation}\nonumber
\rm R_{FeII} = F_{FeII}/F_{H\beta}
\end{equation}

\noindent
where $\rm F_{FeII}$ and $\rm F_{H\beta}$
are the fluxes of the 
FeII emission in the 4434-4684$\rm \AA$ range
and broad $\rm H\beta$ line, respectively.
A sample of 2098 sources with measurements of these parameters 
and reliable spectral fits ($\rm 0 \le \chi^{2}_{\nu,H\beta} \le 1.2$)
was selected. The left panel of figure~\ref{4DE1} shows 
the distribution of this sample 
in the 4DE1 parameter space (grey).
It is expected that a reliable measurement of the FeII component 
will be difficult for many of the lower S/N sources
\citep[see][]{2003ApJS..145..199M}.
For this reason, the subset of sources in figure~\ref{4DE1} 
with a median S/N greater than or equal to 20 per resolution element is also shown (blue).
The right panel of figure~\ref{4DE1} presents the higher S/N sources, 
colour-coded as a function of Eddington ratio.
The expected trend of increasing Eddington ratio 
towards smaller $\rm FWHM\,H_{\beta}^{BC}$ and larger R$\rm_{FeII}$ 
is observed for this sample of high-S/N sources.
Typically, sources with both high $\rm R_{FeII}$
and high $\rm FWHM \, H_{\beta}^{BC}$ are not observed. 
If these sources exist, they may be difficult to detect 
since strong FeII emission might conceal a faint $\rm H_{\beta}$ broad component.
The potential bias in the 4DE1 plane source distribution 
due to model limitations and spectral S/N 
is discussed in sections~\ref{sim} and \ref{sim_bias}.

The grey dashed line in the right panel of figure~\ref{4DE1}
indicates the division between 
population A ($\rm FWHM\,H_{\beta}^{BC}\leq4000\,km\,s^{-1}$) 
and population B ($\rm FWHM\,H_{\beta}^{BC}\geq4000\,km\,s^{-1}$)
sources in the 4DE1 context \citep[see][]{2011BaltA..20..427S}.
Population A sources often possess Lorentzian broad line profiles,
and it has been suggested that Gaussian fits to population A broad lines 
will result in an underestimation of the BH mass \citep[see][]{2014AdSpR..54.1406S}.

\subsection{Relationship Between AGN X-ray and Optical Emission}\label{aox_section}

\begin{figure}
	\centering
                \includegraphics[width=0.49\textwidth]{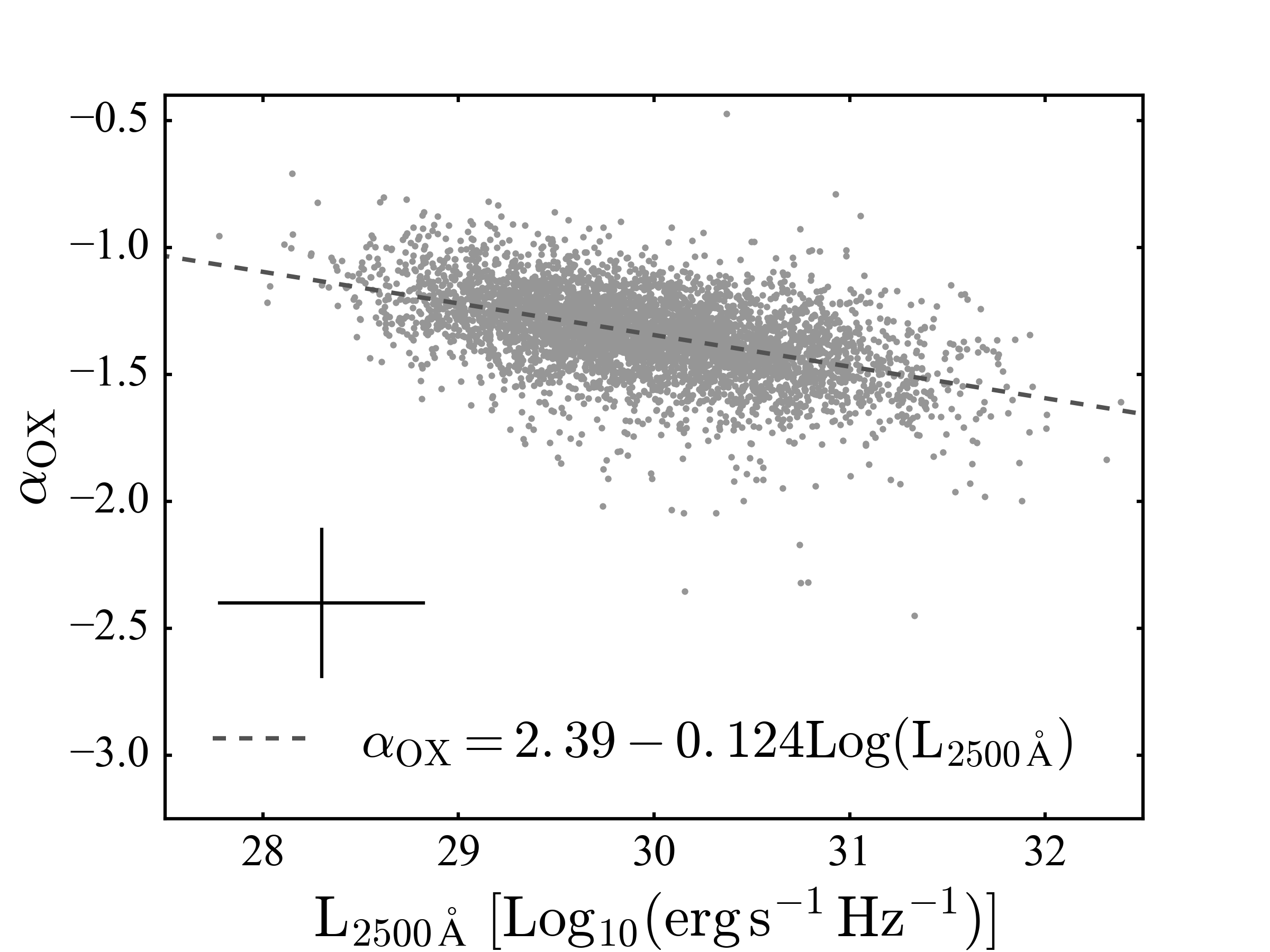}
\caption{\footnotesize
$\rm \alpha_{OX}$ versus UV luminosity.
The dashed line is the best linear fit to the distribution.
The size of the typical uncertainties in both variables is also shown.}
\label{aox}
\end{figure}

\noindent
Quasars exhibit a non-linear relationship between their X-ray and UV emission,
usually represented by the $\rm \alpha_{OX}$ parameter

\begin{equation}\nonumber
\rm \alpha_{OX} = \frac{Log(L_{2\,keV}/L_{2500\,\AA})}{Log(\nu_{2\,keV}/\nu_{2500\,\AA})}
\end{equation}

\noindent
where $\rm L_{2\,keV},\, L_{2500\,\AA},\, \nu_{2\,keV},\, and \, \nu_{2500\,\AA}$
are the monochromatic luminosities and frequencies at 
2\,keV and $\rm 2500\,\AA$, respectively
\citep{2003AJ....125.2876V, 2005AJ....130..387S, 2006AJ....131.2826S, 2007ApJ...665.1004J, 2008ApJS..176..355K, 2009ApJ...690..644G, 2009ApJS..183...17Y, 2010A&A...512A..34L}.
The $\rm \alpha_{OX}$ parameter is considered to be 
a proxy for the relative contribution of
the UV accretion disk emission and the 
X-ray emission from the surrounding corona
to the total luminosity.
In order to study this relationship, 
a sample of sources with measurements of the 
2keV,  2500\AA, and 5100\AA \, luminosities was selected.
For lower redshift sources without spectral coverage of 2500\AA,
equation~\ref{flux_conversion} was used to estimate the 2500\AA\, luminosity
from the 5100\AA\, luminosity.
Extended sources were removed in order to prevent additional scatter in the relationship due to the contribution of the host galaxy.
This was done by requiring that the SDSS g band ``stellarity''\footnote{
For a description of how cModelMag\_g and psfMag\_g are measured see
https://www.sdss.org/dr12/algorithms/magnitudes/} 
(defined as S(g) = cModelMag\_g- psfMag\_g) 
lies between $\pm$0.1. 
This sample does not contain X-ray sources with 
more than one potential AllWISE counterpart and therefore avoids cases where 
the X-ray detection includes emission from more than one object.
This selection process resulted in a sample of 4777 sources.
Figure~\ref{aox} shows the $\rm \alpha_{OX}$ parameter versus 
the monochromatic luminosity at 2500$\AA$.
The $\rm \alpha_{OX}-L_{2500\AA}$ relation
was fit using LINMIX, which gave the following best-fit result

\begin{equation}\nonumber
\rm \alpha_{OX} = 2.39\pm0.16 - (0.124\pm0.005) Log(L_{2500\AA})
\end{equation}

\noindent
with a regression intrinsic scatter of 0.0034.
This slope is consistent with previous results from the literature
\citep[e.g.][]{2008ApJS..176..355K}.

\section{Interpreting the Data and Limitations}\label{limits}

\noindent
In this section, the reliability and limitations of the sample will be discussed.

\subsection{Measuring the FeII Emission}

\noindent
Distinguishing the FeII component from the continuum emission 
becomes more difficult when using low S/N spectra.
In addition, for a given S/N, it may also be more difficult to detect FeII emission 
if the intrinsic broadening of the FeII lines is large, 
since broader, blended FeII emission lines are more likely to be fit 
by the model as continuum emission \citep[see][]{2003ApJS..145..199M}.
Using simulated AGN spectra, \citet{2003ApJS..145..199M} 
estimate the minimum detectable optical FeII emission 
as a function of H$\beta$ width for different bins of S/N.

A poor fit to the FeII emission may affect the accuracy of the BH mass estimates,
since FeII emission can influence measurements of both the broad line width (see section~\ref{feII_FWHM}) and the continuum luminosity.
FeII emission may also conceal a broad $\rm H \beta$ component
thus biasing a source's position in the 4DE1 plane (figure~\ref{4DE1}).
These potential issues are tested in the following three sections.

\subsubsection{Accuracy of the Broad Emission Line FWHM Measurements 
for Sources with FeII Continuum Emission}\label{feII_FWHM}

\begin{figure}[h]
\centering
\includegraphics[width=0.45\textwidth]{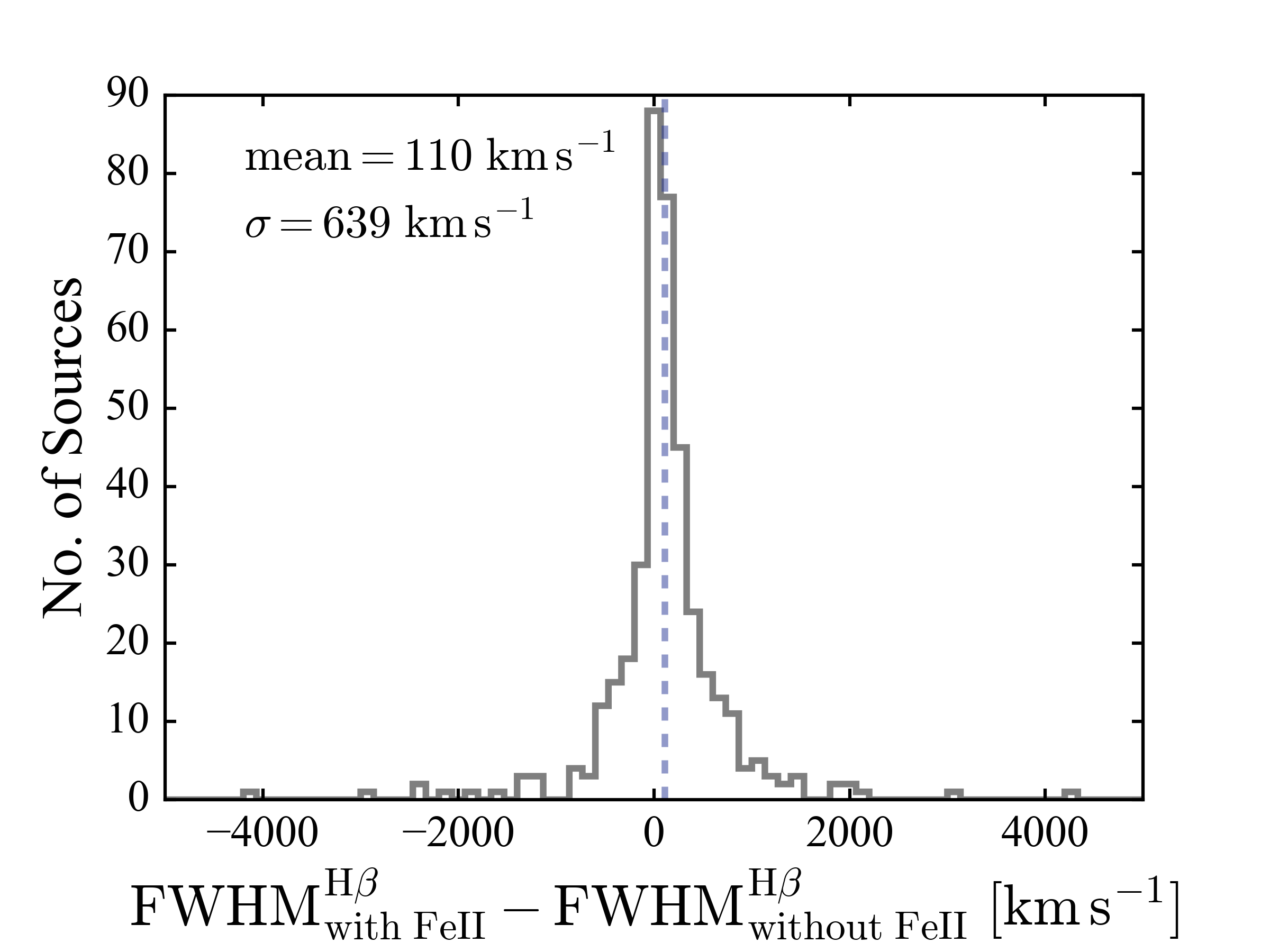}
\caption{\footnotesize
Comparison of the $\rm H \beta$ FWHM measurements
derived using a model with and without an FeII template.
The vertical dashed line shows the mean value of the distribution.}
\label{feII_test_plot}
\end{figure}

A poor fit to the FeII emission may affect the measurement of 
the broad emission line width.
To quantify the magnitude of this effect, the $\rm H\beta$ fitting script 
(using four Gaussians to fit $\rm H\beta$) was run with and without the FeII template
on a sample of $\rm \sim 400$ randomly selected sources.
The fit without the FeII template represents the most extreme case 
where the FeII emission is completely ignored by the model.
Therefore, the change in line widths measured by these two models 
should be the upper limit on what can be expected 
for cases where the FeII fit is inadequate.
Figure~\ref{feII_test_plot} shows that the line width dispersion induced by 
ignoring the presence of FeII emission is $\rm \simeq 640 \,km\,s^{-1}$.

\subsubsection{Model Limitations in Detecting Sources in the 4DE1 Plane}\label{sim}

\noindent
Sources with both large $\rm R_{FeII}$ and large $\rm FWHM\,H_{\beta}^{BC}$
are typically not observed, however,
this absence may be due to model limitations;
at high $\rm R_{FeII}$ the broad $\rm H\beta$ component 
may be concealed beneath the FeII emission, 
and therefore may not be detected.
The experiment outlined in this section was carried out in order to determine 
whether the spectral fitting code used in this work
would return accurate measurements 
for sources with high $\rm R_{FeII}$ and $\rm FWHM\,H_{\beta}^{BC}$ values.

A parameter space defined by 
$\rm 0.1 \leq R_{FeII} < 5$ and
$\rm 1000\,\,km\,s^{-1} \leq FWHM\,H_{\beta}^{BC} < 15000 \,\,km\,s^{-1}$
was divided into a 12$\times$12 grid.
10 S/N bins between 5 and 50 
(a representative range for the samples presented in this work)
were selected for each point on the grid,
and 10 spectra were simulated for each 
$\rm R_{FeII}-\rm FWHM \, H_{\beta}^{BC}-S/N$ combination, 
resulting in 14400 simulated spectra.
For the parameters that were fixed in this experiment, 
the interquartile mean of the best-fit values for the type 1 AGN in this sample were used.
The $\rm H\beta$ line profile was modelled with one narrow and one broad Gaussian.
The wavelength range was set to $\rm 4420 - 5500 \AA$ (as in section~\ref{HB_fit_method})
and the logarithmic wavelength 
spacing\footnote{https://www.sdss.org/dr12/spectro/spectro\_basics/} 
was set to be equal to that of SDSS spectra;

\begin{equation}\nonumber
\rm Log_{10}\lambda_{i+1} - Log_{10}\lambda_{i} = 0.0001
\end{equation}

These spectra were fit using a version of the $\rm H\beta$ fitting script 
which used one narrow and one broad Gaussian component to fit $\rm H\beta$.
The minimum S/N required for the fitting script to return the correct 
$\rm R_{FeII}$ and $\rm FWHM \, H_{\beta}^{BC}$ combinations
was then determined.
In order to consider an $\rm R_{FeII}$ and $\rm FWHM \, H_{\beta}^{BC}$
combination detectable at a given S/N,
at least  7/10 spectra were required to have best fit 
$\rm R_{FeII}$ and $\rm FWHM \, H_{\beta}^{BC}$ values 
that agreed with the input values.

\begin{figure}
\centering
\includegraphics[width=0.45\textwidth]{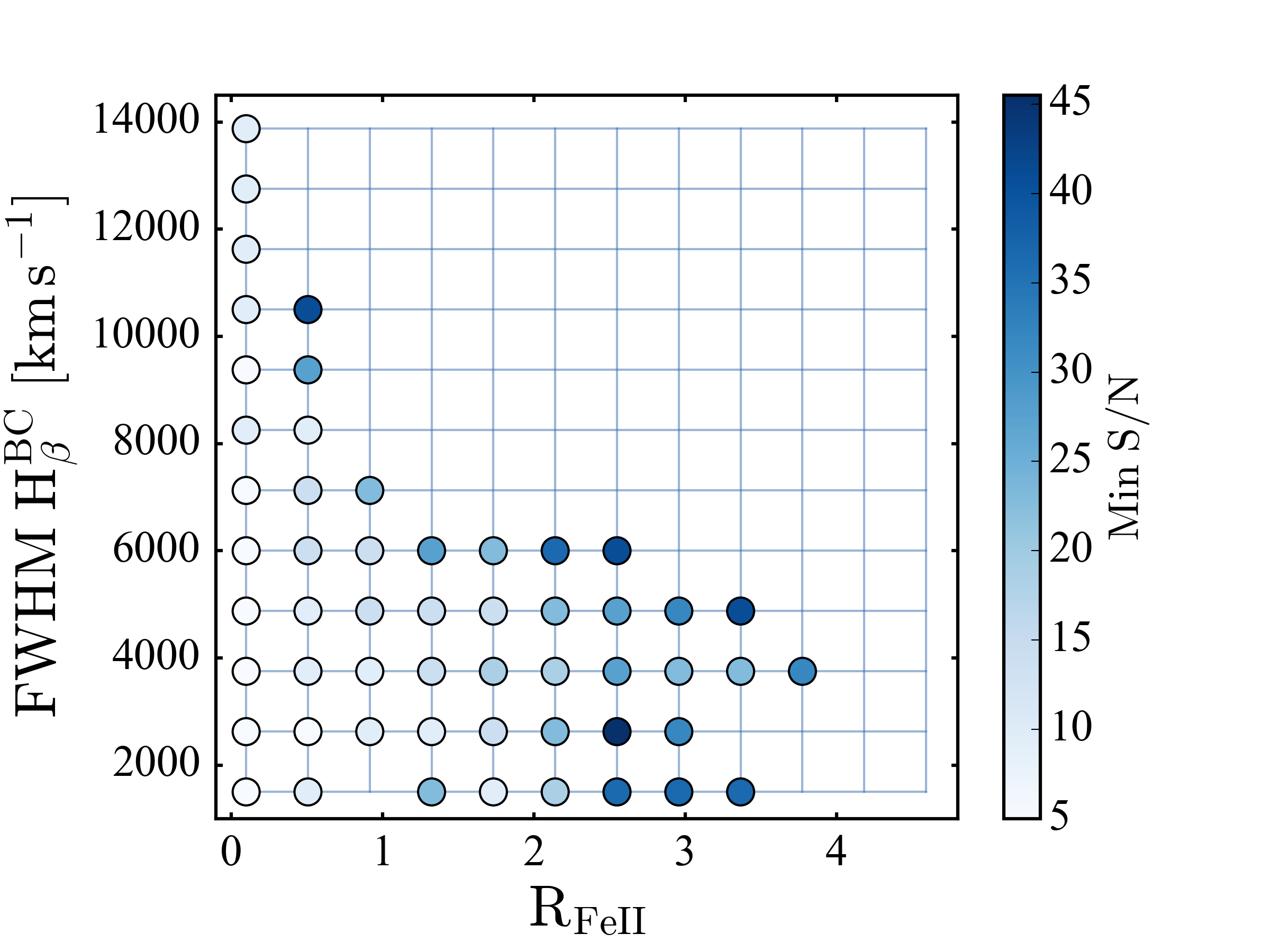}
\caption{\footnotesize
Minimum S/N required to detect a range of 
$\rm R_{FeII}$ and $\rm FWHM \, H_{\beta}^{BC}$ combinations.
The blue grid indicates the range of the parameter space covered
in the simulation described in section~\ref{sim}.
Points on the grid that do not have a minimum S/N indicator represent 
$\rm R_{FeII}$ and $\rm FWHM \, H_{\beta}^{BC}$ combinations
that are not detectable even at the highest S/N used in this experiment.
Sources detected at these $\rm R_{FeII}$ and $\rm FWHM \, H_{\beta}^{BC}$ 
combinations are likely to be spurious (see figure~\ref{sn_slice}).}
\label{4DE1_sim_1}
\end{figure}

Figure~\ref{4DE1_sim_1} shows the detectable 
$\rm R_{FeII}$ and $\rm FWHM \, H_{\beta}^{BC}$
combinations for each point on the grid, along with the minimum S/N required to 
detect that combination. 
It is clear from figure~\ref{4DE1_sim_1} that 
at the S/N levels available in this sample,
a large region of the 4DE1 parameter space would not be detected.

\begin{figure*}
\centering
\includegraphics[width=0.49\textwidth]{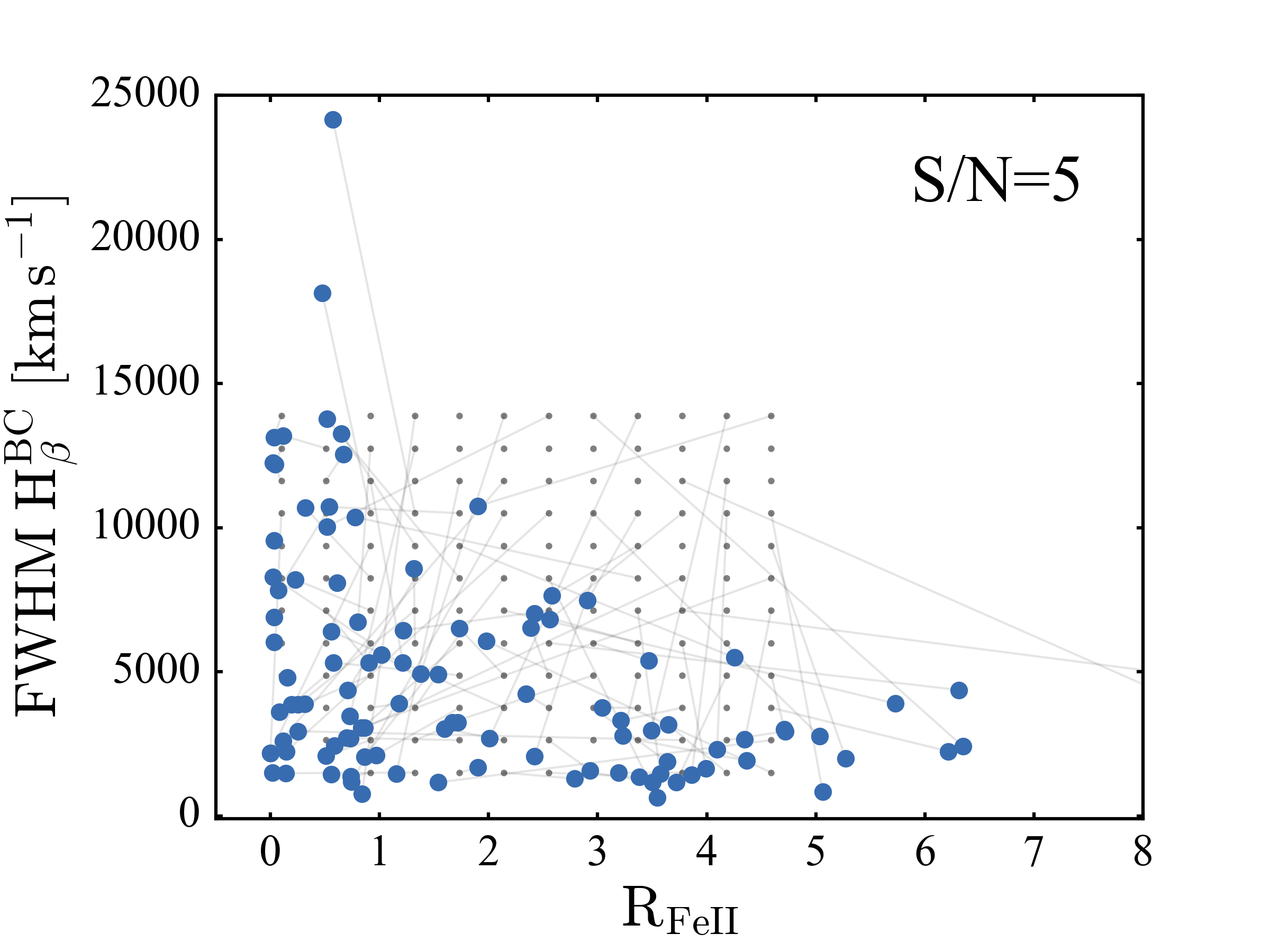}
\includegraphics[width=0.49\textwidth]{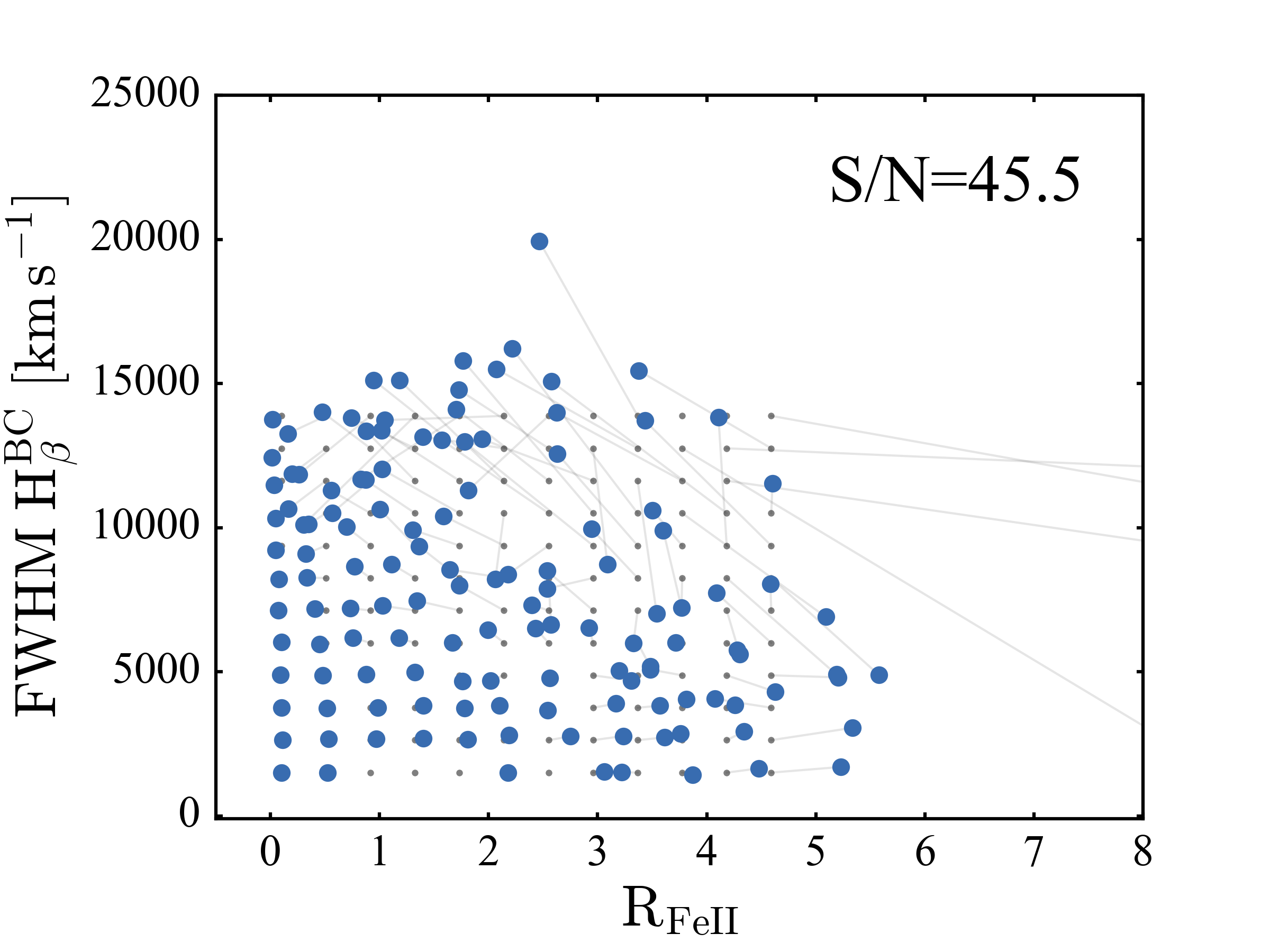}
\caption{\footnotesize
Comparison between the measured and simulated 
$\rm R_{FeII}$ and $\rm FWHM \, H_{\beta}^{BC}$ values
for the highest and lowest S/N bins used in section~\ref{sim}.
For clarity, each figure displays only one of the ten sets of spectra
for that S/N.
The values used to simulate the spectra (grey points)
are connected to the corresponding best fit measurements
(except for cases where either the broad $\rm H\beta$
or FeII components were not detected).
For clarity, the figures do not show a small number of 
unphysically high $\rm R_{FeII}$ measurements.}
\label{sn_slice}
\end{figure*}

\subsubsection{Bias in the $\rm R_{FeII}$ and $\rm FWHM \, H_{\beta}^{BC}$
Distribution due to Model Limitations}\label{sim_bias}

\noindent
Figure~\ref{sn_slice} shows the comparison between the simulated and measured
$\rm R_{FeII}$ and $\rm FWHM \, H_{\beta}^{BC}$ values 
for the highest and lowest S/N bins used in section~\ref{sim}.
The left panel of figure~\ref{sn_slice} shows that at low S/N
the results are clearly biased against high  
$\rm R_{FeII}$ and $\rm FWHM \, H_{\beta}^{BC}$ values.
At higher S/N (figure~\ref{sn_slice}, right panel), 
the accuracy of the lower left quadrant measurements is significantly improved.
However, even at S/N=45.5 
(which is approximately the upper end of the S/N distribution 
of the samples presented in this work)
the high $\rm R_{FeII}$ - $\rm FWHM \, H_{\beta}^{BC}$ measurements 
deviate significantly from the corresponding ``true'' values.
This may suggest that the L-shaped distribution of sources
in the 4DE1 plane (e.g. figure~\ref{4DE1})
is at least in part due to model limitations.

\subsection{Reliability of the Single-Epoch Method for Mass Estimation}\label{reliability}

\noindent
Assuming that AGN broad emission lines are produced by gas 
whose motion is dominated by the gravitational potential of the central SMBH,
the single-epoch method is expected to produce 
reliable mass estimates when compared to RM \citep[see][]{2006ApJ...641..689V},
with a systematic uncertainty of 0.3-0.4 dex.
However, it is not clear how to measure 
the virial FWHM of lines that deviate from this norm.

\begin{figure*}
	\centering
                \includegraphics[width=0.49\textwidth]{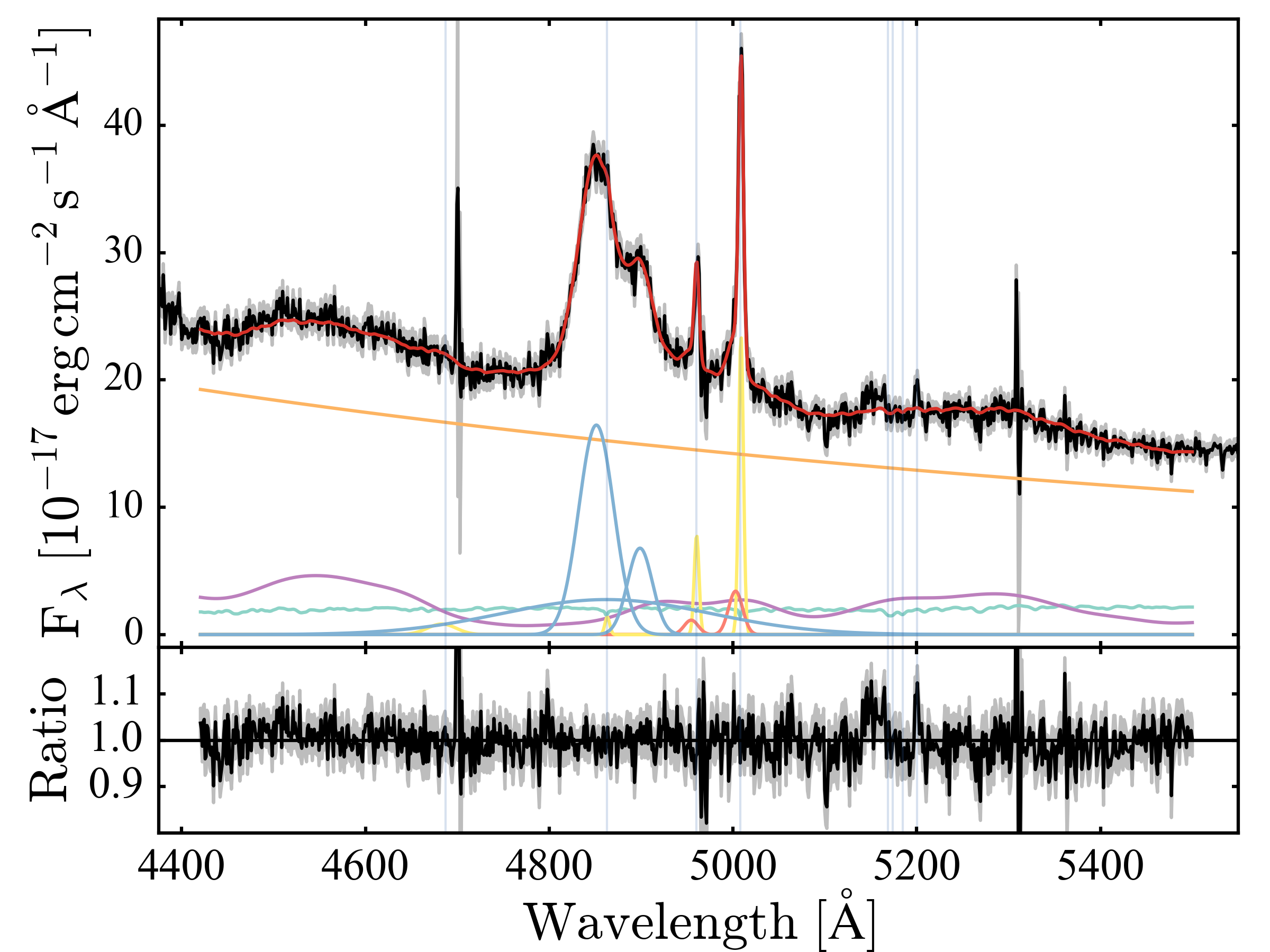}
                \includegraphics[width=0.49\textwidth]{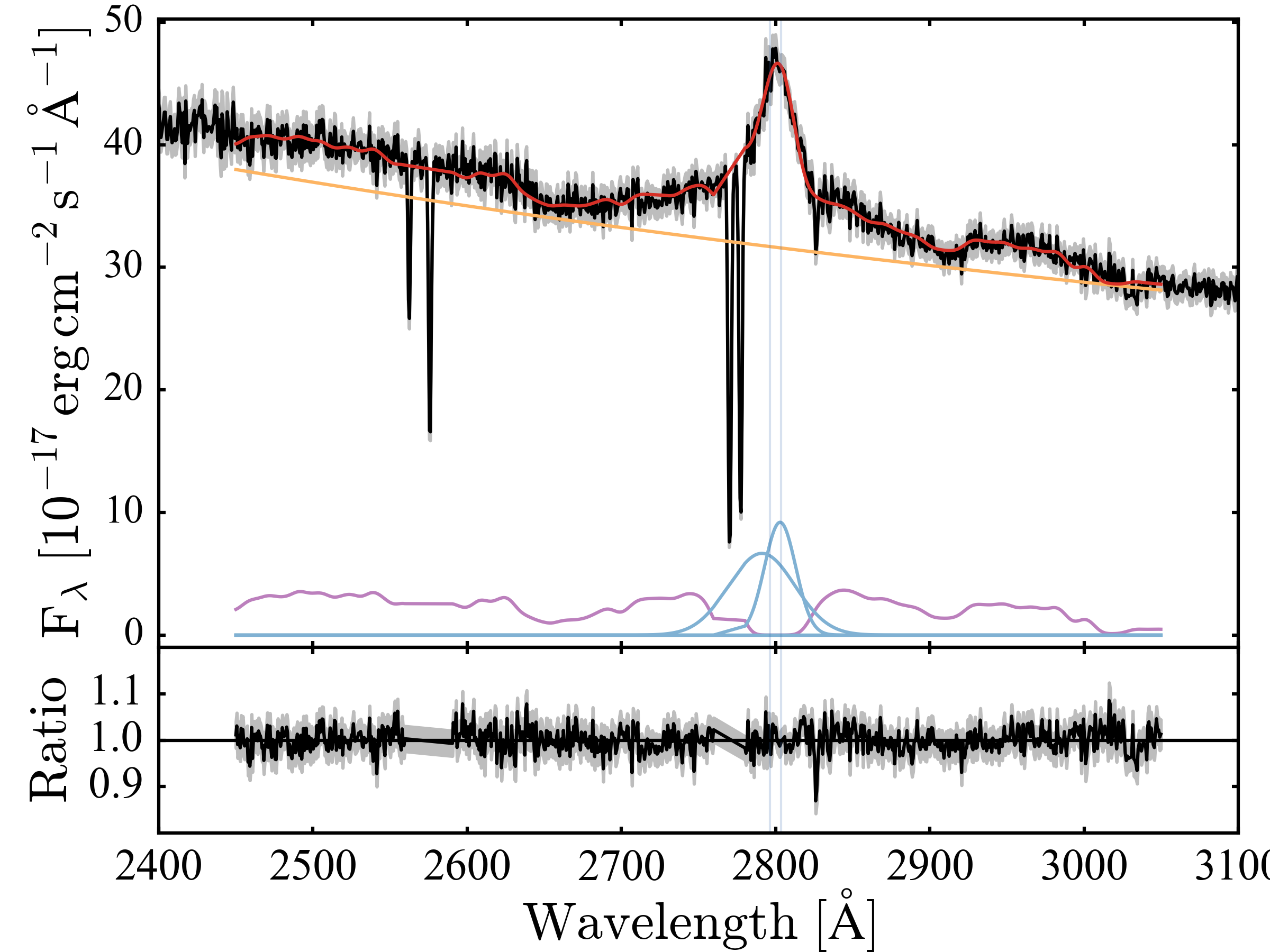}
\caption{\footnotesize
Left panel: Example of a source with a double-peaked H$\rm \beta$ line profile
(plate=7512, MJD=56777, fiber=321).
Right panel: Example of a source showing narrow UV absorption features
which have been masked when fitting the model
(plate=8188, MJD=57348, fiber=946). 
See section~\ref{reliability} for details.}
\label{unusual_fig}
\end{figure*}

The spectrum shown in the left panel of figure~\ref{unusual_fig}
is an example of a source which exhibits 
a double-peaked H$\beta$ line profile,
where a clear inflection point is visible 
between two velocity-shifted broad line components.
Double-peaked broad line profiles in AGN 
are expected to be the result of emission from the accretion disk
\citep{1988MNRAS.230..353P, 1989ApJ...339..742C, 1994ApJS...90....1E, 2003AJ....126.1720S, 2003ApJ...599..886E}.
\citet{2007MNRAS.376.1335Z} have found that single-epoch BH mass estimates 
obtained from double-peaked line profiles 
are significantly larger than BH mass estimates 
derived from stellar velocity dispersion measurements.
\citet{2007MNRAS.376.1335Z} suggest that this discrepancy 
is the result of an overestimation of the BLR radius
by the single-epoch mass calibrations for these objects.
Therefore, the BH mass estimates provided in this work for sources 
which exhibit double-peaked broad emission lines 
should be treated with caution.

The right panel of figure~\ref{unusual_fig} shows
an example of narrow absorption in the UV portion of the spectrum
caused by intervening absorbing material along the line of sight to the AGN.
Sources identified during the visual inspection 
as having narrow absorption lines 
have been flagged in the catalogue (column 189; flag\_abs).
These sources were fit using the model described in section~\ref{mgII_fit} 
with the absorption line regions masked.
However, in many cases, the absorption features distort the broad MgII line, 
and therefore the resulting BH mass estimates may not be reliable.

\section{Conclusions}

\noindent
This work presents a catalogue of spectral properties for
all SPIDERS type 1 AGN up to SDSS DR14.
Visual inspection results were used to select a reliable subsample for spectral analysis,
and the spectral regions around H$\beta$ and MgII were fit with a multicomponent model.
Using the single-epoch method, BH masses, bolometric luminosities, 
Eddington ratios, along with additional spectral parameters were measured. 
A catalogue containing these results has been made available
as part of a set of SDSS DR14 value added catalogues.
This catalogue also includes the results of a visual inspection of the sample,
and is available at 
http://www.mpe.mpg.de/XraySurveys/SPIDERS/SPIDERS\_AGN/.

\section{Acknowledgements}
DC has participated in the International Max Planck Research School 
on Astrophysics at the Ludwig Maximilians University Munich.
DC also acknowledges financial support from the Max Planck Society.
DC would also like to thank Riccardo Arcodia and Julien Wolf
for many helpful discussions.
The authors would like to thank Josephine Reisinger 
for her contribution to the visual inspection of sources in this sample.
Finally, the authors would like to thank 
the anonymous referee for providing a thorough critique of the paper 
which greatly improved its content.

Funding for the Sloan Digital Sky Survey IV has been provided by the Alfred P. Sloan Foundation, 
the U.S. Department of Energy Office of Science, and the Participating Institutions. SDSS-IV acknowledges
support and resources from the Center for High-Performance Computing at
the University of Utah. The SDSS web site is www.sdss.org.

SDSS-IV is managed by the Astrophysical Research Consortium for the 
Participating Institutions of the SDSS Collaboration including the 
Brazilian Participation Group, the Carnegie Institution for Science, 
Carnegie Mellon University, the Chilean Participation Group, the French Participation Group, Harvard-Smithsonian Center for Astrophysics, 
Instituto de Astrof\'isica de Canarias, The Johns Hopkins University, 
Kavli Institute for the Physics and Mathematics of the Universe (IPMU) / 
University of Tokyo, Lawrence Berkeley National Laboratory, 
Leibniz Institut f\"ur Astrophysik Potsdam (AIP),  
Max-Planck-Institut f\"ur Astronomie (MPIA Heidelberg), 
Max-Planck-Institut f\"ur Astrophysik (MPA Garching), 
Max-Planck-Institut f\"ur Extraterrestrische Physik (MPE), 
National Astronomical Observatories of China, New Mexico State University, 
New York University, University of Notre Dame, 
Observat\'ario Nacional / MCTI, The Ohio State University, 
Pennsylvania State University, Shanghai Astronomical Observatory, 
United Kingdom Participation Group,
Universidad Nacional Aut\'onoma de M\'exico, University of Arizona, 
University of Colorado Boulder, University of Oxford, University of Portsmouth, 
University of Utah, University of Virginia, University of Washington, University of Wisconsin, 
Vanderbilt University, and Yale University.

Funding for SDSS-III has been provided by the Alfred P. Sloan Foundation,
the Participating Institutions, the National Science Foundation, and the U.S. Department of Energy Office of Science. 
The SDSS-III web site is http://www.sdss3.org/.

SDSS-III is managed by the Astrophysical Research Consortium for the Participating Institutions of the SDSS-III Collaboration including 
the University of Arizona, the Brazilian Participation Group, Brookhaven National Laboratory, Carnegie Mellon University, University of Florida, 
the French Participation Group, the German Participation Group, Harvard University, the Instituto de Astrof\'isica de Canarias, 
the Michigan State/Notre Dame/JINA Participation Group, Johns Hopkins University, Lawrence Berkeley National Laboratory, 
Max Planck Institute for Astrophysics, Max Planck Institute for Extraterrestrial Physics, New Mexico State University, 
New York University, Ohio State University, Pennsylvania State University, University of Portsmouth, Princeton University, 
the Spanish Participation Group, University of Tokyo, University of Utah, Vanderbilt University, 
University of Virginia, University of Washington, and Yale University.

Funding for the SDSS and SDSS-II has been provided by the Alfred P. Sloan Foundation, 
the Participating Institutions, the National Science Foundation, the U.S. Department of Energy, 
the National Aeronautics and Space Administration, the Japanese Monbukagakusho, 
the Max Planck Society, and the Higher Education Funding Council for England. 
The SDSS Web Site is http://www.sdss.org/.

The SDSS is managed by the Astrophysical Research Consortium for the Participating Institutions. 
The Participating Institutions are the American Museum of Natural History, Astrophysical Institute Potsdam, 
University of Basel, University of Cambridge, Case Western Reserve University, University of Chicago, Drexel University, Fermilab, 
the Institute for Advanced Study, the Japan Participation Group, Johns Hopkins University, the Joint Institute for Nuclear Astrophysics, 
the Kavli Institute for Particle Astrophysics and Cosmology, the Korean Scientist Group, the Chinese Academy of Sciences (LAMOST), 
Los Alamos National Laboratory, the Max-Planck-Institute for Astronomy (MPIA), the Max-Planck-Institute for Astrophysics (MPA), 
New Mexico State University, Ohio State University, University of Pittsburgh, University of Portsmouth, 
Princeton University, the United States Naval Observatory, and the University of Washington.

This publication makes use of data products from the Wide-field Infrared Survey Explorer, which is a joint project of the University of California, Los Angeles, and the Jet Propulsion Laboratory/California Institute of Technology, and NEOWISE, which is a project of the Jet Propulsion Laboratory/California Institute of Technology. WISE and NEOWISE are funded by the National Aeronautics and Space Administration.

Plot colours were in part based on www.ColorBrewer.org, 
by Cynthia A. Brewer, Penn State.
The TOPCAT tool \citep{2005ASPC..347...29T} was used during this work.

\appendix

\section{Catalogue Column Description}\label{columns}

\noindent
The results from the spectral fitting code have been compiled into a single dataset 
published as an SDSS DR14 value added catalogue.
A description of the columns included in the catalogue is given below.
Catalogue entries that are either undetermined or not relevant 
for a given source are set to ``-99''.
Columns that provide information on the X-ray detections 
have been taken from \citet{2008A&A...480..611S} and \citet{2016A&A...588A.103B}.

\begin{enumerate}
\item xray\_detection: Flag indicating whether the X-ray source was detected in the 2RXS or XMMSL1 survey. \\
\item name: IAU name of the X-ray detection.\\
\item RA: Right ascension of the X-ray detection (J2000) [degrees].\\
\item DEC: Declination of the X-ray detection (J2000) [degrees].\\
\item ExiML: Existence likelihood of the 2RXS detection.\\
\item ExpTime\_2RXS: Exposure time (2RXS) [s].\\
\item DET\_ML: Detection likelihood of the XMMSL1 detection in the 0.2-12\,keV range.\\
\item ExpTime\_XMMSL: Exposure time (XMMSL1) [s].\\

\item f\_2RXS: Flux in the 0.1-2.4\,keV range (2RXS) [$\rm erg\,cm^{-2}\,s^{-1}$].\\
\item errf\_2RXS: Uncertainty in the flux in the 0.1-2.4\,keV range (2RXS) [$\rm erg\,cm^{-2}\,s^{-1}$].\\
\item fden\_2RXS: Flux density at 2\,keV (2RXS) [$\rm erg\,cm^{-2}\,s^{-1}\,Hz^{-1}$].\\
\item errfden\_2RXS: Uncertainty in the flux density at 2\,keV (2RXS) [$\rm erg\,cm^{-2}\,s^{-1}\,Hz^{-1}$].\\

\item l\_2RXS: Luminosity in the 0.1-2.4\,keV range (2RXS) [$\rm erg\,s^{-1}$].\\
\item errl\_2RXS: Uncertainty in the luminosity in the 0.1-2.4\,keV range (2RXS) [$\rm erg\,s^{-1}$].\\
\item l2keV\_2RXS: Monochromatic luminosity at 2\,keV (2RXS) [$\rm erg\,s^{-1}\,Hz^{-1}$].\\
\item errl2keV\_2RXS: Uncertainty in the monochromatic luminosity at 2\,keV (2RXS) [$\rm erg\,s^{-1}\,Hz^{-1}$].\\

\item f\_XMMSL: Flux in the 0.2-12\,keV range (XMMSL1; from \citet{2008A&A...480..611S}) [$\rm erg\,cm^{-2}\,s^{-1}$].\\
\item errf\_XMMSL: Uncertainty in the flux in the 0.2-12\,keV range (XMMSL1; from \citet{2008A&A...480..611S}) [$\rm erg\,cm^{-2}\,s^{-1}$].\\
\item l\_XMMSL: Luminosity in the 0.2-12\,keV range (XMMSL1; from \citet{2008A&A...480..611S}) [$\rm erg\,s^{-1}$].\\
\item errl\_XMMSL: Uncertainty in the luminosity in the 0.2-12\,keV range (XMMSL1; from \citet{2008A&A...480..611S}) [$\rm erg\,s^{-1}$].\\

\item Plate: SDSS plate number.\\
\item MJD: MJD that the SDSS spectrum was taken.\\
\item FiberID: SDSS fiber identification.\\
\item DR14\_RUN2D: Spectroscopic reprocessing number.\\
\item DR14\_PLUGRA: Right ascension of the drilled fiber position [degrees].\\
\item DR14\_PLUGDEC: Declination of the drilled fiber position [degrees].\\
\item redshift: Source redshift based on the visual inspection results.\\
\item CLASS\_BEST: Source classification based on the visual inspection results.\\
\item CONF\_BEST: Visual inspection redshift and classification confidence flag.\\
\item DR14\_ZWARNING: Warning flag for SDSS spectra.\\
\item DR14\_SNMEDIANALL: Median S/N ratio per pixel of the spectrum.\\
\item Instrument: Flag indicating which spectrograph was used 
(SDSS or BOSS) to measure the spectrum. \\

\item norm1\_mgII: Normalisation of the first Gaussian used to fit the MgII line [$\rm 10^{-17}\,erg\,cm^{-2}\,s^{-1}\,\AA^{-1}$].\\
\item errnorm1\_mgII: Uncertainty in the normalisation of the first Gaussian used to fit the MgII line [$\rm 10^{-17}\,erg\,cm^{-2}\,s^{-1}\,\AA^{-1}$].\\
\item peak1\_mgII: Wavelength of the peak of the first Gaussian used to fit the MgII line [$\AA$].\\ 
\item errpeak1\_mgII: Uncertainty in the wavelength of the peak of the first Gaussian used to fit the MgII line [$\AA$].\\ 
\item width1\_mgII: Width of the first Gaussian used to fit the MgII line [\AA].\\
\item errwidth1\_mgII: Uncertainty in the width of the first Gaussian used to fit the MgII line [\AA].\\
\item fwhm1\_mgII: FWHM of the first Gaussian used to fit the MgII line [$\rm km\,s^{-1}$].\\
\item errfwhm1\_mgII: Uncertainty in the FWHM of the first Gaussian used to fit the MgII line [$\rm km\,s^{-1}$].\\
\item shift1\_mgII: Wavelength shift of the peak of the first Gaussian used to fit the MgII line relative to the rest-frame wavelength [$\AA$].\\

\item norm2\_mgII: Normalisation of the second Gaussian used to fit the MgII line [$\rm 10^{-17}\,erg\,cm^{-2}\,s^{-1}\,\AA^{-1}$].\\
\item errnorm2\_mgII: Uncertainty in the normalisation of the second Gaussian used to fit the MgII line [$\rm 10^{-17}\,erg\,cm^{-2}\,s^{-1}\,\AA^{-1}$].\\
\item peak2\_mgII: Wavelength of the peak of the second Gaussian used to fit the MgII line [$\AA$].\\ 
\item errpeak2\_mgII: Uncertainty in the wavelength of the peak of the second Gaussian used to fit the MgII line [$\AA$].\\ 
\item width2\_mgII: Width of the second Gaussian used to fit the MgII line [\AA].\\
\item errwidth2\_mgII: Uncertainty in the width of the second Gaussian used to fit the MgII line [\AA].\\
\item fwhm2\_mgII: FWHM of the second Gaussian used to fit the MgII line [$\rm km\,s^{-1}$].\\
\item errfwhm2\_mgII: Uncertainty in the FWHM of the second Gaussian used to fit the MgII line [$\rm km\,s^{-1}$].\\
\item shift2\_mgII: Wavelength shift of the peak of the second Gaussian used to fit the MgII line relative to the rest-frame wavelength [$\AA$].\\

\item norm3\_mgII: Normalisation of the third Gaussian used to fit the MgII line [$\rm 10^{-17}\,erg\,cm^{-2}\,s^{-1}\,\AA^{-1}$].\\
\item errnorm3\_mgII: Uncertainty in the normalisation of the third Gaussian used to fit the MgII line [$\rm 10^{-17}\,erg\,cm^{-2}\,s^{-1}\,\AA^{-1}$].\\
\item peak3\_mgII: Wavelength of the peak of the third Gaussian used to fit the MgII line [$\AA$].\\ 
\item errpeak3\_mgII: Uncertainty in the wavelength of the peak of the third Gaussian used to fit the MgII line [$\AA$].\\ 
\item width3\_mgII: Width of the third Gaussian used to fit the MgII line [\AA].\\
\item errwidth3\_mgII: Uncertainty in the width of the third Gaussian used to fit the MgII line [\AA].\\
\item fwhm3\_mgII: FWHM of the third Gaussian used to fit the MgII line [$\rm km\,s^{-1}$].\\
\item errfwhm3\_mgII: Uncertainty in the FWHM of the third Gaussian used to fit the MgII line [$\rm km\,s^{-1}$].\\
\item shift3\_mgII: Wavelength shift of the peak of the third Gaussian used to fit the MgII line relative to the rest-frame wavelength [$\AA$].\\

\item norm\_heII: Normalisation of the Gaussian used to fit the HeII line [$\rm 10^{-17}\,erg\,cm^{-2}\,s^{-1}\,\AA^{-1}$].\\
\item errnorm\_heII: Uncertainty in the normalisation of the Gaussian used to fit the HeII line [$\rm 10^{-17}\,erg\,cm^{-2}\,s^{-1}\,\AA^{-1}$].\\
\item peak\_heII: Wavelength of the peak of the Gaussian used to fit the HeII line [$\AA$].\\ 
\item errpeak\_heII: Uncertainty in the wavelength of the peak of the Gaussian used to fit the HeII line [$\AA$].\\ 
\item width\_heII: Width of the Gaussian used to fit the HeII line [\AA].\\
\item errwidth\_heII: Uncertainty in the width of the Gaussian used to fit the HeII line [\AA].\\
\item fwhm\_heII: FWHM of the Gaussian used to fit the HeII line [$\rm km\,s^{-1}$].\\
\item errfwhm\_heII: Uncertainty in the FWHM of the Gaussian used to fit the HeII line [$\rm km\,s^{-1}$].\\
\item shift\_heII: Wavelength shift of the peak of the Gaussian used to fit the HeII line relative to the rest-frame wavelength [$\AA$].\\

\item norm1\_hb: Normalisation of the first Gaussian used to fit the $\rm H\beta$ line [$\rm 10^{-17}\,erg\,cm^{-2}\,s^{-1}\,\AA^{-1}$].\\
\item errnorm1\_hb: Uncertainty in the normalisation of the first Gaussian used to fit the $\rm H\beta$ line [$\rm 10^{-17}\,erg\,cm^{-2}\,s^{-1}\,\AA^{-1}$].\\
\item peak1\_hb: Wavelength of the peak of the first Gaussian used to fit the $\rm H\beta$ line [$\AA$].\\ 
\item errpeak1\_hb: Uncertainty in the wavelength of the peak of the first Gaussian used to fit the $\rm H\beta$ line [$\AA$].\\ 
\item width1\_hb: Width of the first Gaussian used to fit the $\rm H\beta$ line [\AA].\\
\item errwidth1\_hb: Uncertainty in the width of the first Gaussian used to fit the $\rm H\beta$ line [\AA].\\
\item fwhm1\_hb: FWHM of the first Gaussian used to fit the $\rm H\beta$ line [$\rm km\,s^{-1}$].\\
\item errfwhm1\_hb: Uncertainty in the FWHM of the first Gaussian used to fit the $\rm H\beta$ line [$\rm km\,s^{-1}$].\\
\item shift1\_hb: Wavelength shift of the peak of the first Gaussian used to fit the $\rm H\beta$ line relative to the rest-frame wavelength [$\AA$].\\

\item norm2\_hb: Normalisation of the second Gaussian used to fit the $\rm H\beta$ line [$\rm 10^{-17}\,erg\,cm^{-2}\,s^{-1}\,\AA^{-1}$].\\
\item errnorm2\_hb: Uncertainty in the normalisation of the second Gaussian used to fit the $\rm H\beta$ line [$\rm 10^{-17}\,erg\,cm^{-2}\,s^{-1}\,\AA^{-1}$].\\
\item peak2\_hb: Wavelength of the peak of the second Gaussian used to fit the $\rm H\beta$ line [$\AA$].\\ 
\item errpeak2\_hb: Uncertainty in the wavelength of the peak of the second Gaussian used to fit the $\rm H\beta$ line [$\AA$].\\ 
\item width2\_hb: Width of the second Gaussian used to fit the $\rm H\beta$ line [\AA].\\
\item errwidth2\_hb: Uncertainty in the width of the second Gaussian used to fit the $\rm H\beta$ line [\AA].\\
\item fwhm2\_hb: FWHM of the second Gaussian used to fit the $\rm H\beta$ line [$\rm km\,s^{-1}$].\\
\item errfwhm2\_hb: Uncertainty in the FWHM of the second Gaussian used to fit the $\rm H\beta$ line [$\rm km\,s^{-1}$].\\
\item shift2\_hb: Wavelength shift of the peak of the second Gaussian used to fit the $\rm H\beta$ line relative to the rest-frame wavelength [$\AA$].\\

\item norm3\_hb: Normalisation of the third Gaussian used to fit the $\rm H\beta$ line [$\rm 10^{-17}\,erg\,cm^{-2}\,s^{-1}\,\AA^{-1}$].\\
\item errnorm3\_hb: Uncertainty in the normalisation of the third Gaussian used to fit the $\rm H\beta$ line [$\rm 10^{-17}\,erg\,cm^{-2}\,s^{-1}\,\AA^{-1}$].\\
\item peak3\_hb: Wavelength of the peak of the third Gaussian used to fit the $\rm H\beta$ line [$\AA$].\\ 
\item errpeak3\_hb: Uncertainty in the wavelength of the peak of the third Gaussian used to fit the $\rm H\beta$ line [$\AA$].\\ 
\item width3\_hb: Width of the third Gaussian used to fit the $\rm H\beta$ line [\AA].\\
\item errwidth3\_hb: Uncertainty in the width of the third Gaussian used to fit the $\rm H\beta$ line [\AA].\\
\item fwhm3\_hb: FWHM of the third Gaussian used to fit the $\rm H\beta$ line [$\rm km\,s^{-1}$].\\
\item errfwhm3\_hb: Uncertainty in the FWHM of the third Gaussian used to fit the $\rm H\beta$ line [$\rm km\,s^{-1}$].\\
\item shift3\_hb: Wavelength shift of the peak of the third Gaussian used to fit the $\rm H\beta$ line relative to the rest-frame wavelength [$\AA$].\\

\item norm4\_hb: Normalisation of the fourth Gaussian used to fit the $\rm H\beta$ line [$\rm 10^{-17}\,erg\,cm^{-2}\,s^{-1}\,\AA^{-1}$].\\
\item errnorm4\_hb: Uncertainty in the normalisation of the fourth Gaussian used to fit the $\rm H\beta$ line [$\rm 10^{-17}\,erg\,cm^{-2}\,s^{-1}\,\AA^{-1}$].\\
\item peak4\_hb: Wavelength of the peak of the fourth Gaussian used to fit the $\rm H\beta$ line [$\AA$].\\ 
\item errpeak4\_hb: Uncertainty in the wavelength of the peak of the fourth Gaussian used to fit the $\rm H\beta$ line [$\AA$].\\ 
\item width4\_hb: Width of the fourth Gaussian used to fit the $\rm H\beta$ line [\AA].\\
\item errwidth4\_hb: Uncertainty in the width of the fourth Gaussian used to fit the $\rm H\beta$ line [\AA].\\
\item fwhm4\_hb: FWHM of the fourth Gaussian used to fit the $\rm H\beta$ line [$\rm km\,s^{-1}$].\\
\item errfwhm4\_hb: Uncertainty in the FWHM of the fourth Gaussian used to fit the $\rm H\beta$ line [$\rm km\,s^{-1}$].\\
\item shift4\_hb: Wavelength shift of the peak of the fourth Gaussian used to fit the $\rm H\beta$ line relative to the rest-frame wavelength [$\AA$].\\
 
\item norm1\_OIII4959: Normalisation of first Gaussian used to fit the $\rm [OIII]4959$ line [$\rm 10^{-17}\,erg\,cm^{-2}\,s^{-1}\,\AA^{-1}$].\\
\item errnorm1\_OIII4959: Uncertainty in the normalisation of first Gaussian used to fit the $\rm [OIII]4959$ line [$\rm 10^{-17}\,erg\,cm^{-2}\,s^{-1}\,\AA^{-1}$].\\
\item peak1\_OIII4959: Wavelength of the peak of the first Gaussian used to fit the $\rm [OIII]4959$ line [$\AA$].\\ 
\item errpeak1\_OIII4959: Uncertainty in the wavelength of the peak of the first Gaussian used to fit the $\rm [OIII]4959$ line [$\AA$].\\ 
\item width1\_OIII4959: Width of the first Gaussian used to fit the $\rm [OIII]4959$ line [\AA].\\
\item errwidth1\_OIII4959: Uncertainty in the width of the first Gaussian used to fit the $\rm [OIII]4959$ line [\AA].\\
\item fwhm1\_OIII4959: FWHM of the first Gaussian used to fit the $\rm [OIII]4959$ line [$\rm km\,s^{-1}$].\\
\item errfwhm1\_OIII4959: Uncertainty in the FWHM of the first Gaussian used to fit the $\rm [OIII]4959$ line [$\rm km\,s^{-1}$].\\
\item shift1\_OIII4959: Wavelength shift of the peak of the first Gaussian used to fit the $\rm [OIII]4959$ line relative to the rest-frame wavelength [$\AA$].\\

\item norm2\_OIII4959: Normalisation of second Gaussian used to fit the $\rm [OIII]4959$ line [$\rm 10^{-17}\,erg\,cm^{-2}\,s^{-1}\,\AA^{-1}$].\\
\item errnorm2\_OIII4959: Uncertainty in the normalisation of second Gaussian used to fit the $\rm [OIII]4959$ line [$\rm 10^{-17}\,erg\,cm^{-2}\,s^{-1}\,\AA^{-1}$].\\
\item peak2\_OIII4959: Wavelength of the peak of the second Gaussian used to fit the $\rm [OIII]4959$ line [$\AA$].\\ 
\item errpeak2\_OIII4959: Uncertainty in the wavelength of the peak of the second Gaussian used to fit the $\rm [OIII]4959$ line [$\AA$].\\ 
\item width2\_OIII4959: Width of the second Gaussian used to fit the $\rm [OIII]4959$ line [\AA].\\
\item errwidth2\_OIII4959: Uncertainty in the width of the second Gaussian used to fit the $\rm [OIII]4959$ line [\AA].\\
\item fwhm2\_OIII4959: FWHM of the second Gaussian used to fit the $\rm [OIII]4959$ line [$\rm km\,s^{-1}$].\\
\item errfwhm2\_OIII4959: Uncertainty in the FWHM of the second Gaussian used to fit the $\rm [OIII]4959$ line [$\rm km\,s^{-1}$].\\
\item shift2\_OIII4959: Wavelength shift of the peak of the second Gaussian used to fit the $\rm [OIII]4959$ line relative to the rest-frame wavelength [$\AA$].\\

\item norm1\_OIII5007: Normalisation of first Gaussian used to fit the $\rm [OIII]5007$ line [$\rm 10^{-17}\,erg\,cm^{-2}\,s^{-1}\,\AA^{-1}$].\\
\item errnorm1\_OIII5007: Uncertainty in the normalisation of first Gaussian used to fit the $\rm [OIII]5007$ line [$\rm 10^{-17}\,erg\,cm^{-2}\,s^{-1}\,\AA^{-1}$].\\
\item peak1\_OIII5007: Wavelength of the peak of the first Gaussian used to fit the $\rm [OIII]5007$ line [$\AA$].\\ 
\item errpeak1\_OIII5007: Uncertainty in the wavelength of the peak of the first Gaussian used to fit the $\rm [OIII]5007$ line [$\AA$].\\ 
\item width1\_OIII5007: Width of the first Gaussian used to fit the $\rm [OIII]5007$ line [\AA].\\
\item errwidth1\_OIII5007: Uncertainty in the width of the first Gaussian used to fit the $\rm [OIII]5007$ line [\AA].\\
\item fwhm1\_OIII5007: FWHM of the first Gaussian used to fit the $\rm [OIII]5007$ line [$\rm km\,s^{-1}$].\\
\item errfwhm1\_OIII5007: Uncertainty in the FWHM of the first Gaussian used to fit the $\rm [OIII]5007$ line [$\rm km\,s^{-1}$].\\
\item shift1\_OIII5007: Wavelength shift of the peak of the first Gaussian used to fit the $\rm [OIII]5007$ line relative to the rest-frame wavelength [$\AA$].\\

\item norm2\_OIII5007: Normalisation of second Gaussian used to fit the $\rm [OIII]5007$ line [$\rm 10^{-17}\,erg\,cm^{-2}\,s^{-1}\,\AA^{-1}$].\\
\item errnorm2\_OIII5007: Uncertainty in the normalisation of second Gaussian used to fit the $\rm [OIII]5007$ line [$\rm 10^{-17}\,erg\,cm^{-2}\,s^{-1}\,\AA^{-1}$].\\
\item peak2\_OIII5007: Wavelength of the peak of the second Gaussian used to fit the $\rm [OIII]5007$ line [$\AA$].\\ 
\item errpeak2\_OIII5007: Uncertainty in the wavelength of the peak of the second Gaussian used to fit the $\rm [OIII]5007$ line [$\AA$].\\ 
\item width2\_OIII5007: Width of the second Gaussian used to fit the $\rm [OIII]5007$ line [\AA].\\
\item errwidth2\_OIII5007: Uncertainty in the width of the second Gaussian used to fit the $\rm [OIII]5007$ line [\AA].\\
\item fwhm2\_OIII5007: FWHM of the second Gaussian used to fit the $\rm [OIII]5007$ line [$\rm km\,s^{-1}$].\\
\item errfwhm2\_OIII5007: Uncertainty in the FWHM of the second Gaussian used to fit the $\rm [OIII]5007$ line [$\rm km\,s^{-1}$].\\
\item shift2\_OIII5007: Wavelength shift of the peak of the second Gaussian used to fit the $\rm [OIII]5007$ line relative to the rest-frame wavelength [$\AA$].\\

\item norm\_pl1: Normalisation of the power law fit to the MgII continuum region [$\rm 10^{-17}\,erg\,cm^{-2}\,s^{-1}\,\AA^{-1}$].\\
\item errnorm\_pl1: Uncertainty in the normalisation of the power law fit to the MgII continuum region [$\rm 10^{-17}\,erg\,cm^{-2}\,s^{-1}\,\AA^{-1}$].\\
\item slope\_pl1: Slope of the power law fit to the MgII continuum region.\\
\item errslope\_pl1: Uncertainty in the slope of the power law fit to the MgII continuum region.\\
\item norm\_pl2: Normalisation of the power law fit to the H$\beta$ continuum region [$\rm 10^{-17}\,erg\,cm^{-2}\,s^{-1}\,\AA^{-1}$].\\
\item errnorm\_pl2: Uncertainty in the normalisation of the power law fit to the H$\beta$ continuum region [$\rm 10^{-17}\,erg\,cm^{-2}\,s^{-1}\,\AA^{-1}$].\\
\item slope\_pl2: Slope of the power law fit to the H$\beta$ continuum region.\\
\item errslope\_pl2: Uncertainty in the slope of the power law fit to the H$\beta$ continuum region.\\

\item norm\_gal1: Normalisation of the galaxy template used to fit the MgII continuum region [$\rm 10^{-17}\,erg\,cm^{-2}\,s^{-1}\,\AA^{-1}$].\\
\item errnorm\_gal1: Uncertainty in the normalisation of the galaxy template used to fit the MgII continuum region [$\rm 10^{-17}\,erg\,cm^{-2}\,s^{-1}\,\AA^{-1}$].\\
\item norm\_gal2: Normalisation of the galaxy template used to fit the H$\beta$ continuum region [$\rm 10^{-17}\,erg\,cm^{-2}\,s^{-1}\,\AA^{-1}$].\\
\item errnorm\_gal2: Uncertainty in the normalisation of the galaxy template used to fit the H$\beta$ continuum region [$\rm 10^{-17}\,erg\,cm^{-2}\,s^{-1}\,\AA^{-1}$].\\

\item norm\_feII1: Normalisation of the iron template used to fit the MgII continuum region [$\rm 10^{-17}\,erg\,cm^{-2}\,s^{-1}\,\AA^{-1}$].\\
\item errnorm\_feII1: Uncertainty in the normalisation of the iron template used to fit the MgII continuum region [$\rm 10^{-17}\,erg\,cm^{-2}\,s^{-1}\,\AA^{-1}$].\\
\item norm\_feII2: Normalisation of the iron template used to fit the H$\beta$ continuum region [$\rm 10^{-17}\,erg\,cm^{-2}\,s^{-1}\,\AA^{-1}$].\\
\item errnorm\_feII2: Uncertainty in the normalisation of the iron template used to fit the H$\beta$ continuum region [$\rm 10^{-17}\,erg\,cm^{-2}\,s^{-1}\,\AA^{-1}$].\\
\item fwhm\_feII1: FWHM of the Gaussian kernel convolved with the iron template used to fit the MgII continuum region [$\rm \AA$].\\
\item errfwhm\_feII1: Uncertainty in the FWHM of the Gaussian kernel convolved with the iron template used to fit the MgII continuum region [$\rm \AA$].\\
\item fwhm\_feII2: FWHM of the Gaussian kernel convolved with the iron template used to fit the H$\beta$ continuum region [$\rm \AA$].\\
\item errfwhm\_feII2: Uncertainty in the FWHM of the Gaussian kernel convolved with the iron template used to fit the H$\beta$ continuum region [$\rm \AA$].\\
\item r\_feII: Flux ratio of the 4434-4684$\AA$ FeII emission to the broad component of H$\beta$. \\
\item OIII\_Hbeta\_ratio: Flux ratio of [OIII]5007$\AA$ to H$\beta$.\\

\item virialfwhm\_mgII: FWHM of the MgII broad line profile [$\rm km\,s^{-1}$].\\
\item errvirialfwhm\_mgII: Uncertainty in the FWHM of the MgII broad line profile [$\rm km\,s^{-1}$].\\
\item virialfwhm\_hb: FWHM of the $\rm H \beta$ broad line profile [$\rm km\,s^{-1}$].\\
\item errvirialfwhm\_hb: Uncertainty in the FWHM of the $\rm H \beta$ broad line profile [$\rm km\,s^{-1}$].\\
\item mgII\_chi: Reduced chi-squared ($\rm \chi_{\nu}^{2}$) of the fit to the MgII region.\\
\item hb\_chi: Reduced chi-squared ($\rm \chi_{\nu}^{2}$) of the fit to the $\rm H\beta$ region.\\

\item l\_2500: Monochromatic luminosity at 2500$\AA$ [$\rm erg\,s^{-1}\,\AA^{-1}$].\\
\item errl\_2500: Uncertainty in the monochromatic luminosity at 2500$\AA$ [$\rm erg\,s^{-1}\,\AA^{-1}$].\\
\item l\_3000: Monochromatic luminosity at 3000$\AA$ [$\rm erg\,s^{-1}\,\AA^{-1}$].\\
\item errl\_3000: Uncertainty in the monochromatic luminosity at 3000$\AA$ [$\rm erg\,s^{-1}\,\AA^{-1}$].\\
\item l\_5100: Monochromatic luminosity at 5100$\AA$ [$\rm erg\,s^{-1}\,\AA^{-1}$].\\
\item errl\_5100: Uncertainty in the monochromatic luminosity at 5100$\AA$ [$\rm erg\,s^{-1}\,\AA^{-1}$].\\
\item l\_bol1: Bolometric luminosity derived from the monochromatic luminosity at 3000$\AA$ [$\rm erg\,s^{-1}$].\\
\item errl\_bol1: Uncertainty in the bolometric luminosity derived from the monochromatic luminosity at 3000$\AA$ [$\rm erg\,s^{-1}$].\\
\item l\_bol2: Bolometric luminosity derived from the monochromatic luminosity at 5100$\AA$ [$\rm erg\,s^{-1}$].\\
\item errl\_bol2: Uncertainty in the bolometric luminosity derived from the monochromatic luminosity at 5100$\AA$ [$\rm erg\,s^{-1}$].\\

\item logBHMVP\_hb: BH mass derived from the H$\beta$ line using the \citet{2006ApJ...641..689V} calibration [$\rm Log_{10}(M_{\odot})$].\\
\item errlogBHMVP\_hb: Uncertainty in the BH mass derived from the H$\beta$ line using the \citet{2006ApJ...641..689V} calibration [$\rm Log_{10}(M_{\odot})$].\\
\item logBHMA\_hb: BH mass derived from the H$\beta$ line using the \citet{2011ApJ...742...93A} calibration [$\rm Log_{10}(M_{\odot})$].\\
\item errlogBHMA\_hb: Uncertainty in the BH mass derived from the H$\beta$ line using the \citet{2011ApJ...742...93A} calibration [$\rm Log_{10}(M_{\odot})$].\\
\item logBHMS\_mgII: BH mass derived from the MgII line using the \citet{2012ApJ...753..125S} calibration [$\rm Log_{10}(M_{\odot})$].\\
\item errlogBHMS\_mgII: Uncertainty in the BH mass derived from the MgII line using the \citet{2012ApJ...753..125S} calibration [$\rm Log_{10}(M_{\odot})$].\\
\item l\_edd1: Eddington luminosity based on the BH mass estimate derived using the \citet{2012ApJ...753..125S} calibration [$\rm erg\,s^{-1}$].\\
\item l\_edd2: Eddington luminosity based on the BH mass estimate derived using the \citet{2011ApJ...742...93A} calibration [$\rm erg\,s^{-1}$].\\
\item edd\_ratio1: Eddington ratio defined as l\_bol1/l\_edd1.\\
\item edd\_ratio2: Eddington ratio defined as l\_bol2/l\_edd2.\\
\item flag\_abs: Flag indicating whether or not strong absorption lines 
have been observed in the spectrum. 
flag\_abs is set to either 0 (no absorption present) or 1 (absorption present).\\
\end{enumerate}

\label{lastpage}

\bibliographystyle{aa} 
\bibliography{biblio} 

\end{document}